\documentclass[twocolumn,tighten]{aastex631}

\usepackage{color,comment,multirow,booktabs,upgreek}
\usepackage{multirow,enumitem}
\usepackage[flushleft]{threeparttable}
\usepackage{amsmath}
\usepackage{CJK}
\graphicspath{{./}{images/}}

\newcommand{\cm}{{\rm cm}}
\newcommand{\pcc}{{\rm cm}^{-3}}
\newcommand {\kms}{\ifmmode{\rm km \, s^{-1}}\else{$\rm km \, s^{-1}$}\fi} 
\newcommand {\mo}{{\rm M}_\odot}

\newcommand{\tworef}[2]{\autoref{#1}-\ref{#2}}

\defcitealias{Armillotta+21}{Paper I}
\defcitealias{Armillotta+22}{Paper II}

\begin{document}
\begin{CJK*}{UTF8}{gbsn}

\title{Cosmic-Ray Acceleration of Galactic Outflows in Multiphase Gas}

\correspondingauthor{Lucia Armillotta}
\email{lucia.armillotta@princeton.edu}

\author[0000-0002-5708-1927]{Lucia Armillotta}
\affiliation{Department of Astrophysical Sciences, Princeton University, Princeton, NJ 08544, USA}

\author[0000-0002-0509-9113]{Eve C. Ostriker}
\affiliation{Department of Astrophysical Sciences, Princeton University, Princeton, NJ 08544, USA}

\author[0000-0003-2896-3725]{Chang-Goo Kim}
\affiliation{Department of Astrophysical Sciences, Princeton University, Princeton, NJ 08544, USA}

\author[0000-0002-2624-3399]{Yan-Fei Jiang(姜燕飞)}
\affiliation{Center for Computational Astrophysics, Flatiron Institute,
New York, NY 10010, USA}

\begin{abstract}
We investigate the dynamical interaction between cosmic rays (CRs) and the multiphase interstellar medium (ISM) using numerical magnetohydrodynamic (MHD) simulations with a two-moment CR solver and TIGRESS simulations of star-forming galactic disks.  We previously studied transport of CRs within TIGRESS outputs using a ``post-processing'' approach, and we now assess the effects of the MHD backreaction to CR pressure.  We confirm our previous conclusion that there are three quite different regimes of CR transport in multiphase ISM gas, while also finding that simulations with ``live MHD'' predict a smoother CR pressure distribution.  The CR pressure near the midplane is comparable to other pressure components in the gas, but the scale height of CRs is far larger. Next, with a goal of understanding the role of CRs in driving galactic outflows, we conduct a set of controlled simulations of the extraplanar region above $z=500$ pc, with imposed boundary conditions flowing from the midplane into this region. We explore a range of thermal and kinematic properties for the injected thermal gas, encompassing both hot, fast-moving outflows, and cooler, slower-moving outflows. The boundary conditions for CR energy density and flux are scaled from the supernova rate in the underlying TIGRESS model.  Our simulations reveal that CRs efficiently accelerate extra-planar material if the latter is mostly warm/warm-hot gas, in which CRs stream at the Alfv\'en speed and the effective sound speed increases as density decreases.  In contrast,  CRs have very little effect on fast, hot outflows where the Alfv\'en speed is small, even when the injected CR momentum flux exceeds the injected MHD momentum flux.   
\end{abstract}

\vspace*{1cm}

\section{Introduction}
\label{sec:introduction}

Cosmic rays (CRs) are charged particles moving with relativistic speeds. While their origins are believed to be localized to supernova remnants where they are accelerated in shocks  \citep[e.g.,][]{Bell04, Morlino&Caprioli12, Blasi13}, CRs spread throughout the interstellar medium (ISM) thanks to their quasi-collisionless nature. Direct observations of CRs in the solar system indicate that their spectrum extends over more than ten orders of magnitude in kinetic energy, from at least $10^6$~eV to $10^{20}$~eV. For the protons, which comprise most of the CR energy, the spectrum is well approximated by a broken power law that peaks at kinetic energies near 1 GeV \citep[e.g.,][]{Grenier+15, Cummings+16}. In the Milky Way, the energy density of CRs in the local ISM is comparable to the thermal, turbulent, and magnetic energy densities \citep[e.g.,][]{Boulares&Cox90,Beck01}, suggesting that CRs can significantly contribute to the dynamics of gas in the ISM.

Because of their potential dynamical importance, CRs are thought to be a possible driving mechanism of galactic winds \citep[see reviews by][]{Hanasz+21, Recchia21}. Large-scale outflows extending up to several kpc from the host galaxy have been commonly observed in star-forming galaxies \citep[see reviews by][]{Veilleux+05,Veilleux+20}, revealing a complex multiphase nature: they are composed of hot \citep[$T \sim 10^{6-8}$~K; e.g.,][]{Strickland&Heckman07, McQuinn+18, Hodges-Kluck+20}, warm ionized \citep[$T \sim 10^{4-5}$~K; e.g.,][]{Heckman+15, Chisholm+17, XuHeckman22,Marasco+23}, warm ($T\sim10^4$~K) and cold ($T\sim10^{1-3}$~K) atomic \citep[e.g.,][]{Rupke+05, Martin+05, Teng+13, Martini+18, Concas+19, Noon+23}, and cold molecular \citep[$T\sim10-100$~K; e.g.,][]{Bolatto+13, Leroy+15, DiTeodoro+19, DiTeodoro+20} gas.

Explaining the multiphase nature of galactic outflows is a longstanding theoretical issue \citep[see review by][]{Heckman+17}. While the hot component is interpreted as supernova-heated gas that accelerates under its own thermal pressure gradients out of the host galaxy, the presence of a high-velocity cooler component is still puzzling. One possible explanation is that the warm/cold outflowing gas consists of ISM material that is accelerated via ram pressure from the surrounding hot wind. A caveat for this mechanism is that the hot wind can rapidly destroy the embedded cool gas due to a combination of shocks and hydrodynamical instabilities \citep[e.g.,][]{Cooper+09, Scannapieco+15, Schneider+17, Zhang+17}. However, idealized simulations of cool clouds travelling thorough a hotter medium have shown that, under appropriate cloud and environmental conditions, cool gas can survive its journey by mixing with the surrounding hot gas: for sufficiently high gas density, mixing between the two gas phases reduces the cooling time of the hot gas, triggering its condensation \citep[e.g.,][]{Armillotta+16, Gronke+20, Sparre+20, Banda-Barragan+21}. \citet{Schneider+20} explored this possibility by means of high-resolution simulations of starburst dwarf galaxies, finding that mixing between cool and hot gas and subsequent cooling is an effective way of transferring momentum from the hot to the cool phase of the wind, accelerating the latter to very high velocity. At the same time as more massive cool clouds are accelerated by mixing, lower mass clouds can be destroyed and add to the mass flux of the hot wind \citep[see e.g.][and references therein]{Fielding2022}.

In addition to gaining momentum from the hot wind, cool gas can in principle be accelerated by interactions with CRs. The dynamical role of CRs in driving galactic winds has been investigated in both analytic models \citep[][]{Dorfi&Breitschwerdt12, Recchia+16, Mao&Ostriker18, Quataert+21a, Quataert+21b} and numerical simulations of isolated galaxies or cosmological zoom-ins \citep[e.g.,][]{Uhlig+12, Salem&Bryan13, Pakmor+16, Ruszkowski+17, Chan+21, Girichidis+22, Peschken+23, Thosmas+23} and portions of ISM \citep[e.g.,][]{Girichidis+16, Simpson+16, Farber+18, Girichidis+18, Rathjen2021}. All these studies found that CR pressure gradients can drive galactic outflows; however, the efficiency of this process is strongly dependent on
the CR transport prescription adopted in the model because this affects the CR pressure gradient and therefore the momentum transfer to thermal gas.

Modeling CR transport on galactic scale is a challenging task due to the complex physical processes that couple CRs to the thermal gas, which are not yet fully understood \citep[see review by][]{Amato&Blasi18}. The interaction between CRs and thermal gas is primarily mediated by magnetic fields. CRs stream along magnetic field lines, while scattering off small-scale (of order the CR gyroradius) magnetic fluctuations. This scattering process couples CRs with the gas and sets the effective diffusive propagation speed of the CR distribution. Current understanding suggests that the origin of waves is energy dependent \citep{Blasi+12, Zweibel17, Evoli+18}, with CRs at moderate (GeV) and low (sub-GeV) energy scattered by Alfv\'{e}n waves excited by the CRs themselves via streaming instability (the self-confinement scenario; e.g., \citealt{Kulsrud&Pearce69, Wentzel74}), while high-energy (ultra-GeV) CRs are scattered by extrinsic turbulent fluctuations cascading down to small scales (the extrinsic turbulence scenario; e.g., \citealt{Chandran00, Yan&Lazarian02}). Nevertheless, to date these ideas have not yet been fully tested with realistic physical models of the underlying multiphase and magnetized ISM.

In most studies of ISM dynamics, the CR mean free path is much smaller than the spatial scales of interest, and it is reasonable to treat CRs as a fluid. Since the CR-wave interaction is not resolved on macroscopic scales, this interaction is treated in the transport equations via a scattering (or diffusion) coefficient. Most commonly, a constant diffusion coefficient which ignores the multiphase structure of the underlying gas is adopted, with a value that is motivated by observational constraints. Considering the vastly varying conditions within the multiphase ISM, however, a constant diffusion coefficient is unrealistic. To address this issue, we recently developed a more detailed physical prescription for the transport of the CR fluid, in which the scattering coefficient varies with the properties of the ambient gas, with a functional form motivated by the theory of self-confinement \citep[][hereafter \citetalias{Armillotta+21}]{Armillotta+21}.
Our model focuses on GeV CRs as they contain most of the energy and momentum of the CR population and are therefore more relevant for the gas dynamics. 

According to the self-confinement paradigm, a CR distribution with a bulk drift speed greater than the Alfv\`{e}n speed can excite Alfv\`{e}n waves through gyro-resonance, and individual CRs scatter off these waves as the CR distribution drifts in the direction of decreasing CR density. The processes of wave excitation through the streaming instability and isotropization via scattering off the waves so produced have been validated directly via MHD-PIC simulations \citep[e.g.,][]{Bai2019}. In principle, scattering by resonant Alfv\'{e}n waves can prevent the CR drift (or ``streaming'') speed from exceeding the local Alfv\'{e}n speed if wave amplitudes are sufficiently large. However, wave amplitudes and therefore scattering rates are reduced by wave damping, and this damping varies with the local properties of the ISM \citep[e.g.,][]{Kulsrud05, Plotnikov2021,Bambic2021}. 

In line with the self-confinement picture, our model treats CR fluid transport as a combination of advection by the thermal gas, streaming along the magnetic field at the local ion Alfv\`{e}n speed, and diffusion relative to the wave frame. The CR scattering coefficient is determined by the local balance between wave excitation and damping mediated by local gas properties (considering both ion-neutral damping and non-linear Landau damping, \citealt{Kulsrud&Cesarsky1971, Kulsrud05}). \citetalias{Armillotta+21} describes the incorporation of this prescription in the algorithm for CR transport implemented by \citet{Jiang&Oh18} in the \textsc{Athena}++ MHD code \citep{Stone+20}, as well as our model for ionization of warm and cold gas as set by CRs in the tens of MeV regime. In \citetalias{Armillotta+21} and in \citet[][hereafter \citetalias{Armillotta+22}]{Armillotta+22} we present the application of our model to computing the propagation of CRs in the TIGRESS MHD simulations modeling kpc-sized portions of star-forming galactic disks for a range of conditions \citep{Kim&Ostriker17, Kim+20}. The advantage of the TIGRESS simulations is that star formation and feedback (including both supernovae and photoelectric heating) are modeled in a self-consistent manner, thus providing a realistic representation of the multiphase star-forming ISM. Our work demonstrates that the transport of CRs is quite different in different thermal phases of the gas, with the CR scattering coefficient varying over more than four orders of magnitude depending on the properties of the underlying gas (e.g., density and ionization fraction). This challenges the common assumption of uniform scattering, and highlights the importance of an accurate representation of the multiphase ISM in CR transport modeling. 

One limitation of our previous studies is that they took a postprocessing approach instead of self-consistently computing the simultaneous evolution of thermal gas, magnetic fields, and CRs. This prevented us from examining the effects of the MHD backreaction to the CR pressure. In the present work, we move beyond the postprocessing approach and investigate the impact of CRs on multiphase ISM dynamics, with a particular focus on the  role of CRs in accelerating galactic outflows. We pursue our investigation in two steps. First, starting from the TIGRESS solar-neighborhood snapshots postprocessed for CR transport in \citetalias{Armillotta+22}, we perform new simulations in which MHD and CRs are evolved together. These simulations allow us to analyse how the transport and distribution of CRs in realistic multiphase gas are affected by ``live'' MHD, thereby testing the conclusions of our previous studies. 

In the second part of this paper, we perform MHD simulations of galactic outflows including CRs. For the MHD variables, the initial conditions are obtained from the original TIGRESS solar-neighborhood simulation outputs, with a domain extracted from the extra-planar region ($\vert z \vert > 500$~pc). During the simulation, thermal gas and CRs are injected as boundary conditions at the bottom of the simulation box. We explore different thermal and kinematic conditions of the injected gas, covering the typical properties of hot and fast-moving outflows, as well as cool and slow-moving outflows, both of which are expected to be driven by supernovae into the extra-planar region \citep{Kim&Ostriker18,Vijayan+20}.  
Our goal is to understand differences in the ways CR pressure gradients act to accelerate gas within different phases; the extra-planar regions are where this is expected to occur, since the CR scale height is larger than that of the gas. 

The paper is organized as follows. In \autoref{sec:equations}, we briefly summarize our numerical algorithms. In \autoref{sec:fullboxsim}, we present the results of the simulations with self-consistent MHD and CRs that use the TIGRESS snapshots postprocessed in \citetalias{Armillotta+22} as initial conditions. In \autoref{sec:windsim}, we describe the setup of the galactic outflow simulations; outcomes of these models for hot wind and warm winds are presented in \autoref{sec:hotwind}
and \autoref{sec:warmwind}, respectively. In \autoref{sec:discussion}, we discuss our work in relation to observational findings and other recent theoretical works. Finally, in \autoref{sec:conclusions}, we summarize our main results.

\section{MHD equations with CR transport}
\label{sec:equations}

In this work, we use the MHD code \textsc{Athena}++ \citep {Stone+20} including the two-moment algorithm for CR transport developed by \citet{Jiang&Oh18}. The full set of ideal MHD equations including CR transport is 

\begin{equation}
\begin{split}
\frac{\partial \rho}{\partial t}  + \nabla \cdot (\rho \mathbf{v}) =  0 \;,
\end{split}
\label{eq:masseq}    
\end{equation}

\begin{equation}
\begin{split}
& \frac{\partial (\rho \mathbf{v})}{\partial t} + \nabla \cdot \left (\rho \mathbf{v} \mathbf{v} + P_\mathrm{t} \tensor{\mathbf {I}} + \frac{B^2}{8\pi} \tensor{\mathbf {I}} - \frac{\mathbf{B} \mathbf{B}}{4 \pi} \right) \\ & =  - \rho \nabla \Phi +\tensor{\mathrm{\sigma}}_\mathrm{tot} \cdot [  \mathbf{F_\mathrm{c}} - \mathbf{v} \cdot (\tensor{{\mathbf{P}}}_\mathrm{c} + e_\mathrm{c} \tensor{\mathbf{I}}) ]\;,
\end{split}
\label{eq:momeq}    
\end{equation}

\begin{equation}
\begin{split}
& \frac{\partial e}{\partial t} + \nabla \cdot \left [ \left(e + P_\mathrm{t} + \frac{B^2}{8\pi}  \right) \mathbf{v} - \frac{\mathbf{B} (\mathbf{B} \cdot \mathbf{v})}{4\pi} \right] =  -\rho \mathcal{L}\\ 
- & \rho \mathbf{v} \cdot \nabla \Phi   + (\mathbf{v} + \mathbf{v_\mathrm{s}}) \cdot 
\tensor{\mathrm{\sigma}}_\mathrm{tot} \cdot   [  \mathbf{F_\mathrm{c}} - \mathbf{v} \cdot (\tensor{{\mathbf{P}}}_\mathrm{c} + e_\mathrm{c} \tensor{\mathbf{I}}) ] \;,
\end{split}
\label{eq:eneq}    
\end{equation}

\begin{equation}
\begin{split}
\frac{\partial \mathbf{B}}{\partial t}  - \nabla \times (\mathbf{v} \times \mathbf{B}) =  0 \;,
\end{split}
\label{eq:Beq}    
\end{equation}

\begin{equation}
\begin{split}
\frac{\partial e_\mathrm{c}}{\partial t} + \mathbf{\nabla} \cdot \mathbf{F_\mathrm{c}} =  - (\mathbf{v} \, + \, & \mathbf{v_\mathrm{s}} ) \cdot 
\tensor{\mathrm{\sigma}}_\mathrm{tot} \cdot   [  \mathbf{F_\mathrm{c}} - \mathbf{v} \cdot (\tensor{{\mathbf{P}}}_\mathrm{c} + e_\mathrm{c} \tensor{\mathbf{I}}) ] \\
& - \Lambda_\mathrm{coll} n_\mathrm{H} e_\mathrm{c}
\;,
\end{split}
\label{eq:CRenergy}
\end{equation}

\begin{equation}
\begin{split}
\frac{1}{v_\mathrm{m}^2} \frac{\partial \mathbf{F_\mathrm{c}}}{\partial t} + \mathbf{\nabla} \cdot \tensor{\mathbf{P}}_\mathrm{c} = & - \tensor{\mathrm{\sigma}}_\mathrm{tot} \cdot [  \mathbf{F_\mathrm{c}} - \mathbf{v} \cdot (\tensor{{\mathbf{P}}}_\mathrm{c} + e_\mathrm{c} \tensor{\mathbf{I}}) ] \\
& - \frac{\Lambda_\mathrm{coll} n_\mathrm{H}}{v_\mathrm{p}^2} \mathbf{F}_\mathrm{c} \;.
\label{eq:CRflux}
\end{split}
\end{equation}

Here, $\rho$ is the gas density, \textbf{v} is the gas velocity, $\mathbf{B}$ is the magnetic field, $e = (1/2) \rho v^2 + P_\mathrm{t}/(\gamma -1) + B^2/(8\pi)$ is the gas energy density with $P_\mathrm{t}$ the gas thermal pressure and $\gamma = 5/3$ the gas adiabatic index,  $e_\mathrm{c}$ is the CR energy density, $\mathbf{F}_\mathrm{c}$ is the CR energy flux, and $\tensor{{\mathbf{P}}}_\mathrm{c}$ is the CR pressure tensor. We assume approximately isotropic pressure, so that $\tensor{\mathbf{P}}_\mathrm{c} \equiv P_\mathrm{c}\tensor{\mathbf{I}}$, with $P_\mathrm{c} = (\gamma_\mathrm{c} -1) \,e_\mathrm{c} = e_\mathrm{c}/3$, where $\gamma_\mathrm{c} = 4/3$ is the adiabatic index of the relativistic fluid, and $\tensor{\mathbf{I}}$ is the identity tensor. In \autoref{eq:CRflux}, the speed $v_\mathrm{m}$ represents the maximum velocity for CRs propagation. In principle, $v_\mathrm{m}$ is equal to the speed of light $c$ for relativistic CRs. However, here we adopt the ``reduced speed of light'' approach \citep[see e.g.][for the corresponding two-moment radiation implementation]{SkinnerOstriker2013} with $v_\mathrm{m} = 10^4 \, \kms \ll c$, as it has been demonstrated that the result is not sensitive to the exact value of $v_\mathrm{m}$ as long as $v_\mathrm{m}$ is much larger than any other speed in the simulation \citep[][]{Jiang&Oh18}. This enables larger numerical timesteps based on the CFL condition for this set of hyperbolic equations.  

In \tworef{eq:momeq}{eq:eneq}, $\Phi$ is the ``external'' gravitational potential from the old stellar disk and dark matter halo \citep[][Equation~6]{Kim&Ostriker17}. Self-gravity is not included in the simulations performed for this paper. $\rho \mathcal{L} = n_\mathrm{H} (n_\mathrm{H} \Lambda(T) - \Gamma)$ is the net cooling function, where $n_\mathrm{H}$ is the hydrogen number density. The cooling coefficient $\Lambda(T)$ is computed using the fitting formula in \citet{Koyama&Inutsuka02} for $T<2\times10^{4}$, and the tabulated values in \citet{Sutherland&Dopita93} with solar metallicity for $T>2\times10^{4}$. For warm and cold gas we apply heating to represent the photoelectric effect on grains. The heating rate $\Gamma$ scales with the instantaneous far-ultraviolet luminosity from star particles. For reference solar-neighborhood values, we adopt a heating rate of $\Gamma_0 = 2 \times 10^{-26}$~erg~s$^{-1}$, and a mean FUV intensity of $4 \pi J_\mathrm{FUV,0} = 2.7 \times 10^{-3}$~erg~s$^{-1}$~cm$^{-2}$ (see \citealt{Kim+20} for further details).

In \autoref{eq:momeq} and \autoref{eq:CRflux}, the term $\tensor{\mathrm{\sigma}}_\mathrm{tot} \cdot [ \mathbf{F_\mathrm{c}} - (4/3) e_\mathrm{c} \mathbf{v}]$ represents the rate of momentum density exchanged between the CR population and the thermal gas. In the energy equations, \autoref{eq:eneq} and \autoref{eq:CRenergy}, $\mathbf{v} \cdot  \tensor{\mathrm{\sigma}}_\mathrm{tot} \cdot   [\mathbf{F_\mathrm{c}} - (4/3)e_\mathrm{c} \mathbf{v} ] $ represents the direct CR pressure work done on or by the gas, while $\mathbf{v_\mathrm{s}} \cdot  \tensor{\mathrm{\sigma}}_\mathrm{tot} \cdot [ \mathbf{F_\mathrm{c}} - (4/3) e_\mathrm{c}\mathbf{v} ] $ represents the rate of energy transferred to the gas via damping of gyro-resonant Alfv\'{e}n waves. In the above, $\mathbf{v_\mathrm{s}}$ is the CR streaming velocity,
\begin{equation}
\mathbf{v_\mathrm{s}} \equiv - \mathbf{v_\mathrm{A,i}}
\, \frac{\mathbf{B} \cdot (\nabla \cdot \tensor{\mathbf{P}}_\mathrm{c})}{\vert \mathbf{B} \cdot (\nabla \cdot \tensor{\mathbf{P}}_\mathrm{c})\vert}
= -\mathbf{v_\mathrm{A,i}} 
\frac{\hat{B} \cdot \nabla P_\mathrm{c}}{\vert \hat{B} \cdot \nabla P_\mathrm{c}\vert}
\;,
\label{eq:vs}
\end{equation}
defined to have the same magnitude as the local Alfv\'{e}n speed in the ions $\mathbf{v_{\rm A,i}} \equiv \mathbf{B}/\sqrt{4\pi\rho_i}$, oriented along the local magnetic field and pointing down the CR pressure gradient. Here, $\rho_\mathrm{i} $
is the ion mass density, where  the fractional ionization is calculated assuming collisional ionization equilibrium for $T>2\times 10^4$ and assuming CR ionization for $T<2\times 10^4$ (see Section 2.2.5 of \citetalias{Armillotta+21}). The diagonal tensor $\tensor{\mathbf{\sigma}}_\mathrm{tot}$ is the wave-particle interaction coefficient, defined to allow for both scattering and streaming along the direction parallel to the magnetic field, 
\begin{equation}
    \sigma_{\rm tot,\parallel}^{-1}= \sigma_\parallel^{-1} + \frac{v_\mathrm{A,i}}{|\hat B \cdot \nabla P_\mathrm{c} |} (P_\mathrm{c} + e_\mathrm{c}) \, ,
\label{eq:sigmatotpar}    
\end{equation}
and only scattering in the directions perpendicular to the magnetic field, 
\begin{equation}
    \sigma_{\rm tot,\perp}= \sigma_{\perp}\, .
\label{eq:sigmatotperp}    
\end{equation}
For the relativistic case, $\sigma_\parallel=\nu_\parallel/c^2$, 
where $\nu_\parallel$ is the scattering rate parallel to the magnetic field direction due to Alfv\'en waves that are resonant with the CR gyro-motion.
In the simulation, we compute $\sigma_\parallel$ in a self-consistent manner (see below) and set $\sigma_\perp = 10 \, \sigma_\parallel$. The latter can be understood as scattering by unresolved fluctuations of the mean magnetic field. In \citetalias{Armillotta+21}, we explored the transport of CRs in the absence of perpendicular scattering ($\sigma_\perp \gg \sigma_\parallel$) as well as the case $\sigma_\perp = 10 \, \sigma_\parallel$, and did not found any substantial difference in the CR distribution.

The scattering coefficient parallel to the magnetic field direction $\sigma_\parallel$ is derived based on the predictions of the self-confinement picture and assuming that, in steady state, the excitation of Alfv\'en waves by streaming CRs  is balanced by some form of wave damping \citep{Kulsrud&Pearce69,Kulsrud&Cesarsky1971}. We consider nonlinear Landau damping (NLL) and ion-neutral damping (IN). The former occurs when thermal ions have a resonance with the beat wave formed by the interaction of two resonant Alfv\`{e}n waves, while the latter arises from friction between ions and neutrals in partially ionized gas, where neutrals are not tied to magnetic fields. We refer to Section~2.2.3 in \citetalias{Armillotta+21} and Section~2.2.2 in \citetalias{Armillotta+22} for a detailed description of the derivation of $\sigma_\parallel$. Here, we only report the final solution. The scattering coefficient $\sigma_\parallel$ reduces to
\begin{equation}
\sigma_\mathrm{\parallel,nll} = \sqrt{\frac{\pi}{16} \, \frac{\vert \mathbf{{\hat{B}}} \cdot \nabla P_\mathrm{c}\vert}{v_\mathrm{A,i} P_\mathrm{c}} \frac{\Omega_0 c}{ 0.3 v_\mathrm{t,i} v_\mathrm{p}^2} \frac{m_\mathrm{p}}{m_\mathrm{i}} \frac{n_\mathrm{1}}{n_\mathrm{i}}}
\label{eq:sigmaNLL}
\end{equation}
in well ionized, low-density gas where nonlinear Landau dominates, and 
\begin{equation}
\sigma_\mathrm{\parallel,in} =  \frac {\pi}{8} \, \frac{\vert \mathbf{\hat{B}} \cdot \nabla  P_\mathrm{c}\vert}{v_\mathrm{A,i} P_\mathrm{c}}  \frac{\Omega_0}{ \langle \sigma v \rangle_\mathrm{in}} \, \frac{m_\mathrm{p} (m_\mathrm{n} + m_\mathrm{i})}{ n_\mathrm{n} m_\mathrm{n}  m_\mathrm{i}}  \frac{n_1}{n_\mathrm{i}} 
\label{eq:sigmaIN}
\end{equation}
in primarily neutral, denser gas where ion-neutral damping dominates. 
In \autoref{eq:sigmaNLL}, $\Omega_0 = e \vert \mathbf{B} \vert / (m_\mathrm{p} c)$ is the cyclotron frequency for $e$ the electron charge and $m_\mathrm{p}$ the proton mass, $v_\mathrm{t,i}$ is the ion thermal velocity (which we set equal to the gas sound speed $c_\mathrm{s}$), $m_\mathrm{i}$ is the ion mass, $n_\mathrm{i}$ is the ion number density, $v_\mathrm{p} = \sqrt{1-(m_\mathrm{p} c^2/E)^2}$ is the CR proton velocity, where $m_\mathrm{p}$ is the proton mass and $E \equiv E_\mathrm{k} + m_\mathrm{p} c^2$ is the total relativistic energy, with $E_\mathrm{k}$ the kinetic energy. For CRs with $E_\mathrm{k} \simeq 1$~GeV, which are the focus of this work, $v_\mathrm{p} \approx c$. 
The quantity $n_1$, which has units of a density, depends on the CR distribution function $F(p)$ in momentum space as
\begin{equation}
n_1 \equiv 4 \pi p_1 \int_{p_1}^\infty p F(p) dp\;,
\label{eq:n1}
\end{equation}
where $p_1=m_p \Omega_0/k$ is the resonant momentum for wavenumber $k$. In Appendix A1 of \citetalias{Armillotta+21}, we demonstrate that, assuming $F(p) \propto p^{-4.7}$ \citep[e.g.,][]{Aguilar+14, Aguilar+15}, $n_1/\cm^{-3} = 1.1 \times 10^{-10} \, [e_\mathrm{c}(E_\mathrm{k} \simeq 1\,\mathrm{GeV})/(1\mathrm{eV\,cm}^{- 3})] $ for $E_\mathrm{k}=1\,\mathrm{GeV}$. In \autoref{eq:sigmaIN}, $x_i=n_\mathrm{i}/n_\mathrm{H}$ is the ion fraction, $m_\mathrm{n}$ is the neutral mass, $n_\mathrm{n}$ the neutral number density (see Section 2.2.5 of \citetalias{Armillotta+21} for the derivation of $x_\mathrm{i}$, $n_\mathrm{n}$, $m_\mathrm{n}$, and $m_\mathrm{i}$), $\langle \sigma v \rangle_\mathrm{in} \sim 3\times10^{-9}$~cm$^3$~s$^{-1}$ is the rate coefficient for collisions between H and H$^+$ \citep[][Table 2.1]{Draine11}.
 
Finally, in \tworef{eq:CRenergy}{eq:CRflux}, the terms $\Lambda_\mathrm{coll} n_\mathrm{H} e_\mathrm{c}$ and $\Lambda_\mathrm{coll} n_\mathrm{H} \mathbf{F}_\mathrm{c}/v_\mathrm{p}^2$ represent, respectively, the rates of CR energy density and CR flux energy lost due to collisional interactions with the ambient gas. GeV CRs interact with the ambient gas through ionization of the neutral atomic/molecular gas and hadronic collisions leading to decays of pions into $\gamma$-rays \citep[e.g.,][]{Padovani+20}. In the above, $\Lambda_\mathrm{coll}$ is the CR collisional coefficient. In \citetalias{Armillotta+21}, we estimate that $\Lambda_\mathrm{coll}$ is equal to $4 \times 10^{-16}$ cm$^{3}$~s$^{-1}$ for CRs with $E_\mathrm{k} = 1$~GeV \citep[see also][]{Padovani+20}.

\begin{figure*}
\centering
\includegraphics[width=\textwidth]{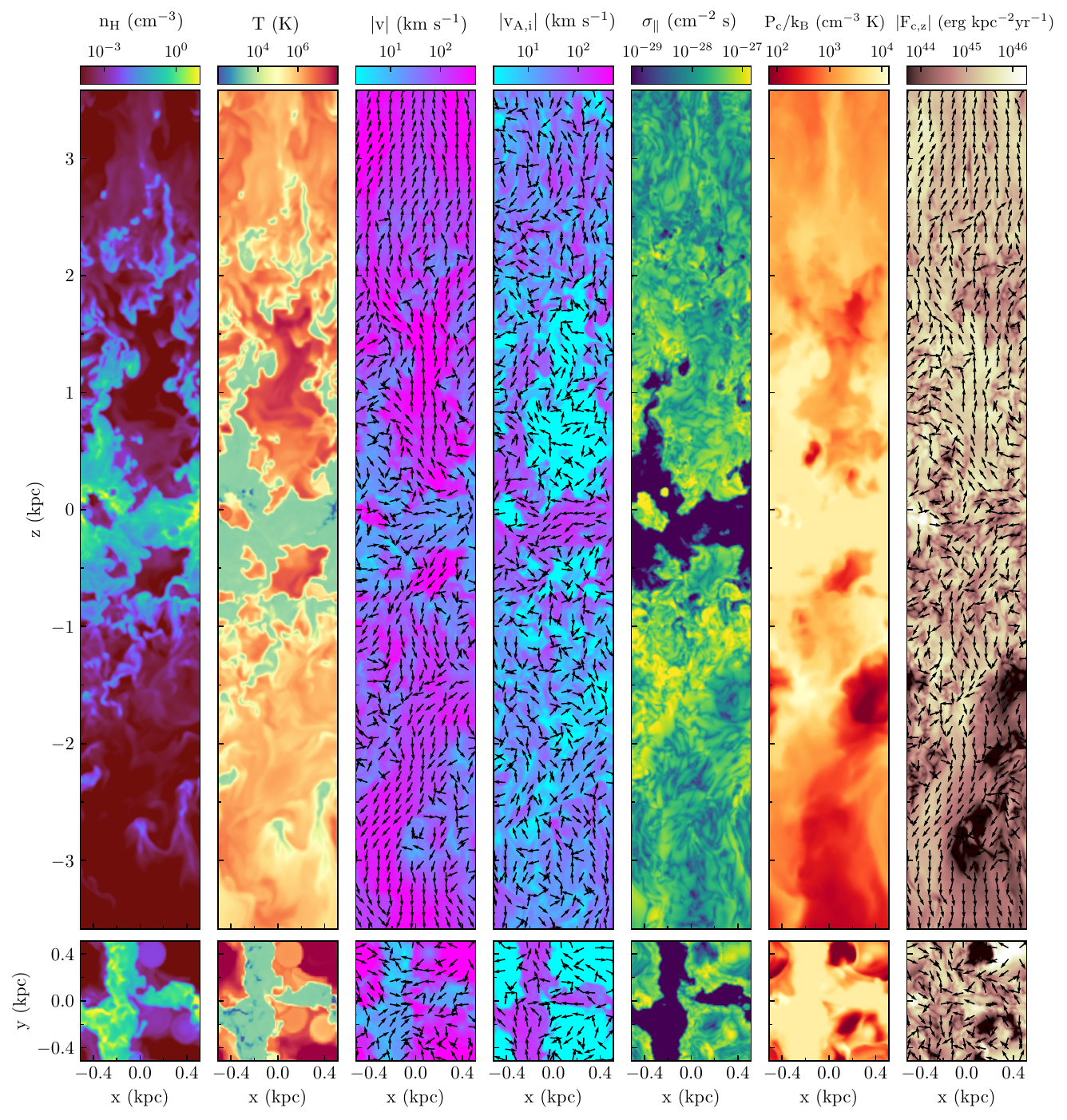}
\caption{Sample TIGRESS snapshot post-processed with the algorithm for CR transport \citep{Jiang&Oh18, Armillotta+21}. The upper (lower) row of panels shows $x$-$z$ ($x$-$y$) slices through the center of the simulation 
box, where $x$, $y$, and $z$ are the local radial, azimuthal, and vertical directions.  
From left to right, columns show hydrogen number density $n_\mathrm{H}$, gas temperature $T$, gas speed $\vert v \vert$, ion Alfv\'{e}n speed $\vert v_\mathrm{A,i} \vert$, scattering coefficient $\sigma_\parallel$, CR pressure $P_\mathrm{c}$, and magnitude of the vertical CR flux $\vert F_{\rm c,z} \vert$. The arrows overlaid on the gas velocity, Alfv\'{e}n speed, and vertical CR flux slices indicate the projected directions of the gas velocity, Alfv\'{e}n speed, and CR flux, respectively, in each slice. The CR pressure is divided by the Boltzmann constant $k_\mathrm{B} = 1.38 \times 10^{-16}$~erg~K$^{-1}$.}
\label{fig:FBsnapIni}
\end{figure*} 

\begin{figure*}
\centering
\includegraphics[width=\textwidth]{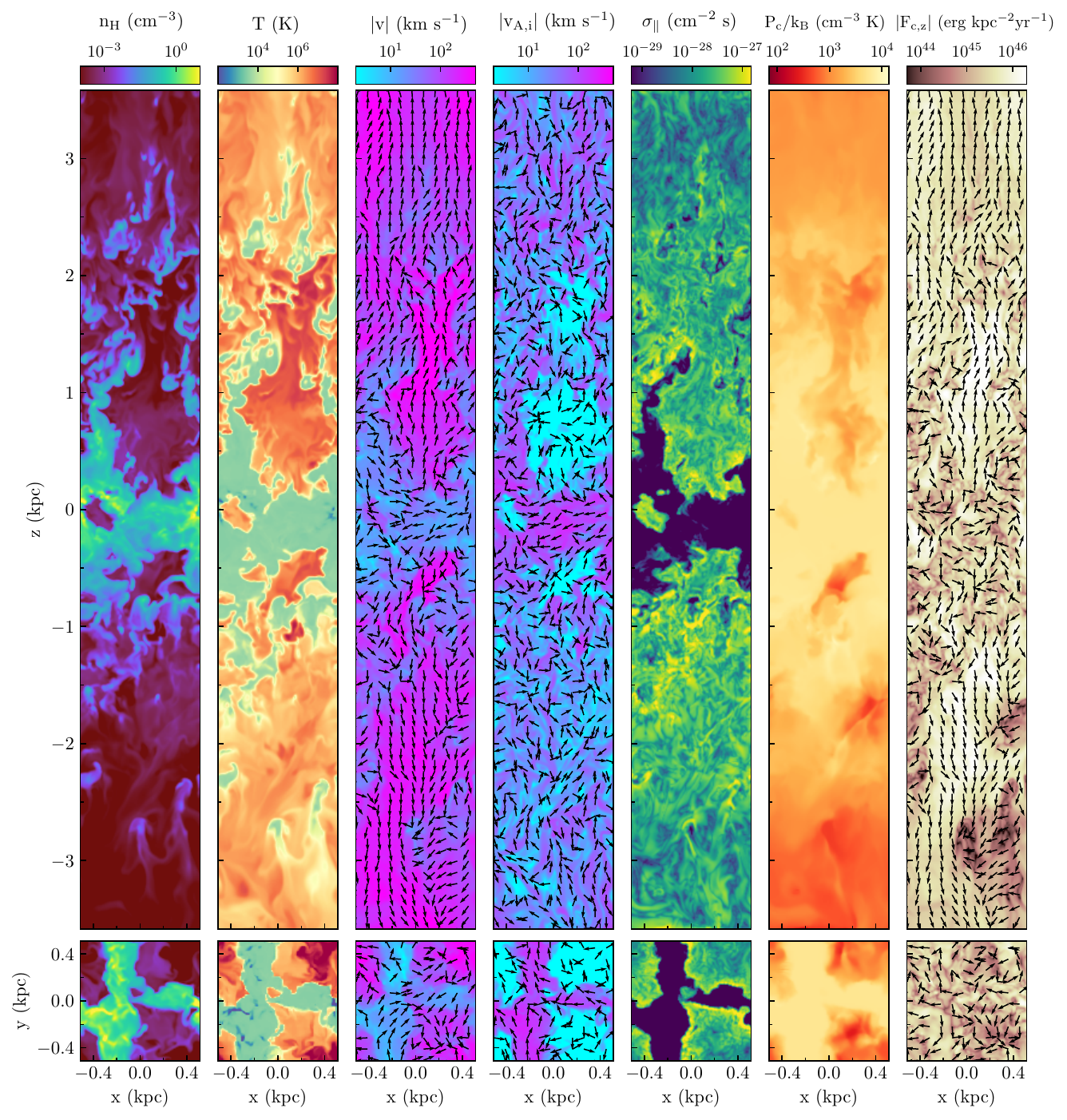}
\caption{Same as \autoref{fig:FBsnapIni}, but for one of the simulations computing MHD and CR physics together. The initial conditions of this simulation are extracted from the post-processed TIGRESS snapshot displayed in \autoref{fig:FBsnapIni}. The snapshot shown here is taken after $t = 3$~Myr.}
\label{fig:FBsnapMHD}
\end{figure*} 

\section{From postprocessed to self-consistent simulations}
\label{sec:fullboxsim}

With the goal of incorporating the MHD backreaction to the CR force, in this paper we move from the postprocessing approach employed in \citetalias{Armillotta+21} and \citetalias{Armillotta+22} to a self-consistent approach computing the MHD together with the CRs. Among the three TIGRESS models investigated in \citetalias{Armillotta+22}, here we focus on the model representative of the solar neighborhood environment only. 

In the postprocessing simulations, the MHD quantities are frozen in time, while the energy and flux density of CRs are evolved through space and time given the background distribution of thermal gas, magnetic field, and star cluster particles extracted from the TIGRESS simulation snapshots. During the post-processing simulations, CR energy is injected near star cluster particles to model effects from supernova events. For each star cluster particle, the rate of injected energy is computed as 
\begin{equation}\label{eq:Ecsp}
\dot{E}_\mathrm{c, sp}=\epsilon_\mathrm{c} \, E_\mathrm{SN} \,\dot{N}_\mathrm{SN}\,,
\end{equation}
where $\epsilon_\mathrm{c}$ is the fraction of supernova energy that goes into production of CRs, assumed to be equal to 0.1 \citep[e.g.][]{Morlino&Caprioli12,Ackermann+13}, $E_\mathrm{SN} = 10^{51}$~erg is the energy released by an individual supernova event, and $\dot{N}_\mathrm{SN}$ is the number of supernovae per unit time determined from the \textsc{Starburst99} code based on the age and mass of the star cluster.  We note that the current postprocessing simulation approach differs from that in \citetalias{Armillotta+21} in that we omit the term  $-\mathbf{v} \cdot 
\tensor{\mathrm{\sigma}}_\mathrm{tot} \cdot ( \mathbf{F_\mathrm{c}} - 4/3 \mathbf{v} e_\mathrm{c} ) $ in the energy equation (\autoref{eq:CRenergy}); we explain the reason for this modification in \autoref{sec:cr_pressure}.

\autoref{fig:FBsnapIni} displays the distribution on the grid of several MHD and CR quantities in one sample postprocessed TIGRESS snapshot. The simulation has box size $L_\mathrm{x} = L_\mathrm{y} = 1024$~pc and $L_\mathrm{z} = 7168$~pc, with a uniform spatial resolution $\Delta x = 8$~pc, sufficient to achieve robust convergence of several ISM and outflow properties, as well as convergence of CR properties \citep{Kim&Ostriker17, Armillotta+21}. The first four columns of \autoref{fig:FBsnapIni} show slices at $y=0$ (upper panels) and $z=0$ (lower panels) of hydrogen number density $n_\mathrm{H}$, gas temperature $T$, magnitude of gas velocity $v$, and magnitude of ion Alfv\'{e}n speed ${v_{\rm A,i}}$. 
This particular snapshot is representative of an outflowing phase of the original TIGRESS simulation, in which part of the gas heated and accelerated by supernova blast waves breaks out of the galactic plane, generating large-scale outflows in the coronal region. These outflows are composed of warm-cold fountain gas reaching several kpc from the midplane, and hot winds that escape from the disk \citep{Kim&Ostriker18, Vijayan+20, Kim+20}. As a consequence, most of the computational volume is occupied by hot ($T > 10^6$~K) and rarefied gas, with a decrease in the hot-gas volume filling factor near the midplane, mostly composed of warm/cold ($T \lesssim 10^4$~K) gas. We note that the gas velocity $v$ exceeds the ion Alfv\'{e}n speed $v_\mathrm{A,i}$ in the hot phase of the gas, while $v_\mathrm{A,i}$ exceeds $v$ in the warm phase (especially in the neutral gas where $x_i\ll 1)$. 

The three rightmost panels of \autoref{fig:FBsnapIni} display some outputs of the CR transport algorithm: scattering coefficient $\sigma_\parallel$, CR pressure\footnote{Note that the CR pressure $P_c/k_B$ in units $ {\rm K} \, \cm^{-3}$ can be converted to CR energy density in units $\mathrm{eV}\, \cm^{-3}$ by multiplying by a factor $2.586 \times 10^{-4}$.  The CR energy flux in units $\mathrm{erg}\,  \mathrm{kpc}^{-2} \mathrm{yr}^{-1}$ can be converted to units $\mathrm{eV}\, \cm^{-3}\, \kms$ by multiplying by a factor $2.077 \times 10^{-44}$.}   $P_\mathrm{c}/k_\mathrm{B}$, with $k_\mathrm{B}$ the Boltzmann constant, and magnitude of CR flux in the vertical direction, $F_\mathrm{c,z}$.  The scattering coefficient distribution closely follows the distribution of the background MHD quantities. In particular, $\sigma_\parallel$ is relatively high (above $10^{-28}$~cm$^{-2}$~s) in hot, highly ionized regions and quite low (below $10^{-29}$~cm$^{-2}$~s) in cooler, neutral regions. Intermediate-density regions at the interface between neutral and
ionized gas are characterized by the highest values of $\sigma_\parallel$ ($\sim 10^{-27}$~cm$^{-2}$~s). Similarly to the scattering coefficient distribution, the distribution of CR pressure also reflects the gas distribution: CRs accumulate in high-density regions, where the relatively-low gas velocities ($v<50\,\kms$) do not foster their removal, while CRs in regions with hot and fast-moving winds ($v \gg 100\,\kms$) are rapidly advected away from the mid-plane. One can note that the CR-flux streamlines mostly align with the velocity streamlines in regions with hot winds, meaning that CRs coupled to the hot gas escape the disk through these ``chimneys''.

Starting from the CR-postprocessed TIGRESS simulation snapshots (as in \autoref{fig:FBsnapIni}), we now perform new simulations where the distributions of thermal gas, magnetic field, and CRs are evolved together (see \autoref{sec:equations}).
While these new simulations properly compute the MHD together with the CRs, they are not fully self-consistent in that they do not include self-gravity to follow new star formation, and they do not include injection of energy and momentum in the thermal gas from radiation and supernova feedback.
Given that new star formation and feedback to create hot gas are not included in the current simulations, we run only for a timescale (a few Myr) shorter than the time for the hot gas to be advected out of the domain. We turn off CR energy injection for these ``MHD relaxation'' restart simulations.  
Our primary goal with these simulations is to allow for the MHD backreaction on the CRs, and to test how this modifies the CR distribution and resulting transport characteristics. 

\autoref{fig:FBsnapMHD} displays how the MHD and CR quantities plotted in \autoref{fig:FBsnapIni} have evolved in 3 Myr. A visual comparison clearly shows that the distribution of CR pressure becomes much smoother once the MHD backreaction is included. In the next section, we investigate what leads to the redistribution of CR pressure.  

\begin{figure*}
\centering
\includegraphics[width=\textwidth]{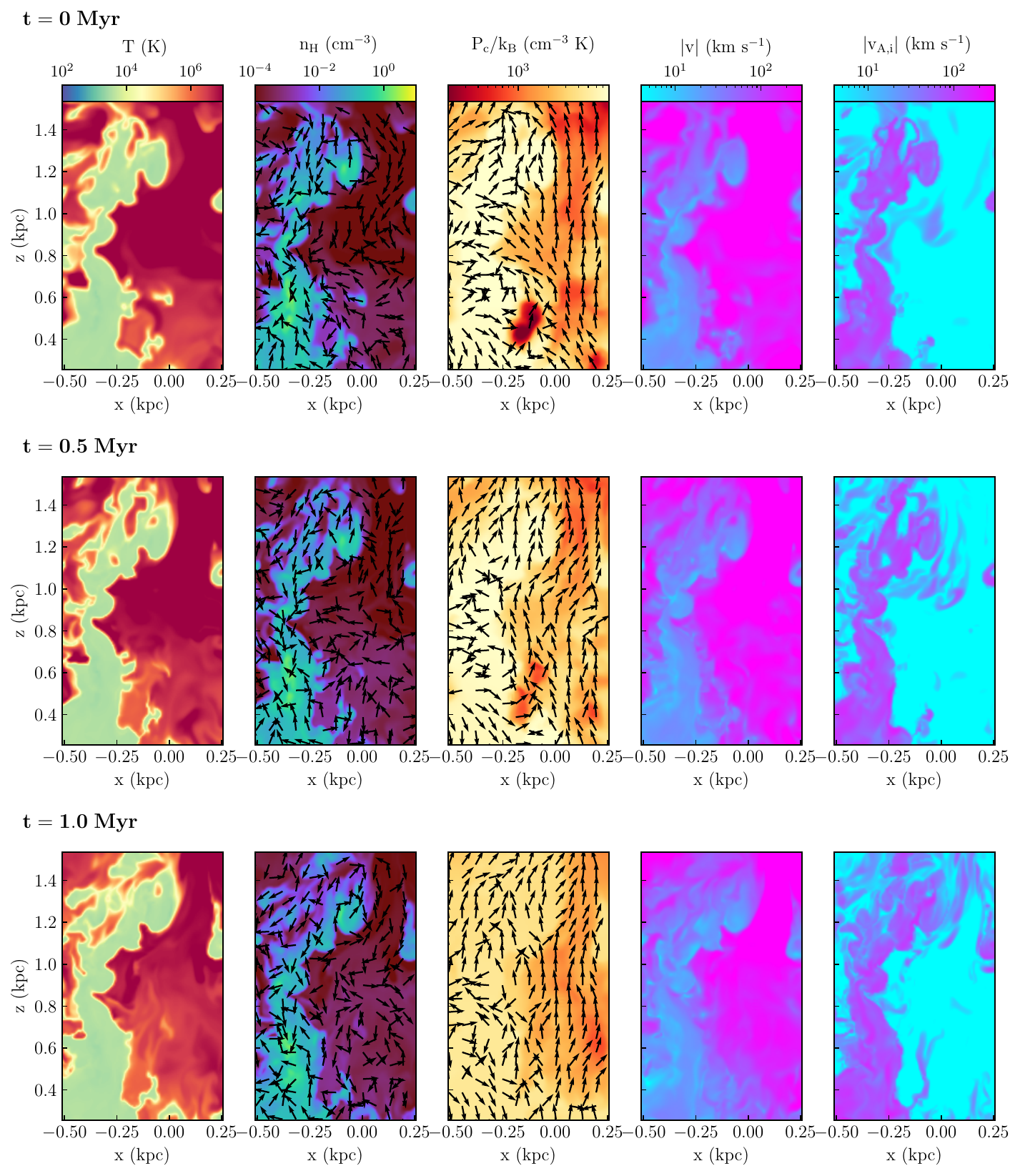}
\caption{Zoom-in on a filament of warm gas located just above the disk in the simulation shown in \autoref{fig:FBsnapMHD}. The snapshots are taken at $t = 0$~Myr (\textit{top panels}), $t = 0.5$~Myr (\textit{middle panels}), and $t = 1$~Myr (\textit{bottom panels}). Each row shows, from left to right, slices of gas temperature $T$, hydrogen number density $n_\mathrm{H}$, CR pressure $P_\mathrm{c}$, magnitude of gas velocity $v$, and magnitude of ion Alfv\'{e}n speed $v_\mathrm{A,i}$. The arrows overlaid on the hydrogen density and CR pressure slices indicate the directions of the magnetic field and gas velocity, respectively.}
\label{fig:FBzoomin}
\end{figure*} 

\begin{figure*}
\centering
\includegraphics[width=\textwidth]{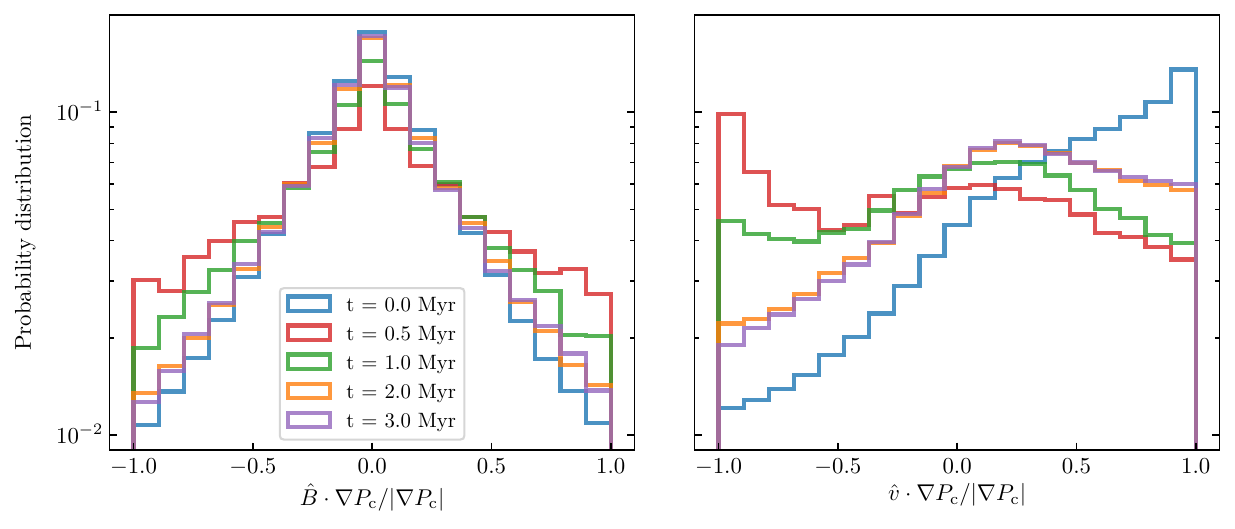}
\caption{Probability distributions of the angle between the magnetic (\textit{left panel}) and velocity field (\textit{right panel}) direction and the CR pressure gradient direction -- $\hat{B} \cdot \nabla P_\mathrm{c}  /|\nabla P_\mathrm{c}|$ and $\hat{v} \cdot \nabla P_\mathrm{c} /|\nabla P_\mathrm{c}|$ respectively -- evaluated at different times in the interval $0-3$~Myr. The distributions are weighted by $\vert \nabla \rho \vert/\rho$, which is largest at interfaces between cold/warm and hot gas.}
\label{fig:FBangles}
\end{figure*} 

\begin{figure*}
\centering
\includegraphics[width=\textwidth]{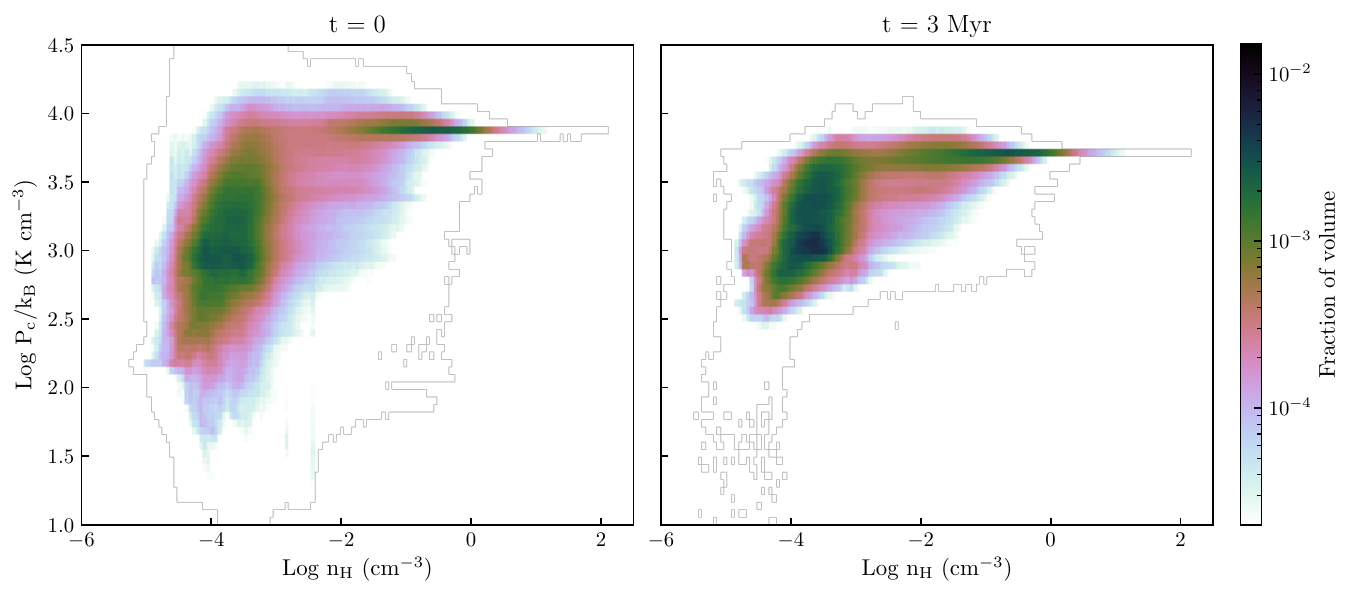}
\caption{Two-dimensional distribution of CR pressure $P_\mathrm{c}$ vs. gas density $n_\mathrm{H}$ at $t=0$ (\textit{left panel}) and $t=3$~Myr (\textit{right panel}). The distributions are computed as a two-dimensional probability density function showing the volume of gas within each logarithmic bin, normalized by the bin area.}
\label{fig:Pc_nh_PDF}
\end{figure*} 

\subsection{CR pressure redistribution}
\label{sec:cr_pressure}

In \autoref{fig:FBzoomin}, we zoom in on a filament of warm ($T \simeq 10^4$~K) and dense gas located above the disk region to show time evolution of the spatial distributions of gas temperature $T$, hydrogen number density $n_\mathrm{H}$, CR pressure $P_\mathrm{c}$, gas velocity magnitude $\vert v \vert$, and ion Alfv\'{e}n speed $\vert v_\mathrm{A,i} \vert$, as well as the orientation of the velocity and magnetic field. At $t = 0$, CRs are trapped in the dense gas. In the postprocessing simulation, the expansion of the hot and fast-moving gas into the warm gas leads to the transport of CRs towards the dense filament (see velocity streamlines in the hot gas). Once in the filament -- where CR streaming dominates over advection ($v_\mathrm{A,i}>v$) -- CRs are unable to escape due to the magnetic field lines mostly parallel to the filament edge.
This CR trapping results in a significant CR pressure gradient at the filament edges. At $t>0$, the backreaction of the CR pressure on the gas rearranges the velocity and magnetic field topology, allowing CRs to propagate out of the dense gas. The velocity field direction and the evolution of the temperature and density distributions clearly indicate that the warm and dense gas is now expanding into the hot gas as a consequence of the CR pressure gradients initially present at the edge of the filament. 

A more quantitative analysis of the evolution of the velocity and magnetic field topology at interfaces between cold/warm and hot gas is presented in \autoref{fig:FBangles}. Here, the left and right panels display the probability distributions of $\hat{B} \cdot \nabla P_\mathrm{c}  /|\nabla P_\mathrm{c}|$ and $\hat{v} \cdot \nabla P_\mathrm{c} /|\nabla P_\mathrm{c}|$, respectively. These are, respectively, 
the cosine of the angle between the magnetic field direction and the CR pressure gradient direction, and 
the cosine of the angle between the velocity direction and the CR pressure gradient direction. In order to track the interface regions, the distributions are weighted by the normalized density gradient $\vert \nabla \rho \vert/\rho$. At $t=0$, the distribution of $\hat{B} \cdot \nabla P_\mathrm{c}  /|\nabla P_\mathrm{c}|$ in peaks at values near zero,
meaning that the CR pressure gradient is mostly perpendicular to the magnetic field direction. As discussed above, this prevents CRs from streaming out of the dense gas in the postprocessing simulations. Once the backreaction of the CR pressure on the gas is included, however, the angle between the CR pressure gradient direction and the magnetic field direction decreases, as indicated by the slight shift of the probability distribution towards larger absolute values of $\hat{B} \cdot \nabla P_\mathrm{c}  /|\nabla P_\mathrm{c}|$.

The CR backreaction is even more evident in the right panel of \autoref{fig:FBangles}, showing how the distribution of the angles between the velocity direction and the CR pressure gradient direction varies with time. At $t=0$, the distribution of $\hat{v} \cdot \nabla P_\mathrm{c}  /|\nabla P_\mathrm{c}|$ peaks at values near 1.  
As shown in \autoref{fig:FBzoomin}, CR pressure gradients are initially aligned with the velocity streamlines at interfaces, and pointing in the same direction. Due to the excess of CR pressure in the cold/warm dense gas, when MHD becomes ``live'', CRs exert a back-reaction force that leads the dense gas to expand into the hot gas. 
As a result, the distribution of $\hat{v} \cdot \nabla P_\mathrm{c} /|\nabla P_\mathrm{c}|$ in  \autoref{fig:FBangles} peaks at values near $-1$ at $t=0.5$~Myr, meaning that the velocity field direction becomes mostly anti-parallel to the CR pressure gradient (see also panels at $t=0.5$~Myr in \autoref{fig:FBzoomin}).
We conclude that the rearrangement of velocity and magnetic field topology enable CRs initially trapped in the dense gas to propagate out of it, leading to a more uniform CR pressure distribution (see \autoref{fig:Pc_nh_PDF}). For $t\geq 1$~Myr, the distribution of $\hat{v} \cdot \nabla P_\mathrm{c}  /|\nabla P_\mathrm{c}|$ peaks near values $\gtrsim 0$, as CR pressure gradients are smoothed out at interfaces (see panels at $t=1$~Myr in \autoref{fig:FBzoomin}). Both the distribution of $\hat{B} \cdot \nabla P_\mathrm{c}  /|\nabla P_\mathrm{c}|$ and $\hat{v} \cdot \nabla P_\mathrm{c}  /|\nabla P_\mathrm{c}|$ achieve convergence for $t\geq 1$~Myr.

\autoref{fig:Pc_nh_PDF} quantitatively shows the redistribution of CR pressure in the different phases of the gas once the backreaction of CRs is included. The left and right plots display the volume-weighted probability distributions of CR pressure vs.\ gas density at $t=0$ and $t=3$~Myr, respectively. Overall, $P_\mathrm{c}$ increases in the low-density hot gas: at $t=0$, the distribution of CR pressure spans across more than two orders of magnitude, from $P_\mathrm{c}/k_\mathrm{B} \sim 10^4$~K~cm$^{-3}$ to as low as $P_\mathrm{c}/k_\mathrm{B} \sim 10^{1.5}$~K~cm$^{-3}$; by contrast, at $t=3$~Myr, the distribution is much narrower and $P_\mathrm{c}/k_\mathrm{B}$ consistently remains above $10^{2.5}$~K~cm$^{-3}$. While increasing in the hot and rarefied gas, $P_\mathrm{c}$ decreases in the cooler high-density gas: for $n_\mathrm{H} \gtrsim 10^{-2}$~cm$^{-3}$, $P_\mathrm{c}$ decreases by a factor 1.5 from $t=0$ to $t=3$~Myr.
 
\begin{figure*}
\centering
\includegraphics[width=\textwidth]{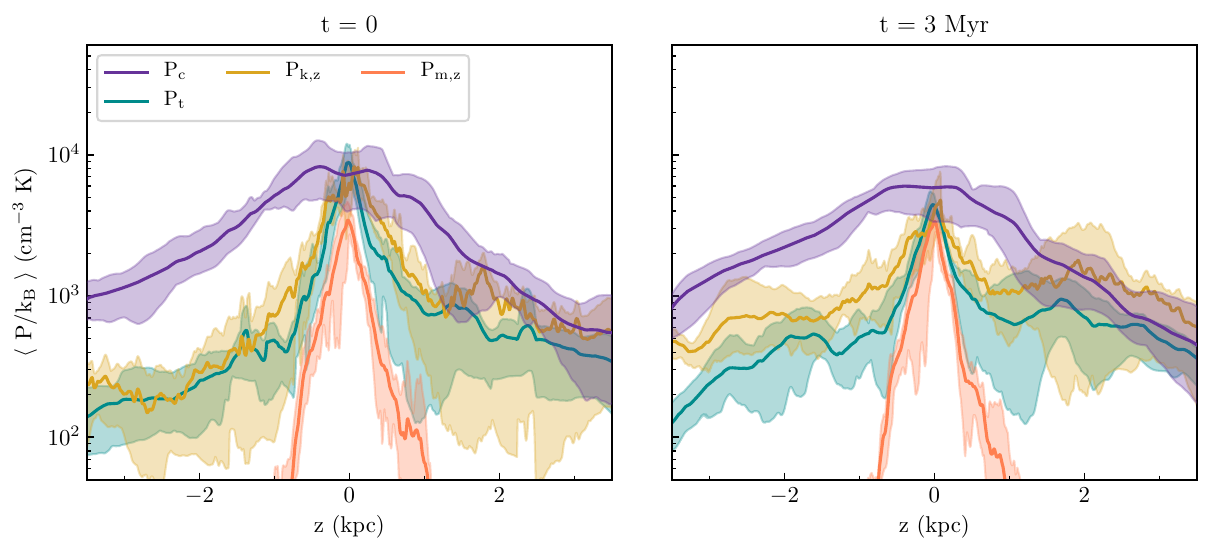}
\caption{Horizontally-averaged vertical profiles of CR pressure $P_\mathrm{c}$ (purple), thermal pressure $P_\mathrm{t}$ (cyan), vertical kinetic pressure $P_\mathrm{k,z}$ (yellow), vertical magnetic stress $P_\mathrm{m,z}$ (orange) at $t = 0$ (\textit{left panel}) and $t = 3$~Myr (\textit{right panel}). The average is computed over 10 snapshots output from TIGRESS and evolved with time-dependent MHD and CR physics in this work. The shaded areas cover the 16th and 84th percentiles of fluctuations.}
\label{fig:FBpressprof}
\end{figure*} 

Finally, we analyse how the mean vertical profiles of CR and MHD pressures vary in ``live MHD'' simulations. We first apply post-processing and then run with ``live MHD'' for ten TIGRESS solar neighborhood snapshots taken between $t = 200$ to $550$~Myr, when the mean surface density is equal to $\Sigma_\mathrm{gas} = 9.5 \,\mo$~pc$^{-2}$ and the mean star formation rate per unit area is $\Sigma_\mathrm{SFR} = 5 \times 10^{-3} \, \mo$~kpc$^{-2}$~yr$^{-1}$ ($8.7 <\Sigma_\mathrm{gas}< 10.2 \,\mo$~pc$^{-2}$, $3 \times 10^{-3} < \Sigma_\mathrm{SFR} < 8 \times 10^{-3} \, \mo$~kpc$^{-2}$~yr$^{-1}$). In order to study mean trends, all averages are taken over a given time in the 10 simulations initiated from these 10 TIGRESS snapshots. 

The left and right plots of \autoref{fig:FBpressprof} display the horizontally and temporally 
averaged profiles of CR pressure $P_\mathrm{c}$, thermal pressure $P_\mathrm{t}$, vertical kinetic pressure $P_\mathrm{k,z}$, and and vertical magnetic stress $P_\mathrm{m,z}$, at $t=0$ and $t=3$~Myr, respectively. The vertical kinetic pressure and magnetic stress are defined as 
\begin{equation}
P_\mathrm{k,z} = \rho v_\mathrm{z}^2
\end{equation}
and 
\begin{equation}
 P_\mathrm{m,z} = \dfrac{B_\mathrm{x}^2+B_\mathrm{y}^2-B_\mathrm{z}^2}{8 \pi}  \; .
\end{equation}
These are the  {$\hat z \hat z$} components of the Reynolds and Maxwell stress tensors, respectively, which appear in the conservation-law form of the momentum equation. With shearing-periodic boundary conditions, 
contributions to the mean vertical force per unit volume are obtained from the vertical gradient of 
horizontal averages over the domain of these terms \citep[e.g.,][]{PiontekOstriker2007}. 
$P_\mathrm{m,z}$ can in principle be negative if the vertical magnetic field dominates, but in practice, the horizontal field dominates in the midplane region where magnetic stresses are dynamically important.  For $z\lesssim 500$~pc, the horizontal average of $P_\mathrm{m,z}$ is positive and is smaller than the horizontal average of the magnetic pressure $ P_\mathrm{m} = ({B_\mathrm{x}^2+B_\mathrm{y}^2 + B_\mathrm{z}^2})/{8 \pi}$ by a factor $1.3-1.5$. At large $|z|$, the magnetic stress is dynamically unimportant compared to other terms, but magnetic fields still mediate the interaction between CRs and the thermal gas.

Near the midplane, both the CR pressure and the MHD pressures slightly decrease (by 20\% and 30\% at the midplane, respectively) going from $t=0$ to $t=3$~Myr. From a detailed examination of the simulations, we find that the decrease of MHD pressures is mostly in the hot gas, as the latter does not gain new energy and momentum in the absence of supernova feedback. By contrast, the decrease of CR pressure is mostly in the warm/cold and dense gas, due to CRs propagating out of it (see above). At higher $\vert z \vert$, $P_\mathrm{k,z}$ increases in time (from $t=0$ to $t=3$~Myr) as the vertical CR pressure-gradient forces accelerate the gas away from the disk. We shall explore the dynamical effect of CRs in the extra-planar region in the next sections. 

By comparing the left panel of \autoref{fig:FBpressprof} with the right panel of Figure~14 in \citetalias{Armillotta+21} or Figure~4 in \citetalias{Armillotta+22}, one can notice that while in the present work the midplane CR pressure is in equipartition with the thermal and kinetic pressures, in \citetalias{Armillotta+21} and \citetalias{Armillotta+22} the midplane CR pressure was a factor of $\sim 2$ higher than the other pressures. This is because in the postprocessing simulations carried out for this paper, the adiabatic work term on the RHS of \autoref{eq:CRenergy} ($-\mathbf{v} \cdot 
\tensor{\mathrm{\sigma}}_\mathrm{tot} \cdot ( \mathbf{F_\mathrm{c}} - 4/3 \mathbf{v} e_\mathrm{c} ) $) is set to zero.
In the present post-processing simulations, we omit this adiabatic work term because it can introduce spurious sources of CR energy at interfaces between cold/warm and hot gas 
when there is no CR backreaction on the gas. To understand this effect, first note that assuming negligible time-dependent term and collisional loss term in \autoref{eq:CRflux}, $\tensor{\mathrm{\sigma}}_\mathrm{tot} \cdot (\mathbf{F_\mathrm{c}} - 4/3 \mathbf{v} e_\mathrm{c}) \approx - \nabla P_\mathrm{c}$.  
Second, at hot/cool interfaces, in post-processing simulations the velocity field and the CR pressure gradient are generally in the same direction (since CRs are flowing from hot gas into cooler gas; see \autoref{fig:FBzoomin} and \autoref{fig:FBangles}).  As a result, $- \mathbf{v} \cdot
\tensor{\mathrm{\sigma}}_\mathrm{tot} \cdot ( \mathbf{F_\mathrm{c}} - 4/3 \mathbf{v} e_\mathrm{c} )$ near these interfaces is positive, and overall the adiabatic work done at these interfaces, where 
$|\nabla P_\mathrm{c}|$ is large, can be considerable. 

In \citetalias{Armillotta+21} and \citetalias{Armillotta+22}, we limited the contribution from adiabatic work by setting 
$-\mathbf{v} \cdot 
\tensor{\mathrm{\sigma}}_\mathrm{tot} \cdot ( \mathbf{F_\mathrm{c}} - 4/3 \mathbf{v} e_\mathrm{c} ) = 0$ in cells with $e_\mathrm{c} > e_\mathrm{k}$.
Even with this limit, the total adiabatic work (integrated over the whole simulation domain) was comparable to the total injected CR energy (see Table 2 of \citetalias{Armillotta+22}). As noted above, however, 
as soon as we turn on 
CR backreaction in ``live MHD'' simulations, 
the velocity vectors reorient and the adiabatic work becomes initially negative, and then within a few Myr the velocity field becomes preferentially perpendicular to $\nabla P_\mathrm{c}$  (\autoref{fig:FBangles}). In ``live MHD'' simulations, CRs lose energy overall by doing work on the gas. 
By setting the adiabatic work term to zero to avoid unphysical enhancement of CR energy in postprocessing simulations, for a 10\% energy injection rate from SNe to CRs, the CR midplane pressure is comparable to the other MHD pressures, similar to the observed situation.  

\subsection{Role of Streaming, Diffusive and Advective Transport}
\label{sec:transport}

\begin{figure*}
\centering
\includegraphics[width=\textwidth]{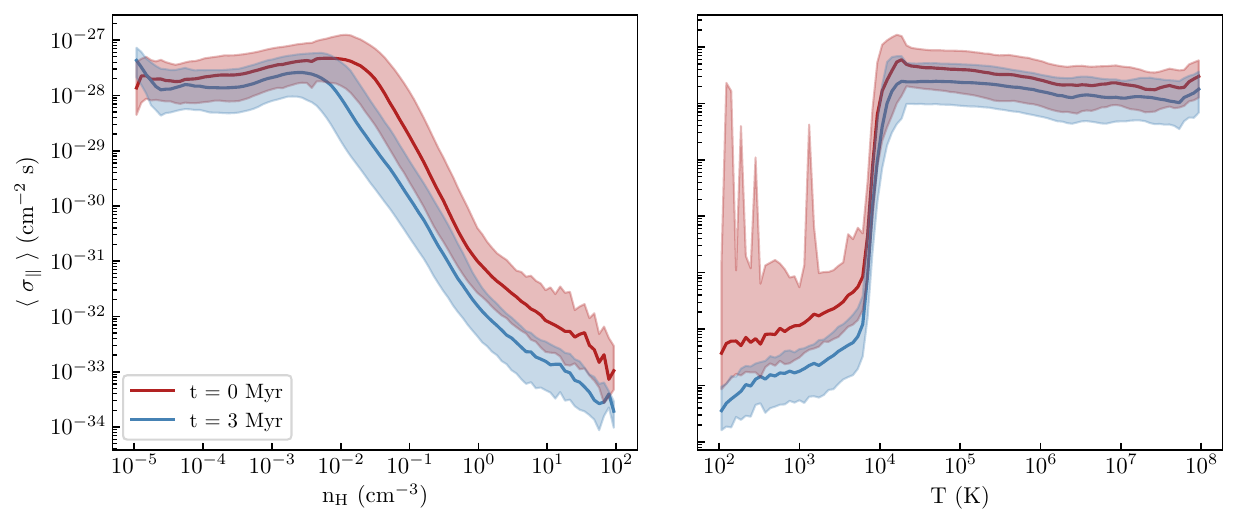}
\caption{Median of the scattering coefficient $\sigma_\parallel$ as a function of hydrogen density $n_\mathrm{H}$ (\textit{left panel}) and gas temperature $T$ (\textit{right panel}) at $t = 0$ (red) and $t = 3$~Myr (blue). For both panels, the shaded areas cover the 16th to 84th percentiles of fluctuations.}
\label{fig:FBsigma}
\end{figure*} 

\begin{figure*}
\centering
\includegraphics[width=\textwidth]{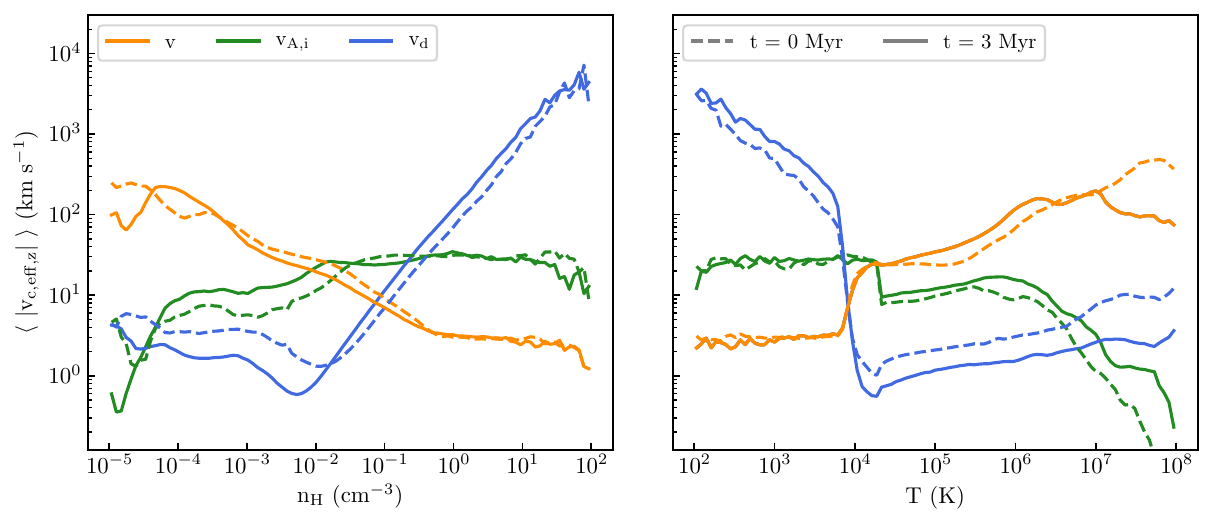}
\caption{Medians of the three vertical components of the effective CR propagation speed -- advection speed $v$ (orange), ion Alfv\'{e}n speed $v_\mathrm{A,i}$ (green), and diffusive speed $v_\mathrm{d}$ (blue) -- as a function of hydrogen density $n_\mathrm{H}$ (\textit{left panel}) and gas temperature $T$ (\textit{right panel}) at $t = 0$ (dashed lines) and $t = 3$~Myr (solid lines).}
\label{fig:FBvel}
\end{figure*} 

In this section, we verify that the main findings of our previous work regarding transport still hold in the ``live MHD'' simulations. Those are 1) the scattering coefficient varies over more than four orders of magnitude depending on the properties of the background gas, with the maximum value reached at intermediate densities, and 2) CR advection dominates in the high-velocity, low-density hot gas, while CR diffusion and streaming are more important in higher-density, cooler gas. 
We emphasize that the results for scattering coefficients we report on here are valid for CRs with $E_\mathrm{kin}=1$~GeV only. In fact, $\sigma_\parallel$ depends on $E_\mathrm{kin}$ (and $p$) through the $n_\mathrm{1}$ term (see \autoref{eq:n1}): $n_\mathrm{1}$ decreases with increasing $E_\mathrm{kin}$ (see Appendix~A.1 in \citetalias{Armillotta+21}), which implies the values of $\sigma_\parallel$ as produced by the streaming instability would be lower for CRs with $E_\mathrm{kin}>1$~GeV.

In \autoref{fig:FBsigma}, we show the variation of $\sigma_\parallel$ with density $n_\mathrm{H}$ (left) and temperature $T$ (right) at $t = 0$ and $t = 3$~Myr
The analysis is based on all the simulations performed in this section (same set of simulations used for \autoref{fig:FBpressprof}); we show median values as well as 16th and 84th percentiles. The profiles predicted by the postprocessing simulations and by the ``live MHD'' simulations are overall similar, thus confirming our finding from \citetalias{Armillotta+21}. In the left panel, $\sigma_\parallel$ is fairly flat at low densities up to $n_\mathrm{H}\sim 10^{-2}\pcc$, where the gas is well ionized and nonlinear Landau damping dominates, 
and then 
rapidly decreases at higher densities, where the gas is mostly neutral and ion-neutral damping becomes stronger than nonlinear Landau damping (see \citetalias{Armillotta+21} for a detailed analysis of the dependence of $\sigma_\parallel$ on $n_\mathrm{H}$).  
In the right panel, at low temperatures ($T < 10^4$~K) where gas is mainly neutral and ion-neutral damping dominates, $\sigma_\parallel$  $<10^{-31} \mathrm{cm}^{-2}\mathrm{s}$ and increases mildly with $T$ (as $n_\mathrm{H}$ decreases). Near $T\approx10^4$~K, where gas becomes mostly ionized \citep[][]{Sutherland&Dopita93},  $\sigma_\parallel$ abruptly increases by more than four orders of magnitude to a few $\times 10^{-28}\mathrm{cm}^{-2}\mathrm{s}$; in this high temperature regime where nonlinear Landau damping dominates, $\sigma_\parallel$ slowly decreases with $T$. 

Despite the overall similar profiles, $\sigma_\parallel$ is smaller at $t = 3$~Myr than at $t=0$. This difference can be attributed to the smaller CR pressure gradients in the ``live MHD'' simulations. In the low-density, high-temperature regime, where $\sigma_\parallel = \sigma_\mathrm{NLL, \parallel} \propto (\vert \hat{\mathbf{B}} \cdot  \nabla {P_\mathrm{c}} \vert)^{1/2}$ (see \autoref{eq:sigmaNLL}),
$\sigma_\parallel$ decreases by a factor $\sim 2$ under ``live'' MHD.   In the high-density, low-temperature regime, where $\sigma_\parallel = \sigma_\mathrm{IN, \parallel} \propto \vert \hat{\mathbf{B}} \cdot  \nabla {P_\mathrm{c}} \vert$ (see \autoref{eq:sigmaIN}), $\sigma_\parallel$ decreases by almost one order of magnitude under ``live'' MHD. As $\sigma_\mathrm{IN, \parallel}$ decreases with density faster than $\sigma_\mathrm{NLL, \parallel}$, the regime where ion-neutral damping dominates extends down to lower densities at $t=3$~Myr compared to $t=0$. This results in slightly different turnovers of the scattering coefficient profiles: this is at $n_\mathrm{H} \simeq 1-2 \times 10^{-2}$~cm$^{-3}$ at $t=0$, and $n_\mathrm{H} \simeq 5-6 \times 10^{-3}$~cm$^{-3}$ at $t=3$~Myr.

In \autoref{fig:FBvel}, we show the relative contributions of advection, streaming, and diffusion to the overall CR transport.  If we define the vertical flux as $F_{\mathrm{c},z} \equiv v_\mathrm{c,eff,z} (4/3) e_\mathrm{c}$, each of these contributions is characterized by a speed, where the first two are the vertical components of the gas flow speed and ion Alfv\'en speed.   
The diffusive flux $\mathbf{F}_\mathrm{d}$ is obtained from \autoref{eq:CRflux} in the limit of negligible  time-dependent term and collisional losses by subtracting the advective and streaming fluxes from the total CR flux: $\mathbf{F}_\mathrm{d} \equiv \mathbf{F}_\mathrm{c} - 4/3(\mathbf{v} + \mathbf{v}_\mathrm{s}) e_\mathrm{c} $. Using \autoref{eq:sigmatotpar}, the diffusive velocity becomes 
\begin{equation}
  \mathbf{v}_\mathrm{d} \equiv \frac{3}{4} \frac{\mathbf{F}_\mathrm{d}}{e_\mathrm{c}} \, \equiv - \frac{3}{4} \, \tensor{\mathbf{\sigma}}^{-1} \cdot \frac{  \nabla P_\mathrm{c}}{e_\mathrm{c}}.  
 \label{eq:vd_def}
\end{equation} 
The left and right panels of \autoref{fig:FBvel} show the median profiles of the vertical components of the advection speed $v$, ion Alfv\'{e}n speed $v_\mathrm{A,i}$, and diffusive speed $v_\mathrm{d}$ as a function of gas density and temperature, respectively. 
The velocity profiles are consistent at $t=0$ and $t=3$~Myr: the advection speed largely dominates in hot, low-density, well ionized gas, while the diffusive speed largely dominates in cold, high-density, mostly neutral gas. Streaming is the most important transport mechanism in warm and intermediate-density regions. Since the overall transport speed is the largest of the three, and this maximum speed is smallest for the streaming-dominated regime, the primary overall limit on CR transport is from the ion Alfv\'en speed in gas at $n_\mathrm{H}\sim 0.01 - 0.1 \,\pcc$ with  $T\sim 10^4$~K. 

We emphasize that our conclusion regarding the overall limit on transport set by $v_{\rm A,i}$ holds only for CRs with $E_\mathrm{kin} \sim 1$~GeV.  At higher CR energy where the scattering rate is expected to decrease, the diffusion velocity (\autoref{eq:vd_def}) would increase, pushing the blue curves in \autoref{fig:FBvel} upward and squeezing the regime where streaming dominates transport.  For CRs of sufficiently high energy, diffusion would dominate in all warm and cold gas.

We point out that even though the transport speed in the colder, higher-density gas is comparable to or larger than the transport speed in the hotter, lower-density gas, the CR residence time is significantly greater in the dense gas, as indicated by the higher mean CR pressure there (see \autoref{fig:Pc_nh_PDF}). In fact, the propagation of CRs out of the higher-density, primarily neutral  gas (which comprises most of the ISM mass) is limited by the low transport speed in the surrounding intermediate-density streaming-dominated gas. Given the low-scattering (but high-column) midplane and high-scattering (but low column) exterior illustrated by the slice depicted in \autoref{fig:FBsnapMHD}, realistic galaxies have physical elements that connect to both traditional ``leaky box'' and diffusive pictures.

As the distribution of CR pressure becomes smoother from $t=0$ to $t=3$~Myr, the diffusive speed decreases in the low-density, high-temperature regime, where $v_\mathrm{d} \propto \sqrt{\nabla P_\mathrm{c}}/e_\mathrm{c}$. By contrast, the diffusive speed increases in the high-density, low-temperature regime, where $v_\mathrm{d} \propto 1/e_\mathrm{c}$, as the energy density decreases in the dense gas (see above). However, these changes do not significantly affect CR transport overall, since the main transport mechanism at low density and high temperature is advection, and since the CR transport is already extremely rapid (producing quite uniform $P_\mathrm{c}$) in dense gas. 

\begin{deluxetable*}{ccccccccccc}
\tablecaption{Horizontally-averaged initial conditions measured at $z = 500$~pc in the absence of CRs. \label{tab:tab1}}
\tablehead{
\colhead{Model} &
\colhead{phase} &
\colhead{$\mathcal{F}_\mathrm{M}$} &
\colhead{$\mathcal{F}_\mathrm{p, MHD}$} &
\colhead{$\mathcal{F}_\mathrm{E, MHD}$} &
\colhead{$f_\mathrm{A}$} &
\colhead{$\tilde{T}$} &
\colhead{$\tilde{n}_\mathrm{H}$} &
\colhead{$\tilde{v}_\mathrm{z}$} &
\colhead{$\tilde{v}_\mathrm{z,out}$} &
\colhead{$|\tilde{B}_\mathrm{z}|$}
\\
\colhead{} &
\colhead{} &
\colhead{\footnotesize{$\left( \dfrac{\rm{M_\odot}}{\rm{kpc^{2} yr}} \right)$}} &
\colhead{\footnotesize{$\left( \dfrac{\rm{M_\odot}}{\rm{kpc^{2} yr}} \dfrac{\rm{km}}{\rm{s}} \right)$}} &
\colhead{\footnotesize{$\left( \dfrac{\rm{erg}}{\rm{kpc^{2} yr}} \right)$}} &
\colhead{} &
\colhead{\footnotesize(K)} &
\colhead{\footnotesize(cm$^{-3}$)} & 
\colhead{\footnotesize{$\left(\dfrac{\rm{km}}{\rm{s}} \right)$}} &
\colhead{\footnotesize{$\left(\dfrac{\rm{km}}{\rm{s}} \right)$}} &
\colhead{$\mu$G}
}
\colnumbers
\startdata
M1 & warm & $-1.9\times10^{-3}$ & $0.29$ & $\;\;\;3.8 \times 10^{43}$ & 0.21 & $9.8 \times 10^3$ & $1.5 \times10^{-1}$ & 5.2 & $1.5 \times 10^1$ & 1.6 \\
 & warm-hot & $\;\;\;1.2\times10^{-3}$ & $0.12$ & $\;\;\;1.8 \times 10^{44}$ & 0.05 & $2.3 \times 10^5$ & $2.1 \times 10^{-2}$ & $3.6 \times 10^1$ & $4.1 \times 10^1$ & $5.7 \times 10^{-1}$ \\
 & hot & $\;\;\;2.4\times10^{-3}$ & $1.89$ & $\;\;\;2.5 \times 10^{46}$ & 0.74 & $1.3 \times 10^7$ & $4.3 \times10^{-4}$ & $3.3 \times 10^2$ & $3.5 \times 10^2$ & $2.7 \times 10^{-2}$ \\
\hline
M2 & warm & $-8.8\times10^{-4}$ & $0.50$ & $\;\;\;5.9 \times 10^{43}$ & 0.59 & $9.8 \times 10^3$ & $7.8 \times10^{-2}$ & 2.6 & $1.7 \times 10^1$ & $7.4 \times 10^{-1}$ \\
 & warm-hot & $\;\;\;6.8\times10^{-4}$ & $0.06$ & $\;\;\;8.6 \times 10^{43}$ & 0.10 & $1.9 \times 10^5$ & $5.8 \times 10^{-3}$ & $3.9 \times 10^1$ & $4.3 \times 10^1$ & $1.2 \times 10^{-1}$ \\
 & hot & $\;\;\;4.2\times10^{-4}$ & $0.16$ & $\;\;\;7.2 \times 10^{44}$ & 0.31 & $4.9 \times 10^6$ & $3.1 \times10^{-4}$ & $1.0 \times 10^2$ & $1.1 \times 10^2$ & $7.3 \times 10^{-3}$ \\
\enddata
\tablecomments{
Columns: (1) model name; (2) gas phase; (3) gas mass flux; (4) MHD momentum flux; (5) MHD energy flux; (6) area fraction; (7) temperature; (8) hydrogen density; (9) gas velocity along the vertical direction; (10) outflow gas velocity along the vertical direction; (11) magnitude of the magnetic field along the vertical direction.
}
\end{deluxetable*}

\begin{deluxetable*}{lccccccccc}
\tablecaption{Model list and inflow boundary conditions. \label{tab:tab2}}
\tablehead{
\colhead{Model} &
\colhead{$T_\mathrm{0}$} &
\colhead{$n_\mathrm{H,0}$} &
\colhead{$v_\mathrm{z,0}$} &
\colhead{$v_\mathrm{A_i,z,0}$} &
\colhead{$P_\mathrm{c,0}/k_\mathrm{B}$} & 
\colhead{$F_\mathrm{c,z,0}$} &
\colhead{$\mathcal{F}_\mathrm{M,0}$} &
\colhead{$\mathcal{F}_\mathrm{p, MHD, 0}$} &
\colhead{$\mathcal{F}_\mathrm{p, c, 0}$} 
\\
\colhead{} &
\colhead{\footnotesize(K)} &
\colhead{\footnotesize(cm$^{-3}$)} & 
\colhead{\footnotesize(km/s)} & 
\colhead{\footnotesize(km/s)} &
\colhead{\footnotesize(K/cm$^{3}$)} &
\colhead{\footnotesize(erg/kpc$^2$/yr)} &
\colhead{\footnotesize{$\left( \dfrac{\rm{M_\odot}}{\rm{kpc^{2} yr}} \right)$}} &
\colhead{\footnotesize{$\left( \dfrac{\rm{M_\odot}}{\rm{kpc^{2} yr}} \dfrac{\rm{km}}{\rm{s}} \right)$}} &
\colhead{\footnotesize{$\left( \dfrac{\rm{M_\odot}}{\rm{kpc^{2} yr}} \dfrac{\rm{km}}{\rm{s}} \right)$}}
}
\colnumbers
\startdata
M1 -- HW -- NoCR & $10^7$ & $4\times10^{-4}$ & 350 & - & - & - & $5.1\times10^{-3}$ & $2.6$ & - \\
M1 -- HW -- LPCR & $10^7$ & $4\times10^{-4}$ & 350 & 2 & $6.8 \times 10^2$ & $3.9 \times 10^{45}$ & $5.1\times10^{-3}$ & $2.6$ & $1.4 \times 10^{-1}$\\
M1 -- HW -- HPCR  & $10^7$ & $4\times10^{-4}$ & 350 & 2 & $9.5 \times 10^3$ & $5.5 \times 10^{46}$ & $5.1\times10^{-3}$ & $2.6$ & 1.9\\
M2 -- HW -- NoCR & $5 \times 10^6$ & $3\times10^{-4}$ & 110 & - & - & - & $1.2\times10^{-3} $ & $4.4 \times 10^{-1}$ & -\\
M2 -- HW -- LPCR & $5 \times 10^6$ & $3\times10^{-4}$ & 110 & 0.5 & $1.1 \times 10^3$ & $1.9 \times 10^{45}$ & $1.2 \times 10^{-3}$ & $4.4 \times 10^{-1}$ & $2.3 \times 10^{-1}$ \\
M2 -- HW -- HPCR  & $5 \times 10^6$ & $3\times10^{-4}$ & 110 & 0.5 & $9.0 \times 10^3$ & $1.6 \times 10^{46}$ & $1.2 \times 10^{-3} $ & $4.4 \times 10^{-1}$ & 1.9\\
M1 -- WW -- HD & $10^4$ & $10^{-1}$ & 15 & 12 & $6.9 \times 10^3$ & $3.9 \times 10^{45}$ & $5.4\times10^{-2} $ & $1.0$ & $1.4$ \\
M1 -- WW -- ID & $10^4$ & $10^{-2}$ & 15 & 20 & $6.9 \times 10^3$ & $3.9 \times 10^{45}$ & $5.4\times10^{-3}$ & $1.1 \times 10^{-1}$ & $1.4$\\
M1 -- WW -- LD & $10^4$ & $10^{-3}$ & 15 & 37 & $6.9 \times 10^3$ & $3.9 \times 10^{45}$ & $5.4\times10^{-4}$& $2.1 \times 10^{-2}$ & $1.4$\\
M2 -- WW -- HD & $10^4$ & $10^{-1}$ & 17 & 16 & $3.4 \times 10^3$ & $1.9 \times 10^{45}$ & $6.1\times10^{-2}$ & $1.2$ & $7.0 \times 10^{-1}$\\
M2 -- WW -- ID & $10^4$ & $10^{-2}$ & 17 & 16 & $3.4 \times 10^3$ & $1.9 \times 10^{45}$ & $6.1\times10^{-3}$ & $1.3 \times 10^{-1}$ & $7.0 \times 10^{-1}$\\
M2 -- WW -- LD & $10^4$ & $10^{-3}$ & 17 & 16 & $3.4 \times 10^3$ & $1.9 \times 10^{45}$ & $6.1\times10^{-4}$ & $2.2 \times 10^{-2}$ & $7.0 \times 10^{-1}$\\
\enddata
\tablecomments{
Columns: (1) model name, where ``LPCR'' and ``HPCR'' stand for Low CR pressure and High CR pressure, and ``HD'', ``ID'' and ``LD'' stand for High Density, Intermediate Density, and Low Density; (2) gas temperature; (3) hydrogen number density; (4) gas velocity along the vertical direction (5) ion Alfv\'{e}n speed along the vertical direction; (6) CR pressure; (7) CR energy flux along the vertical direction; (8) gas mass flux; (9) MHD momentum flux; (10) CR momentum flux.
}
\end{deluxetable*}

\section{Simulations of galactic outflows with cosmic rays}
\label{sec:windsim}

In the second part of this work, we investigate under what physical conditions CRs are able to accelerate galactic outflows. To this end, we use the same code described in \autoref{sec:equations} to perform controlled simulations of galactic outflows with imposed inflow boundary conditions. The initial conditions of these simulations are taken from the original TIGRESS solar neighborhood simulation outputs \citep{Kim+20}, with a domain extracted from the upper extra-planar region ($z > 500$~pc). 

We select two different TIGRESS snapshots -- hereafter called M1 (Model 1) and M2 (Model 2) -- from which to extract the initial conditions. 
M1 is representative of an outflow-dominated period, while  M2 is representative of a quiescent period.  
The total gas surface density and star formation rate (averaged over 40 Myr) for these two snapshots are $\Sigma_\mathrm{gas}= 9.6 \, \mo$~pc$^{-2}$ and $\Sigma_\mathrm{SFR}= 1.1 \times 10^{-2} \,\mo$~kpc$^{-2}$~yr$^{-1}$ for M1, and $\Sigma_\mathrm{gas}= 9.9 \,\mo$~pc$^{-2}$ and $\Sigma_\mathrm{SFR}= 4 \times 10^{-3} \,\mo$~kpc$^{-2}$~yr$^{-1}$ for M2.  
For the two snapshots, the main gas and magnetic field properties, measured at $z = 500$~pc and averaged in the horizontal directions, are listed in \autoref{tab:tab1}. Here, we define three different gas phases based on temperature: warm ($\rm{T}<2\times10^4~\rm{K}$), warm-hot ($2\times10^4~\rm{K}<\rm{T}<5\times10^5~\rm{K}$), and hot ($T>5\times10^5~\rm{K}$) phase. For each phase, the horizontally-averaged quantities are computed as
\begin{equation}
\bar{q}_\mathrm{ph}(z) =   \sum_\mathrm{x,y} \frac{q(x,y,z) \Theta_\mathrm{ph} (T) \Delta x \Delta y}{L_\mathrm{x}L_\mathrm{y}}\;,
\label{eq:average}
\end{equation}
with $\Delta x = \Delta y = 8$~pc the grid resolution, and $\Theta_\mathrm{ph} (T)$ the top-hat function that returns 1 for gas at temperatures within the temperature range of each phase (ph = warm, warm-hot, or hot) or 0 otherwise. 
In columns 3-4-5, the mass, MHD momentum, and MHD energy fluxes of the thermal gas and magnetic fields  along the vertical direction are respectively defined as
\begin{equation}
\mathcal{F}_\mathrm{M} =   \overline{\rho v_\mathrm{z}}\;,
\label{eq:massflux}
\end{equation}
\begin{equation}
\mathcal{F}_\mathrm{p, MHD} =   \overline{\rho v_\mathrm{z}^2 + P_\mathrm{t} + P_\mathrm{m,z}}\;,
\label{eq:momflux}
\end{equation}
\begin{equation}
\mathcal{F}_\mathrm{E, MHD} =    \overline{\left (e + P_\mathrm{t} + \frac{B^2}{2} \right) v_\mathrm{z} - (\mathbf{B}\cdot\mathbf{v})B_\mathrm{z}}\;.
\label{eq:enflux}
\end{equation}
In column 6, $f_\mathrm{A}$, the area fraction occupied by a given phase, is defined as 
\begin{equation}
{f}_\mathrm{A,ph} =   \sum_\mathrm{x,y} \frac{\Theta_\mathrm{ph} (T) \Delta x \Delta y}{L_\mathrm{x}L_\mathrm{y}}\;.
\label{eq:area}
\end{equation}
In columns 7-8-9-10-11, the ``typical'' quantities $\tilde{q}$ are  defined as the horizontally-averaged quantities divided by the fractional area occupied by each gas phase, 
\begin{equation}
\tilde{q}_\mathrm{ph}(z) = \frac{\bar{q}_\mathrm{ph}(z)}{{f}_\mathrm{A,ph}(z)} \;.
\label{eq:typicalvalue}
\end{equation}
We note that $\bar{v}_\mathrm{z}$ averages over all gas, while $\bar{v}_\mathrm{z,out}$ averages just over outflowing ($v_\mathrm{z}>0$) gas.  

From the properties listed in the table, it is clear that M1 and M2 differ in terms of thermal and dynamical gas properties. For M1, the hot gas occupies most of the volume in the extra-planar region ($f_\mathrm{A} (z = 500 \, \rm{pc}) = 0.74$), and is characterized by very high velocities ($\tilde{v}_\mathrm{z,out} (z = 500 \, \rm{pc}) = 350$~\kms). By contrast, in M2 the warm component fills up a considerable part of the volume ($f_\mathrm{A} (z = 500 \, \rm{pc}) = 0.59$), and the hot gas is characterized by a relatively low velocity ($\tilde{v}_\mathrm{z,out} (z = 500 \, \rm{pc}) = 100$~\kms). While the temperature of hot gas is more than a factor of two higher in M1 than in M2, the densities within the hot gas are similar for the two models.  

\begin{figure*}
\centering
\includegraphics[width=0.95\textwidth]
{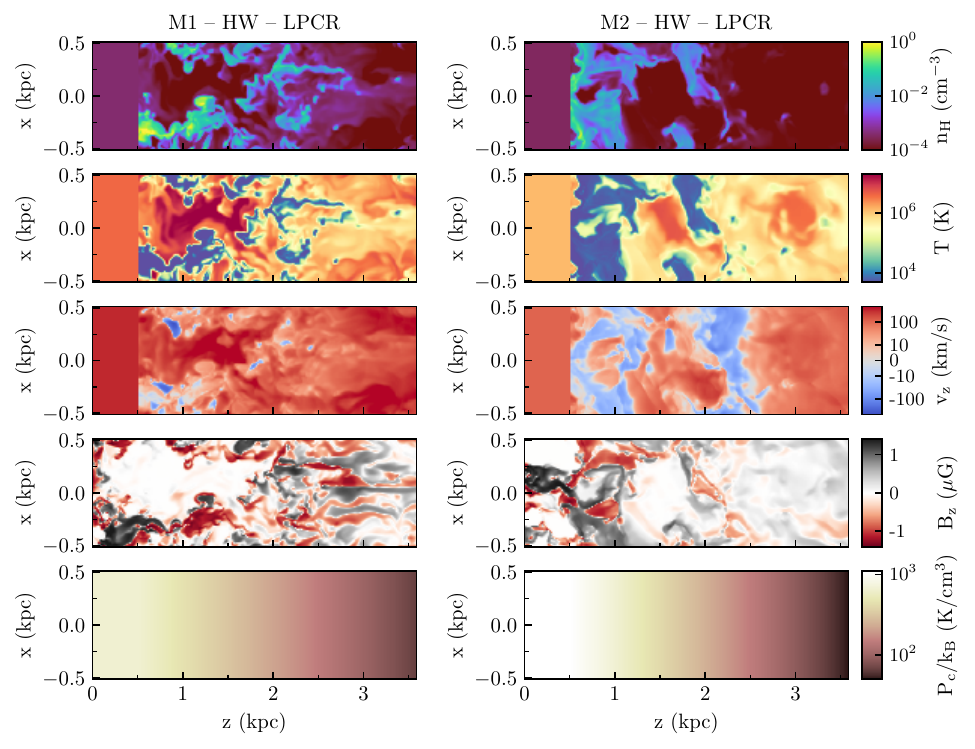}
\caption{Initial conditions of two sample galactic outflow simulations: M1 -- HW -- LPCR (\textit{left panels}) and  M2 -- HW -- LPCR (\textit{right panels}) - see \autoref{tab:tab2}. The initial conditions for the MHD variables are extracted from the M1 and M2 TIGRESS snapshots, respectively. From top to bottom, the panels show the slices through $y = 0$ of hydrogen number density $n_\mathrm{H}$, gas temperature $T$, gas vertical velocity $v_\mathrm{z}$, vertical magnetic field $B_\mathrm{z}$, and CR pressure $P_\mathrm{c}$.}
\label{fig:ICs}
\end{figure*}

\autoref{fig:ICs} shows the initial distribution on a grid of gas density, temperature, velocity, magnetic field, and CR pressure in two of the galactic outflow simulations performed in this section. The initial conditions of the MHD variables in these two simulations are extracted from the M1 and M2 TIGRESS snapshots, respectively. The simulation domain has size $L_\mathrm{x} = L_\mathrm{y} = 1024$~pc in the horizontal directions and extends from $z=0$ to $z=L_\mathrm{z}=3584$~pc in the vertical direction. For $z\geq500$~pc, the initial conditions for the thermal gas are extracted from the TIGRESS snapshot; for $z<500$~pc, the hydro variables are initialized to spatially uniform values. These are the same values assigned to the hydro variables in the ghost zones at the bottom of the $z$-axis throughout the entire simulation (see below and \autoref{tab:tab2}). The initial conditions for the magnetic field are entirely extracted from the TIGRESS snapshot.
The CR pressure is initialized to be uniform in the horizontal directions and decreasing in the vertical direction as $P_\mathrm{c}(z) = P_\mathrm{c,0} \,\mathrm{exp}[-(z-z_\mathrm{i})/H_\mathrm{c}]$, with $z_\mathrm{i} = 500$~pc. Here, the CR scale height $H_\mathrm{c} = 1.5 $~kpc is determined through a linear fit of ln$P_\mathrm{c}$ versus $z$ for $z\geq500$~pc in \autoref{fig:FBpressprof}. For $z<500$~pc, $P_\mathrm{c}$ is initialized to the normalization factor $P_\mathrm{c,0}$.
The CR energy flux is initialized to 0. 

All the simulations employ periodic boundary conditions in the horizontal directions and open boundary conditions at the top of the vertical direction, while the boundary conditions at the bottom of the vertical direction are set to user-defined values. In doing so, we control the injection of gas and CRs in the simulation box. 
\autoref{tab:tab2} lists the model names and the inflow boundary conditions. Our CR simulation model nomenclature is such that the first two letters refer to the TIGRESS snapshot from which the initial conditions are extracted. Depending on the thermal phase of the injected gas, the set of simulations is divided into two main groups, HW and WW, standing for hot wind and warm wind, respectively.
The third letters denote the different CR pressure adopted for the HW models (see column 6) or the different inflow density adopted for the WW models (see column 3). 
We describe the inflow boundary conditions employed in the HW and WW  models in \autoref{sec:hotwind} and \autoref{sec:warmwind}, respectively.

\begin{figure*}
\centering
\includegraphics[width=\textwidth]{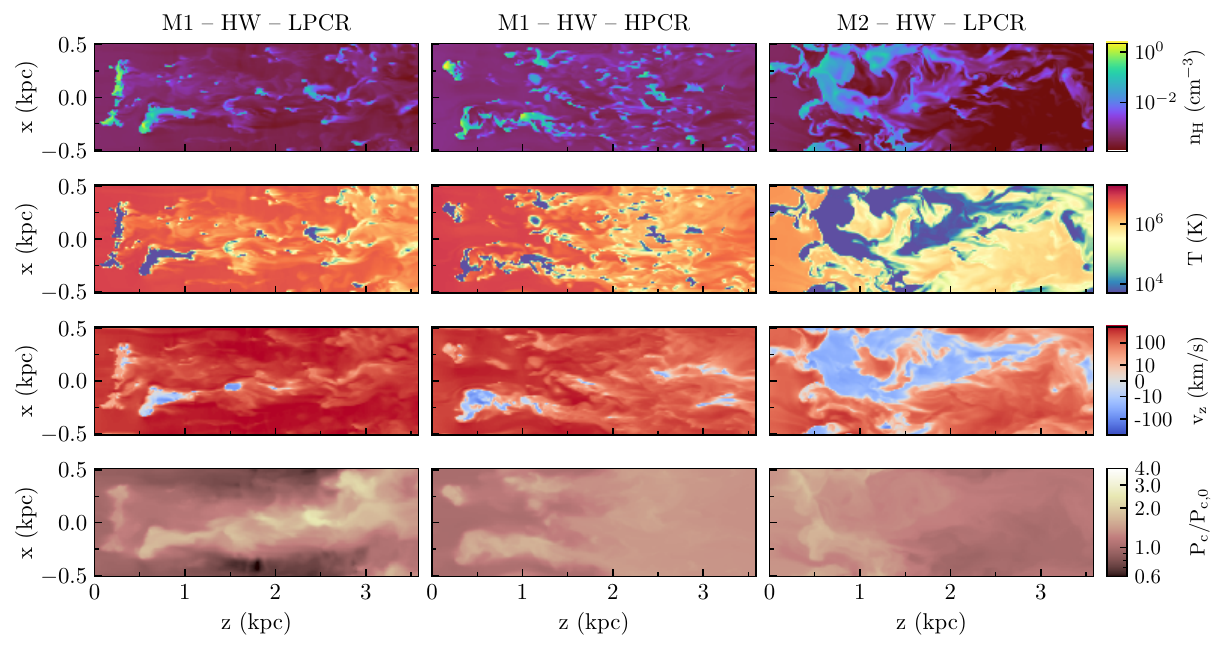}
\caption{Snapshots from three hot wind 
 simulations:  M1 -- HW -- LPCR (fast wind and low CR pressure; \textit{left column}), M1 -- HW -- HPCR (fast wind and high CR pressure; \textit{middle column}), M2 -- HW -- LPCR (relatively slow wind and low CR pressure; \textit{right column}). From top to bottom, the panels show the slices through $y = 0$ of hydrogen number density $n_\mathrm{H}$, gas temperature $T$, gas vertical velocity $v_\mathrm{z}$, and CR pressure $P_\mathrm{c}$ divided by the injected CR pressure $P_\mathrm{c,0}$. The snapshots are taken at $t = 70$~Myr.}
\label{fig:HWsnap}
\end{figure*} 

For convenience in comparing the MHD and CR fluxes, \autoref{tab:tab2} includes for each model the mass flux $\mathcal{F}_\mathrm{M,0}$, the MHD momentum flux $\mathcal{F}_\mathrm{p,MHD,0}$, and the CR momentum flux $\mathcal{F}_\mathrm{p,c,0}$ injected in the boundary conditions (note that $\mathcal{F}_\mathrm{p,c,0} \equiv P_\mathrm{c,0}$).  

\begin{figure*}
\centering
\includegraphics[width=\textwidth]{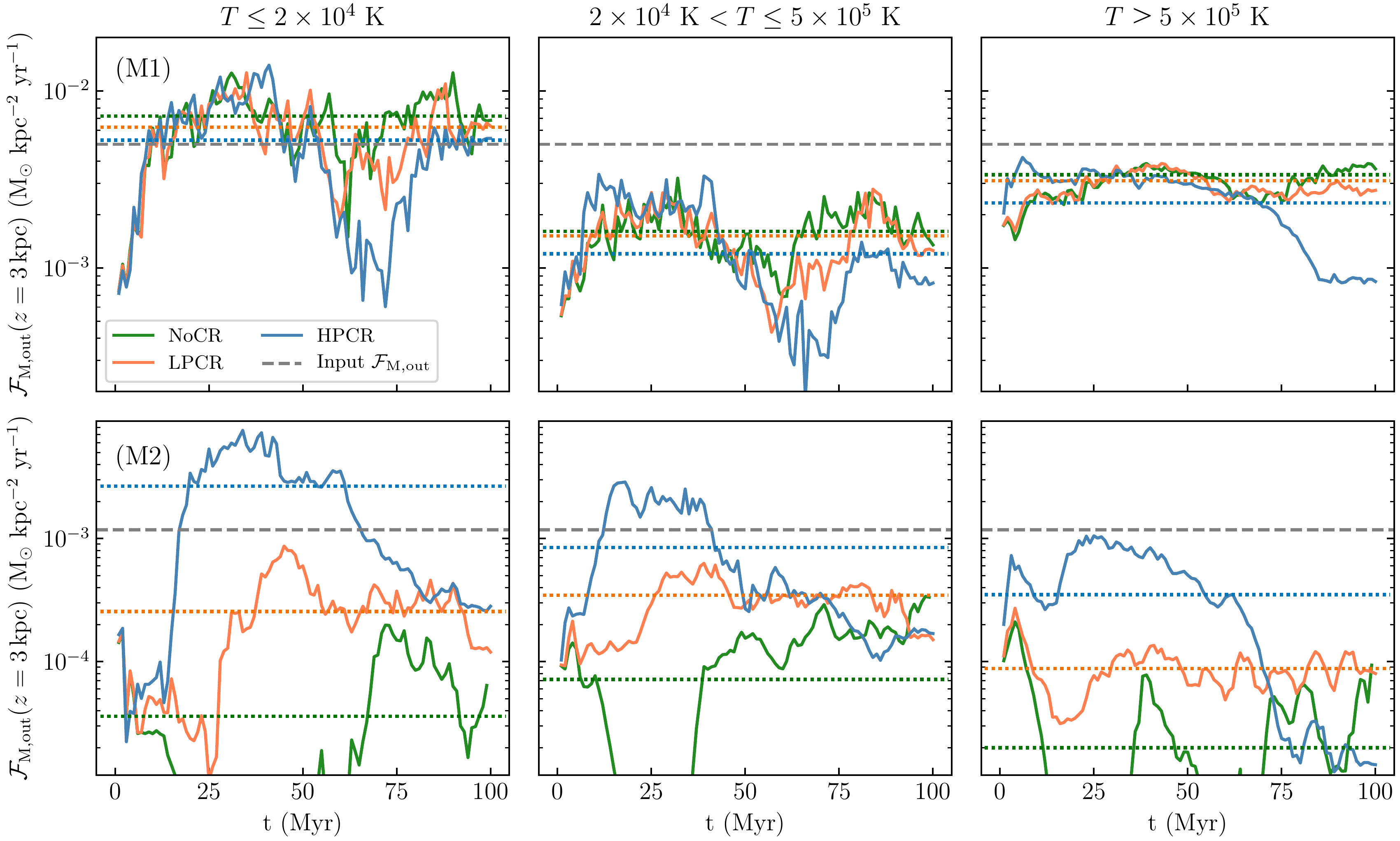}
\caption{Time evolution of outward mass fluxes $\mathcal{F}_\mathrm{M,\,out}$ for the M1 higher velocity hot wind models (\textit{top panels}) and the M2 lower velocity hot wind models (\textit{bottom panels}) simulations; note that upper and lower panels have different scales. The fluxes are taken at $z = 3$~kpc. Each column represents different phases: warm (\textit{left column}), warm-hot (\textit{middle column}), and hot (\textit{right column}). In each panel, different colors refer to different models: green for models without CRs (NoCR), coral for models with low $P_\mathrm{c}$ (LPCR), and blue for models with high $P_\mathrm{c}$ (HPCR). The dashed gray lines indicate the mass flux injected at the bottom of the simulation box ($z = 0$). The dotted lines indicate the time-averaged fluxes, with the averages taken over $t = 20 -100$ Myr.}
\label{fig:HWfluxes}
\end{figure*} 

\section{Cosmic rays and hot winds}
\label{sec:hotwind}

We start with the analysis of the HW models where hot gas is injected at the bottom of the simulation box. 
In the HW models, the inflow boundary conditions of the MHD variables are set to their initial mean values at $z = 500$~pc in the hot outflowing gas (except for the horizontal components of the gas velocity and magnetic field, which are set to 0). This means that the injected gas is hot and fast-moving in the M1 -- HW models ($T_\mathrm{0} = 10^7$~K, $v_\mathrm{z,0} = 350 \, \kms$), and mildly hot and relatively slow in the M2 -- HW models ($T_\mathrm{0} = 5 \times 10^6$~K, $v_\mathrm{z,0} = 110 \, \kms$). 

For both the M1 -- HW and the M2 -- HW cases, we run three different simulations, characterized by different conditions for CRs (\autoref{tab:tab2}). The NoCR models are control models that do not include CR physics, while the LPCR and HPCR models include CR physics, but differ in the inflow boundary conditions for the CR pressure, $P_\mathrm{c,0}$, and the vertical CR energy flux, $F_\mathrm{c,z,0}$ (the horizontal components of the CR energy flux are set to 0). 

In the LPCR models (where LPCR stands for low CR pressure), $F_\mathrm{c,z,0}$ is set to the total CR energy flux injected from supernova events in the associated TIGRESS snapshot:
$F_\mathrm{c,z,0} \equiv F_\mathrm{c,inj} = \sum_\mathrm{sp=1}^{N_\mathrm{sp} } {\dot{E}_\mathrm{c, sp}}/({L_\mathrm{x}L_\mathrm{y}})$, with the sum taken over all the star cluster particles $N_\mathrm{sp}$, and $\dot{E}_\mathrm{c, sp}$ the rate of CR energy injected from a given star cluster particle (\autoref{eq:Ecsp}).
Since the gas advection speed in hot gas dominates over the other CR propagation speeds (see $v_\mathrm{z,0}$ vs. $v_\mathrm{A_i,z,0}$ in \autoref{tab:tab2}), we compute\footnote{Here, we assume that the time-dependent term in \autoref{eq:CRflux} is negligible (for large $v_m$). In this limit, we obtain the canonical expression for the CR energy flux, $\mathbf{F_\mathrm{c}} = 4/3 e_\mathrm{c} (\mathbf{v} + \mathbf{v_\mathrm{s}}) - \tensor{\mathbf{\sigma}}^{-1} \cdot \nabla P_\mathrm{c}$. In the advection-dominated regime, this reduces to $\mathbf{F_\mathrm{c}} = 4/3 e_\mathrm{c} \mathbf{v} = 4 P_\mathrm{c} \mathbf{v}$.} $P_\mathrm{c,0}$ as $F_\mathrm{c,z,0}/(4 v_\mathrm{z,0})$. 
This choice assumes that 1) once injected, CRs do not experience additional gains/losses of energy during their propagation, and 2) CRs propagate at the same velocity of the hot gas. 
The first assumption is generally valid in the region at $z \lesssim 500$~pc, where streaming and collisional losses are smaller than the injected energy density. However, in the disk region, which contains mostly cold/warm gas, the transport of CRs is slower than would be the case if all the gas were hot and high velocity, so the second assumption provides a lower limit to the CR pressure at the disk-halo interface ($z \sim 500$~pc).

In the HPCR models (where HPCR stands for high CR pressure), $P_\mathrm{c,0}$ is set to the mean value of CR pressure (in the total gas) measured at $z = 500$~pc in the corresponding TIGRESS snapshot postprocessed with CRs, while $F_\mathrm{c,z,0} = 4 P_\mathrm{c,0} v_\mathrm{z,0}$. 
This is an upper limit to the actual CR pressure in the hot gas at $z = 500$~pc because hot gas tends to have somewhat lower CR pressure than warm gas (and therefore, lower than the average) even in simulations with time-dependent MHD and CR physics, where the distribution of CR pressure is smoother across different gas phases (see  \autoref{fig:FBsnapMHD}).

We note that the injected MHD momentum flux exceeds the injected CR momentum flux in all models except for M2 -- HW -- HPCR. In particular, in M1 -- HW -- LPCR, the injected MHD momentum flux is more than one order of magnitude larger than the injected CR momentum flux (see \autoref{tab:tab2}).

\autoref{fig:HWsnap} displays the distribution on the grid of hydrogen number density $n_\mathrm{H}$, gas temperature $T$, vertical velocity $v_\mathrm{z}$, and CR pressure $P_\mathrm{c}$ in three different models, M1 -- HW -- LPCR, M1 -- HW -- HPCR, and M2 -- HW -- LPCR, at $t = 70$~Myr. 
We note that compared to the M1 simulations, the M2 simulations are characterized by a slower injected wind (see \autoref{tab:tab2} and \autoref{fig:ICs}).

Despite their difference in CR pressure (the CR pressure imposed in the boundary conditions differs by more than one order of magnitude; see \autoref{tab:tab2}), the two M1 -- HW models exhibit overall similar gas distribution: most of the volume is occupied by hot ($T>10^6 \,{\rm K}$), rarefied and fast-moving gas, which surrounds higher-density cooler clouds/filaments with both positive and negative velocities. In model M2 -- HW -- LPCR, the gas distribution is quite different, with most of the volume occupied by gas with temperatures between $10^4-10^6 {\rm \,K}$ (``warm and warm-hot''). Within this gas, the cooler and denser component flows inward, while the hotter and more rarefied component flows outward. We emphasize that only hot gas is injected in these models. Therefore, the warm/warm-hot gas present in the simulation domain is either warm/warm-hot gas which was already present at $t=0$ and has not escaped the box, or injected gas that has cooled down. 
We estimate the net amount of hot gas that has cooled down throughout the simulation by adding the mass of warm/warm-hot gas that has left the simulation box to the difference between the final ($t=100$~Myr; end of the simulation) and the initial ($t=0$) mass of warm/warm-hot gas within the simulation domain. We find that the amount of cooled-down hot gas is about 30\% of the final mass of warm/warm-hot gas in the M1 models, and less than 20\% in the M2 model, implying that most of the warm/warm-hot gas present in the simulation domain was already present at $t=0$.

The M1 and M2 models also differ in terms of CR pressure distribution. Unlike M2 -- HW -- LPCR (which has hot gas inflowing at $110 \,\kms$), neither M1 -- HW -- LPCR nor M1 -- HW -- HPCR (which both have hot gas inflowing at $350 \, \kms$) present a clear decrease of CR pressure with increasing $z$. Rather, both of the high-velocity HW models exhibit an upward trend in CR pressure with increasing $z$. 
Since the large-scale vertical CR pressure gradient in the M1 -- HW models points away from rather than toward the disk midplane, we expect that CRs will be ineffective in accelerating gas outward.  

\autoref{fig:HWfluxes} shows, for all models, the time evolution of the outward ($v_\mathrm{z} > 0$) mass flux $\mathcal{F}_\mathrm{M, out}$ (defined as in \autoref{eq:massflux}, but just considering cells with $v_\mathrm{z}>0$) for each thermal phase (separated into three columns) taken at $z = 3$~kpc. We show the M1 -- HW and M2 -- HW models in the top and bottom panels, respectively. For all phases, the evolution of the outward mass flux is similar in the three M1 -- HW models, regardless of the presence or absence of CRs. This implies that the contribution of CRs to the acceleration of the ambient gas is negligible in these models. Interestingly, the model with higher CR pressure (M1 -- HW -- HPCR) exhibits the lowest outward mass flux at later times, especially in the hottest component of the gas. This makes the time-averaged (average over $t=20-100$~Myr) mass flux (dotted line) lower in the M1 -- HW model. By contrast, the mass flux evolution is significantly different among the three M2 -- HW models, which have a factor $\sim 3$ lower hot gas injection velocity than the M1 -- HW models ($110\,\kms$ vs. $350\,\kms$). The model without CRs (M2 -- HW -- NoCR) is characterized by the lowest mass flux, indicating that CRs are dynamically important in these models. \autoref{fig:HWfluxes} shows that, at earlier times, for the M2 models, $\mathcal{F}_\mathrm{M, out}$ is about one order of magnitude higher in the model with higher CR pressure (M2 -- HW -- HPCR) than in the model with lower CR pressure (M2 -- HW -- LPCR) for all thermal phases. However, while $\mathcal{F}_\mathrm{M, out}$ remains roughly constant with time in M2 -- HW -- LPCR, it decreases with time in M2 -- HW -- HPCR, becoming comparable to M2 -- HW -- LPCR towards the end of the simulation. 

In the following sections, we quantify to what extent the dynamical impact of CRs on the gas varies across different models, exploring the reasons behind the different trends observed in \autoref{fig:HWfluxes}.

\subsection{Momentum exchange between gas and cosmic rays}
\label{sec:momexchange}

\begin{figure*}
\centering
\includegraphics[width=\textwidth]{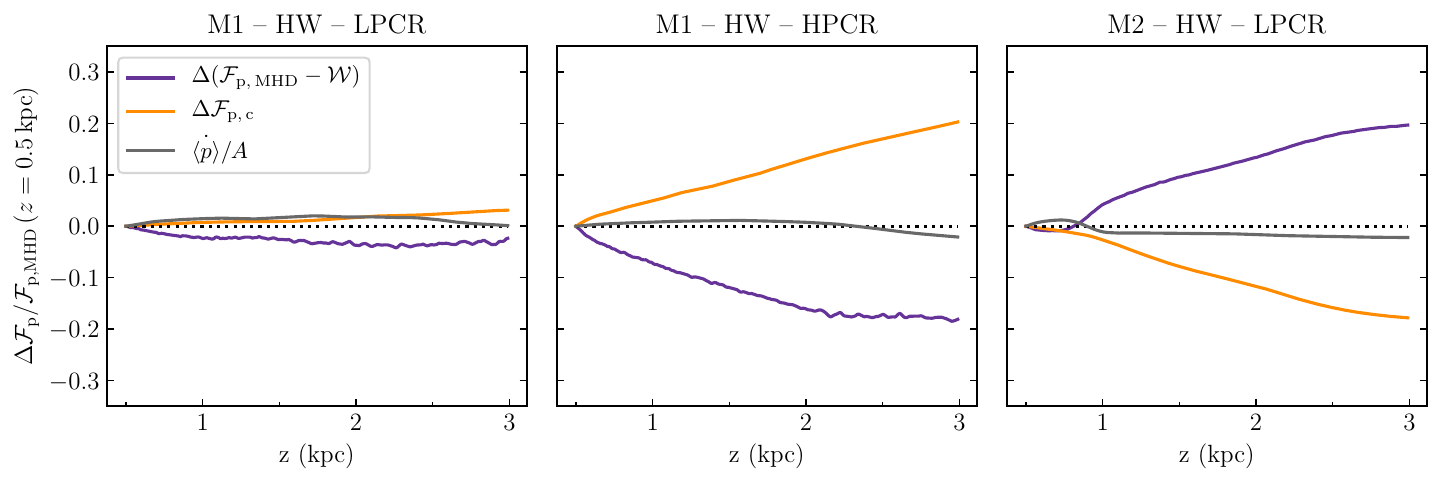}
\caption{Temporally averaged vertical profiles of the ``net'' MHD momentum flux difference $\Delta (\mathcal{F}_\mathrm{p, \,MHD} - \mathcal{W})$ (purple lines) and of the CR momentum flux difference $\Delta \mathcal{F}_\mathrm{p, \,c}$ (orange lines) in the total gas for hot wind models (average is over $t=20-100$~Myr). The difference is calculated between an arbitrary height $z$ and the initial height $z_\mathrm{i} = 0.5$~kpc. Different panels refer to different models: M1 -- HW -- LPCR (\textit{left panel}), M1 -- HW -- HPCR (\textit{middle panel}), and M2 -- HW -- LPCR (\textit{right panel}). In each panel, the gray line shows the momentum change rate per unit area $\langle \dot{p} \rangle$. The vertical profiles are divided by the MHD momentum flux $\mathcal{F}_\mathrm{p,MHD}$ measured at $z_\mathrm{i} = 0.5$~kpc.}
\label{fig:HWfluxdiff}
\end{figure*} 

In this section, we measure the exchange of vertical momentum flux between gas and CRs in different models. 
We start from the momentum equation in the vertical direction (\autoref{eq:momeq}), considering a shearing-periodic box and taking horizontal and temporal averages:
\begin{equation}
\begin{split}
  &\left <  \frac{\partial}{\partial t} ({\rho v_\mathrm{z}}) \right> (z)
+ \frac{d}{dz}  \left < P_\mathrm{k,z} + P_\mathrm{t} + P_\mathrm{m,z} \right >  (z)
\\ & + \frac{d}{dz}  \left < P_\mathrm{c} \right > (z)
=  - \left < \rho \frac{\partial}{\partial z} \Phi_\mathrm{tot}   \right> (z)
\;,
\end{split}
\label{eq:vertmomeq}    
\end{equation}
with $\left < {q} \right >$ the average over time of $\bar{q}(z;t)$, the horizontal average of a quantity $q$. In \autoref{eq:vertmomeq}, we assume that the time-dependent term and the loss term of \autoref{eq:CRflux} are on average negligible, so that $\left < \tensor{\mathrm{\sigma}}_\mathrm{tot} \cdot ( \mathbf{F_\mathrm{c}} - 4/3 e_\mathrm{c} \mathbf{v}) \right >$ reduces to $- \left < \nabla P_\mathrm{c} \right >$.
From \autoref{eq:vertmomeq}, 
$\mathcal{F}_\mathrm{p, MHD}(z) \equiv \langle P_\mathrm{k,z} + P_\mathrm{t} + P_\mathrm{m,z} \rangle$ is the contribution to the momentum flux from the MHD pressures (see \autoref{eq:momflux}),
while $\mathcal{F}_\mathrm{p, c} (z)\equiv \langle P_\mathrm{c}\rangle$ is the contribution to the momentum flux from the CR pressure. We may integrate \autoref{eq:vertmomeq} from  an initial height $z_\mathrm{i}$ to an arbitrary height $z$, and express the momentum equation in terms of momentum flux differences along the vertical direction \citep[see][]{Kim&Ostriker18,Vijayan+20,Armillotta+22}:
\begin{equation}
\langle \dot{p}_\mathrm{z} \rangle (z) +  \Delta_\mathrm{z} \mathcal{F}_\mathrm{p, MHD} (z) + \Delta_\mathrm{z} \mathcal{F}_\mathrm{p, c} (z)  = \Delta_\mathrm{z} \mathcal{W}(z)\;.
\label{eq:vertmomeq2}    
\end{equation}
Here, $\langle \dot{p}_\mathrm{z} \rangle(z)$ is the volume-integrated rate of change in $z$-momentum normalized to the area of the horizontal plane,
\begin{equation}
\langle \dot{p}_\mathrm{z} \rangle (z) = \int_{z_\mathrm{i}}^{z} \left < \frac{\partial}{\partial t}
(\rho v_\mathrm{z})
\right>
d z'\;,
\end{equation}
and the momentum flux and weight differences are defined as
\begin{equation}
\Delta_\mathrm{z} q \equiv q(z) - q(z_\mathrm{i}), 
\label{eq:diff}
\end{equation}
for 
$\mathcal{W}(z)$ the weight of gas above $z$, 
\begin{equation}
\mathcal{W}(z) = \int_z^{L_\mathrm{z}/2} \left < \rho \frac{\partial \Phi_\mathrm{tot} }{\partial z}  \right> d z'\;,
\label{eq:weight}
\end{equation}
with $L_\mathrm{z}/2$ the top of the simulation box.  Note that $-\Delta_\mathrm{z} {\cal W}(z) >0$ is the weight of gas in the range  $[z_i, z]$.

In steady state, $\langle \dot{p}_\mathrm{z} \rangle (z) \approx 0$, and \autoref{eq:vertmomeq2} reduces to:
\begin{equation}
\Delta_\mathrm{z} \mathcal{F}_\mathrm{p, MHD} (z) + \Delta_\mathrm{z} \mathcal{F}_\mathrm{p, c} (z) = \Delta_\mathrm{z} \mathcal{W}(z)\;;
\label{eq:vertmomeq3}    
\end{equation}
this says that in steady state, the momentum flux differences $\Delta_\mathrm{z} \mathcal{F}_\mathrm{p, MHD} (z)$ and $\Delta_\mathrm{z} \mathcal{F}_\mathrm{p, c} (z)$ 
together must balance the weight of gas in a given range of $z$.  

We can rearrange \autoref{eq:vertmomeq3} so that the LHS represents the ``net'' momentum flux difference of the thermal gas that arises from its interaction with CRs:
\begin{equation}
\Delta_\mathrm{z} \mathcal{F}_\mathrm{p,MHD} (z) - \Delta_\mathrm{z} \mathcal{W}(z) =  - \Delta_\mathrm{z} \mathcal{F}_\mathrm{p,c} (z) \;.
\label{eq:vertmomeq4}    
\end{equation}
We note that in the absence of CRs ($\Delta_\mathrm{z} \mathcal{F}_\mathrm{p,c} (z) = 0$), 
since the weight $-\Delta_\mathrm{z}\mathcal{W}(z)>0$ (for $z>z_i$), the momentum flux of thermal gas must always decrease as the flow moves outward ($\Delta_\mathrm{z} \mathcal{F}_\mathrm{p,MHD} (z) < 0$) to compensate for losses in climbing out of the gravitational potential. In the presence of CRs, $\Delta_\mathrm{z} (\mathcal{F}_\mathrm{p,MHD} (z) - \mathcal{W}(z))$ may 
be either positive or negative 
depending on whether CRs give momentum to ($\Delta_\mathrm{z} \mathcal{F}_\mathrm{p,c} (z) < 0$) or receive momentum from ($\Delta_\mathrm{z} \mathcal{F}_\mathrm{p,c} (z) > 0$) the thermal gas. 

\begin{figure*}
\centering
\includegraphics[width=\textwidth]{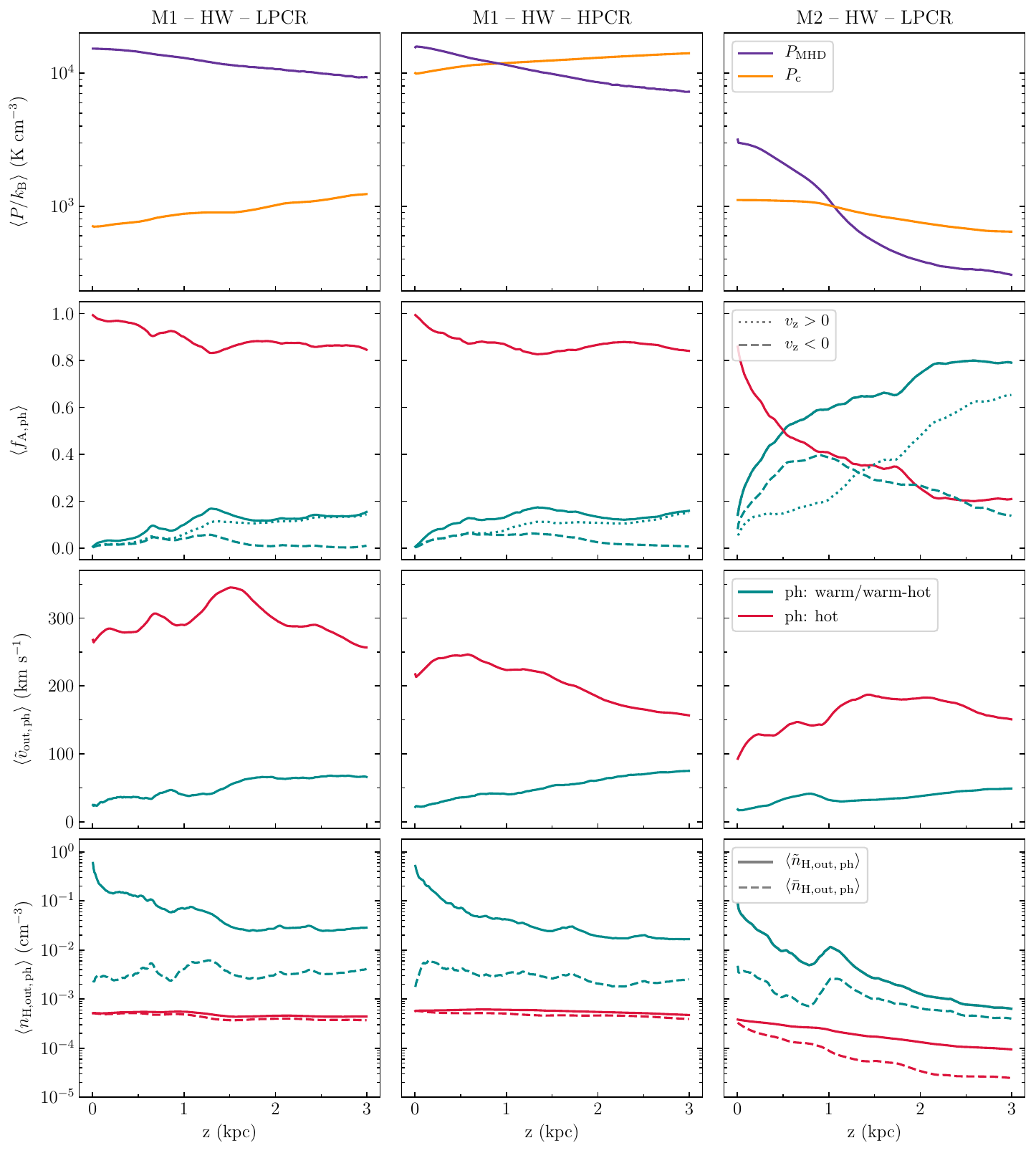}
\caption{Temporally averaged vertical profiles of CR pressure ${P}_\mathrm{c}$ and MHD pressure ${P}_\mathrm{MHD}$ (\textit{first row}), fractional areas for each thermal phase $f_\mathrm{A,ph}$ (\textit{second row}), typical velocities $\tilde{v}_\mathrm{out,ph}$ (\textit{third row}), and hydrogen number densities ${n}_\mathrm{H,out,ph}$ (\textit{forth row}) of the outflowing gas (average is over $t=20-100$~Myr). Different columns refer to different hot wind models: M1 -- HW -- LPCR (\textit{left column}), M1 -- HW -- HPCR (\textit{middle column}), and M2 -- HW -- LPCR (\textit{right column}). In the top panels, the profiles of CR and MHD pressures are indicated with orange and purple lines, respectively. In the central and bottom panels, different colors represent different gas phases: cyan for warm/warm-hot, and crimson for hot. In the second-row panels, different line styles are used to separate outflowing ($v_\mathrm{z} > 0$) from inflowing ($v_\mathrm{z} < 0$) gas: dotted for outflowing, dashed for inflowing, solid for total. In the bottom-row panels, solid and dotted lines represent the typical (\autoref{eq:typicalvalue}) hydrogen densities ${\tilde{n}}_\mathrm{H,out,ph}$ and the horizontally-averaged (\autoref{eq:average}) hydrogen densities, respectively.}
\label{fig:HWprofiles}
\end{figure*} 

\autoref{fig:HWfluxdiff} shows the time-averaged vertical profiles of the LHS and RHS terms of \autoref{eq:vertmomeq4} as measured in the three hot wind models displayed in \autoref{fig:HWsnap}, M1 -- HW -- LPCR, M1 -- HW -- HPCR, and M2 -- HW -- LPCR. The momentum flux and weight differences are calculated from an initial height $z_\mathrm{i}=500$~pc (\autoref{eq:diff}), and the temporal average is over $t=20-100$~Myr. In \autoref{fig:HWfluxdiff}, we also plot $\langle \dot{p}_\mathrm{z} \rangle$ to show that this term is negligible and the system is in quasi-steady state in the three displayed models. We omit M2 -- HW -- HPCR from this analysis as this model does not exhibit a steady behaviour, that is $\vert \langle \dot{p}_\mathrm{z} \rangle \vert \gg 0$. 

In both of the M1 simulations, $\Delta_\mathrm{z} \mathcal{F}_\mathrm{p,c}$ is \textit{positive} and increases towards larger $z$, meaning that CRs gain momentum from the gas ($\Delta_\mathrm{z}  (\mathcal{F}_\mathrm{p,MHD} (z) - \mathcal{W}(z)) < 0$) while moving outward. Over the height range we consider here, CRs gain only a few \% of the 
initial (i.e., at $z=500$pc)
MHD momentum flux in the M1 model with lower CR pressure (M1 -- HW -- LPCR), and about 20\% in the M1 model with higher CR pressure (M1 -- HW -- HPCR); for M1 -- HW -- HPCR this amounts to a 25\% increase in the CR pressure. As a consequence, the MHD momentum flux decreases faster in the latter model, thus explaining why the outward mass flux is lower here than in M1 -- HW -- LPCR (see \autoref{fig:HWfluxes}). 
In the M2 simulation, the exchange of momentum flux between gas and CRs goes in the opposite direction: $\Delta_\mathrm{z} \mathcal{F}_\mathrm{p,c}$ is  negative and decreases towards larger $z$ as a consequence of CRs transferring momentum to the gas  ($\Delta_z (\mathcal{F}_\mathrm{p,MHD} (z) - \mathcal{W}(z)) > 0$). The change in $\mathcal{F}_\mathrm{p,c}$ from $z=500$~pc to $z=3$~kpc is about 20\% of the $z=500$~pc MHD momentum flux and about 30\% of the $z=500$~pc CR momentum flux.

From this analysis, we conclude that the dynamical impact of CRs on hot outflows varies depending on the conditions of the thermal gas: in the M2 -- HW -- LPCR model, characterized by a relatively slow hot wind, CRs efficiently transfer momentum flux to the gas. By contrast, in the M1 -- HW models, where the velocities of the wind are much higher, momentum is transferred from the thermal gas to the CRs, causing the gas to decelerate. In the next section, we investigate how the different gas properties in the two groups of models control the different contribution of CRs to the overall gas dynamics.

\subsection{Vertical profiles}

In \autoref{fig:HWprofiles}, we present the time-averaged vertical profiles of the HW simulations, showing MHD pressure, CR pressure, fractional area occupied by individual thermal phases, outward gas velocity, and gas density in the simulations shown in \autoref{fig:HWfluxdiff}. 
The top panels of \autoref{fig:HWprofiles} plot the time-averaged vertical profiles of MHD pressure $P_\mathrm{MHD}$ and CR pressure $P_\mathrm{c}$ in the total gas.
In M1 -- HW -- LPCR, $P_\mathrm{c}$ is much lower than $P_\mathrm{MHD}$, thus explaining the negligible contribution of CRs to the gas dynamics in this model (see \autoref{fig:HWfluxes} and \autoref{fig:HWfluxdiff}). In M1 -- HW -- HPCR, the two pressures are comparable.  The MHD pressure decreases outward, while the CR pressure  increases outward (in agreement with \autoref{fig:HWfluxdiff}, where $\mathcal{F}_\mathrm{p,c} = \langle P_\mathrm{c} \rangle$). With a CR pressure gradient pointing upward along the $z$ axis, CRs exert a negative vertical force ($d\langle P_\mathrm{c}\rangle/dz > 0$ in \autoref{eq:vertmomeq}) that deprives the surrounding gas of momentum (see analysis in \autoref{sec:momexchange}), causing its deceleration (see third row). Finally, in M2 -- HW -- LPCR,  both $P_\mathrm{MHD}$ and $P_\mathrm{c}$ decrease with $z$, with the CR pressure profile flatter than the MHD pressure profile. In this model, CRs exert an overall positive vertical force that increases the momentum of the gas. In the range $z=[0, \,1.5\,\mathrm{kpc}]$, the hot gas accelerates, and beyond this, the velocity decreases slightly. At larger $z$, most of the volume is filled with warm and warm-hot temperature gas (see  also \autoref{fig:HWsnap}).  

\begin{figure*}
\centering
\includegraphics[width=\textwidth]{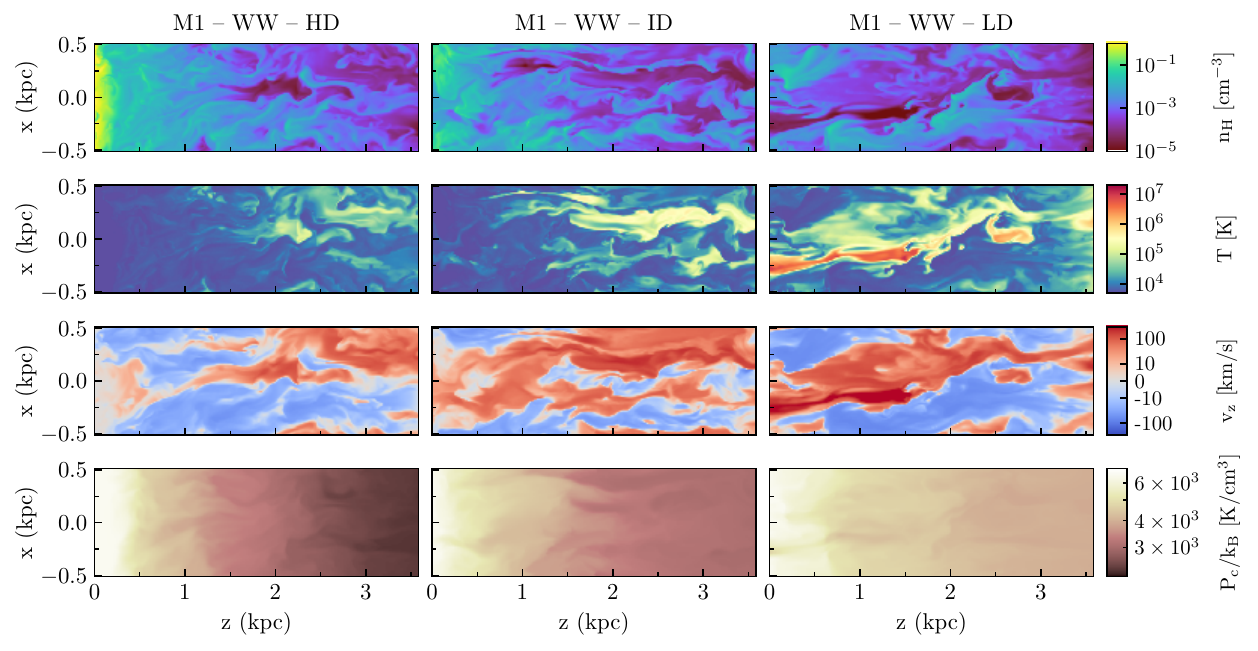}
\caption{Same as \autoref{fig:HWsnap}, but for the warm wind models M1 -- WW -- HD (\textit{left column}), M1 -- WW -- ID (\textit{middle column}), and M1 -- WW -- LD (\textit{right column}).}
\label{fig:WWsnap}
\end{figure*} 

The different CR distribution in M1 -- HW 
and M2 -- HW -- LPCR depends on the different thermal and dynamical properties of the gas. In M1 -- HW, the fast-moving hot gas helps to push the lower-velocity, warmer gas present in the simulation domain towards higher altitudes. The interaction with a faster-moving fluid shreds the dense cloudlets during their motion, triggering mixing between the two fluids (see \autoref{fig:HWsnap}). This results in a small increase in the mean velocity of warm/warm-hot gas with z, although this remains $<100\, \kms$.  As shown in the second row of \autoref{fig:HWprofiles}, the fractional area occupied by the warm/warm-hot gas increases from about $0.01 \%$ to 20\% within $z = 3$~kpc. 
Since the volume remains predominantly occupied by the hot gas in both M1 -- HW models (with $f_\mathrm{A,w/w-h} < 0.2$), the CR transport is primarily controlled by hot gas advection. The decrease in the hot gas velocity at higher $z$ in the M1 -- HW -- HPCR model leads to an increase in CR pressure with $z$ (in the advection-dominated limit, $P_c \propto \rho^{4/3} \propto v_z^{-4/3}$), and at the same time the positive CR pressure gradient and interaction with warm gas decelerates the hot gas. Because CR transport is slower in warm gas than hot gas, an additional factor contributing to CR pressure buildup at large $z$ is the increase of $f_\mathrm{A,w/w-h}$ with height.    

In M2 -- HW -- LPCR, the relatively slow hot wind is less able to accelerate the warm and warm-hot gas. Also, the fraction of warm and dense gas initially present in the simulation domain is much higher than in M1. \autoref{fig:HWprofiles} shows that for $z \sim 0.5 - 1$~kpc, $\sim 30\%$ of the volume is occupied by inflowing ($v_\mathrm{z}<0$) warm/warm-hot temperature gas. CRs stall in the high-density region at low altitudes slightly increasing their pressure. 
Similarly to the M1 models, $f_\mathrm{A,w/w-h}$ increases with increasing $z$ in M2 -- HW -- LPCR. In this model, though, 
warm/warm-hot gas occupies most of the volume for $z>1$~kpc, and therefore it is this material that primarily governs the CR transport. This warm/warm-hot gas is mainly outflowing, with an average velocity increasing with $z$ (see the right panel in the third row of \autoref{fig:HWprofiles}). This implies that the advection of CRs out of the simulations domain increases with $z$, thus explaining the decrease of $P_c$ with $z$. We note that the large fraction of warm gas present in M2 -- HW -- LPCR makes this model akin to the warm wind models discussed in the next section. We therefore refer to \autoref{sec:warmwind} for a detailed discussion of the role of CRs in accelerating outflows in the warm gas. 

The middle panel in the third row of \autoref{fig:HWprofiles} shows that the typical (see \autoref{eq:typicalvalue}) outflow velocity of the hot gas $\tilde{v}_\mathrm{out,hot}$ decreases with $z$ in M1 -- HW -- HPCR, where CRs act to decelerate the wind. Because of this deceleration, despite the identical initial gas conditions, the typical outflow velocities are lower in  M1 -- HW -- HPCR than in  M1 -- HW -- LPCR, where the CR pressure is negligible compared to the other pressures. By contrast, in M2 -- HW -- LPCR, $\tilde{v}_\mathrm{out,hot}$ increases with $z$ for $z<1.5$~kpc, while flattening at higher altitudes. The typical outflow velocity of the warm/warm-hot gas always increases with $z$, even in the M1 -- HW models where CRs decelerate the hot gas. This happens both because the warm/warm-hot gas gains momentum from the hotter and faster-moving gas, and because lower-velocity cloudlets increasingly ``drop out'' of the flow with height \citep[see also][]{Vijayan+20, Armillotta+22}.

Finally, the bottom panels of \autoref{fig:HWprofiles} show that, in the M2 -- HW -- LPCR model, the warm/warm-hot outflowing gas exhibits lower typical densities compared to the M1 -- HW models. 
Part of the reason for the lower densities in the M2 model is that, with a hot gas mass injection rate four times smaller than in the M1 -- HW models (see \autoref{tab:tab2}), the warm/warm-hot gas can expand in area, dropping its density. Furthermore, unlike the M1  -- HW models, in the M2 model the horizontally-averaged density of the warm/warm-hot gas (dashed line) decreases outward with $z$ due to the inefficiency of the lower-velocity hot gas in sweeping up the warm cloudlets.
With the lower-velocity gas dropping out, the higher-velocity warm/warm-hot gas expands in area, further reducing its density.

\section{Cosmic-ray driven warm winds}
\label{sec:warmwind}

In this section, we analyse the simulations in which slow-moving warm ($T = 10^4$~K) gas is injected at the bottom of the simulation box (referred to as WW models in \autoref{tab:tab2}). 
In the WW models, the injected gas is warm ($T_\mathrm{0} = 10^4$~K) and slow-moving ($v_\mathrm{z,0} \sim 10-20 \,\kms$). The inflow boundary conditions for the vertical components of gas velocity and magnetic field are set to their initial mean values in the warm outflowing gas at $z=500$~pc. 
As in the HW--LCPR models, we fix $F_\mathrm{c,z,0}$ to the total CR energy flux injected in the associated postprocessed TIGRESS snapshot.
$P_\mathrm{c,0} $ is then computed as $F_\mathrm{c,z,0}/[4(v_\mathrm{z,0}+v_\mathrm{A_i,z,0})]$, 
assuming that both streaming and advection contribute to the propagation of CRs in the warm gas (diffusion is negligible at the gas densities explored here, except for the high-density HD model). We note that for the WW models, $P_\mathrm{c,0}$ is comparable to the average value of CR pressure measured in the postprocessed snapshot at $z=500$~pc. Hence, there is no need for a second ``HPCR'' model defined using the postprocessed CR pressure for $P_\mathrm{c,0}$, as in the HW models. 

For the WW models, we explore the effects of three different densities for the injected gas: $n_\mathrm{H,0} = 10^{-1}, 10^{-2}, 10^{-3}$~cm$^{-3}$ for models HD, ID, and LD, respectively, which stand for high, intermediate, and low density. The HD models have density closest to that of the warm fountain gas in the 
TIGRESS simulation (see \autoref{tab:tab1}).
For each set of M1 and M2 simulations, we employ the same value of $P_\mathrm{c,0}$. This is computed for $v_\mathrm{A_i,z,0}$ from the ID model, rather than for the value of $v_\mathrm{A_i,z,0}$ of the corresponding model (in that case, $P_\mathrm{c,0}$ would vary by less than a factor 1.5).
In the M2 -- WW models, the injected CR pressure is almost a factor of two lower compared to the M1 -- WW models. 
We note that the injected CR momentum flux is larger than the MHD momentum flux for all models except for M2 -- WW -- HD.
We find that our conclusions regarding the effect of varying the injected density in WW models are independent of the initial conditions employed in the simulation (M1 vs M2). Hence, in the following, we mostly focus on the analysis of the M1 simulations.

\autoref{fig:WWsnap} displays the distribution on the grid of hydrogen number density $n_\mathrm{H}$, gas temperature $T$, vertical velocity $v_\mathrm{z}$, and CR pressure $P_\mathrm{c}$, in the three M1 -- WW models at t = 100 Myr. 
Unlike the HW models, $P_\mathrm{c}$ now decreases outward along the vertical direction in all models, suggesting a non negligible  contribution of CRs to counteracting gravity and accelerating gas in this extraplanar region. A comparison between the three panels shows that the large-scale vertical gradient of CR pressure becomes smaller going from the model with higher injected gas density (M1 -- WW -- HD) to the model with lower injected density (M1 -- WW -- LD). At $z>1$~kpc, the gas density and velocity distributions are qualitatively similar in the three models: lower-density gas flows outward, while higher-density gas mostly flows inward. The outflowing gas is clearly accelerated by CR pressure forces as thermal pressure forces are negligible in the warm gas populating most of the simulation box. A difference that emerges from a visual comparison is that the fraction of gas at $T \gtrsim 10^5$~K (color-coded in yellow and red in the second row of \autoref{fig:WWsnap}) increases going from M1 -- WW -- HD to M1 -- WW -- ID to M1 -- WW -- LD. At the time the snapshots are taken ($t = 100$~Myr), the fast-moving hot gas initially present in the simulation domain has already escaped the box. As we shall discuss in \autoref{sec:heating}, gas at $T>10^4$~K currently present in the simulation box is low-density gas heated up by CRs. 

\begin{figure*}
\centering
\includegraphics[width=\textwidth]{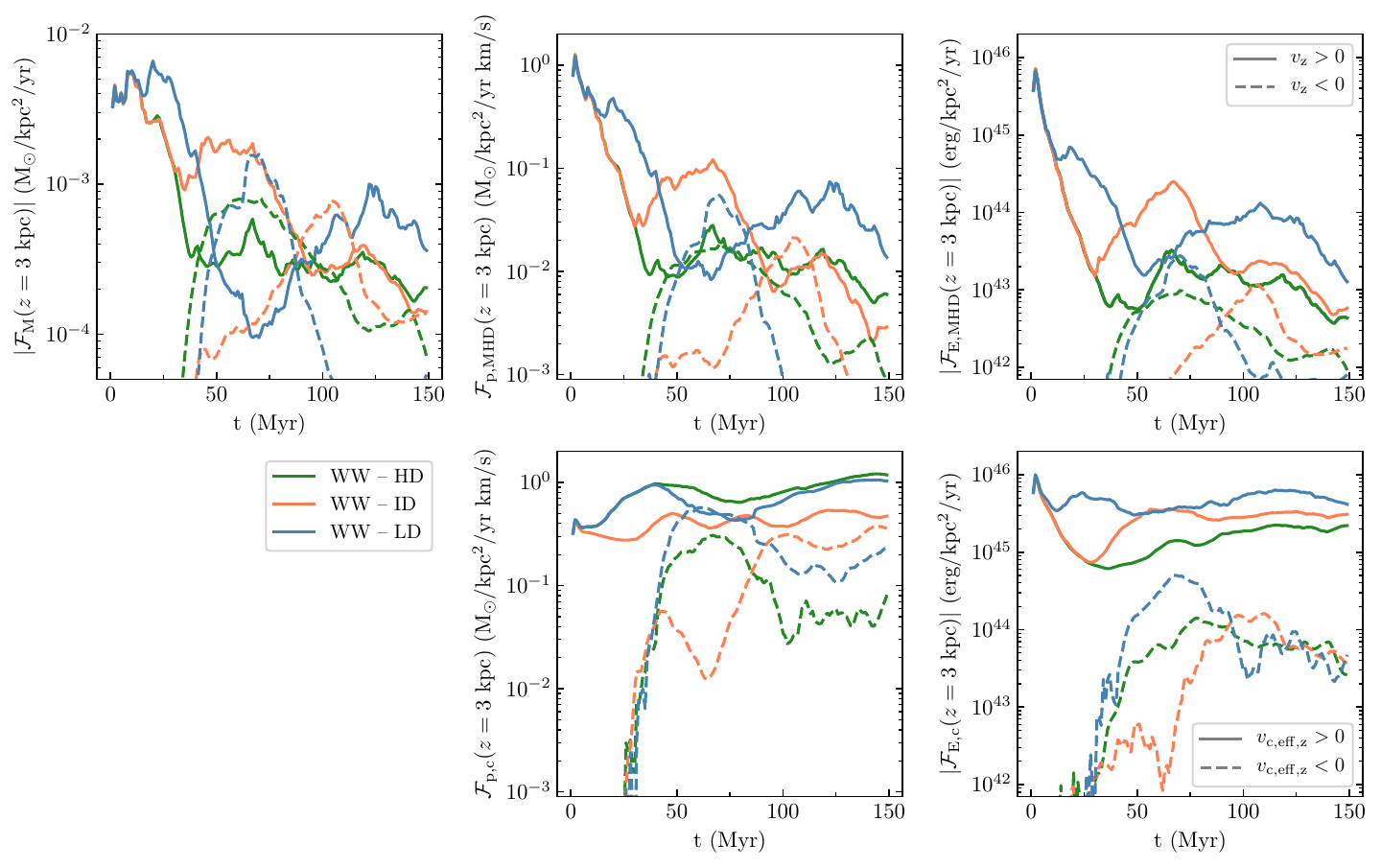}
\caption{Time evolution of mass fluxes $\mathcal{F}_\mathrm{M}$ (\textit{upper left panel}), MHD momentum fluxes $\mathcal{F}_\mathrm{p, MHD}$ (\textit{upper middle panel}), MHD energy fluxes $\mathcal{F}_\mathrm{E, MHD}$ (\textit{upper right panel}), CR momentum fluxes $\mathcal{F}_\mathrm{p, c}=\langle P_\mathrm{c}\rangle$ (\textit{lower middle panel}), CR energy fluxes $\mathcal{F}_\mathrm{E, c}$ (\textit{lower right panel}) taken at $z = 3$~kpc for the three M1 -- WW warm wind models. In each panel, different colors refer to different models: green for the model with high inflow density (WW -- HD), coral for the model with intermediate inflow density (WW -- ID), and blue for the model with low inflow density (WW -- LD). In top panels, the solid and dashed lines represent the fluxes measured in the outflowing ($v_\mathrm{z}>0$) and inflowing ($v_\mathrm{z}<0$) gas, respectively; in the bottom panels, the solid and dashed lines represent the fluxes measured in the outflowing ($v_\mathrm{c,eff,z}\equiv F_\mathrm{c}/4P_\mathrm{c} >0$) and inflowing ($v_\mathrm{c,eff,z}<0$) CR fluid, respectively.}
\label{fig:WWfluxes}
\end{figure*} 

\begin{figure*}
\centering
\includegraphics[width=\textwidth]{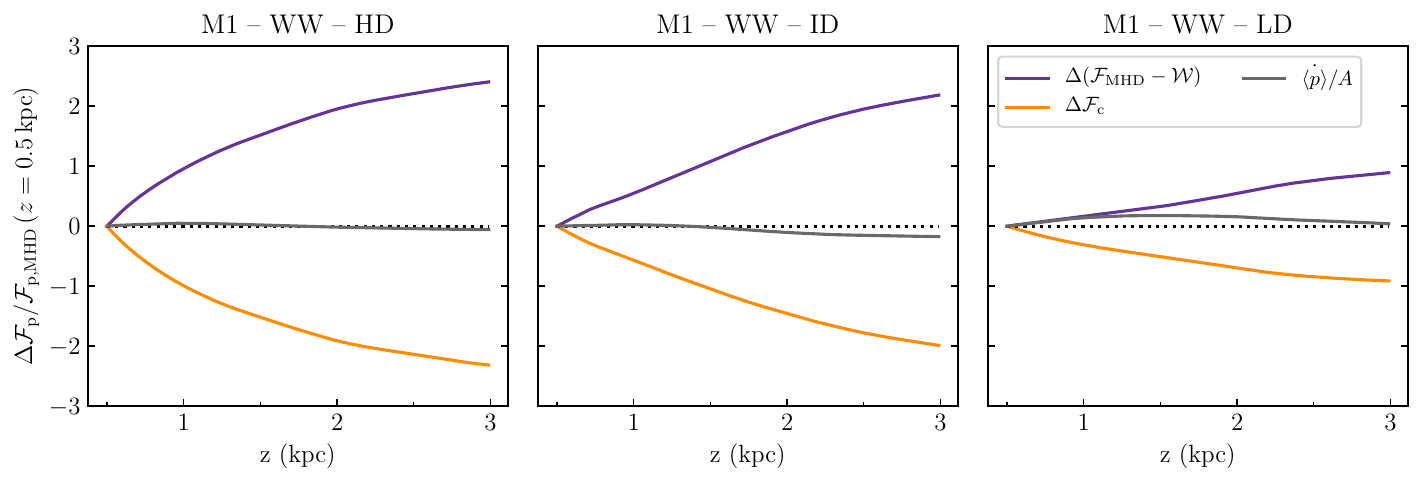}
\caption{Temporally averaged vertical profiles of momentum flux difference for the warm wind simulations M1 -- WW -- HD (\textit{left panel}), M1 -- WW -- ID (\textit{middle panel}), and M1 -- WW -- LD (\textit{right panel}); average is over $t=50-150$~Myr. The purple line shows the ``net'' MHD momentum flux difference $\Delta (\mathcal{F}_\mathrm{p, \,MHD} - \mathcal{W})$ taking into account the flux loss due to gas climbing out of the potential, the orange line shows the CR momentum flux difference $\Delta \mathcal{F}_\mathrm{p, \,c}$, while the gray line shows the momentum change rate per unit area $\langle \dot{p} \rangle/A$. The difference is calculated between an arbitrary height $z$ and the initial height $z_\mathrm{i} = 0.5$~kpc.  In all panels, the profiles are divided by the MHD momentum flux $\mathcal{F}_\mathrm{p, \,MHD}$ at $z_\mathrm{i} = 0.5$~kpc. The time average is taken starting from $t = 30$~Myr to $t = 150$~Myr. On average, the system is in quasi-steady state ($\langle \dot{p} \rangle$ is much smaller than the other terms).}
\label{fig:WWfluxdiff}
\end{figure*} 

In the top panels of \autoref{fig:WWfluxes}, we plot the time evolution of the outward ($v_\mathrm{z}>0$) and inward ($v_\mathrm{z}<0$) mass flux $\mathcal{F}_\mathrm{M}$ (\autoref{eq:massflux}), MHD momentum flux $\mathcal{F}_\mathrm{p,MHD}$ (\autoref{eq:momflux}), and MHD energy flux $\mathcal{F}_\mathrm{E,MHD}$ (\autoref{eq:enflux}) taken at $z=3$~kpc in the three models. 
Unlike our approach in \autoref{fig:HWfluxes}, here we do not separate the fluxes of the three different thermal phases since the fraction of hot gas is negligible (in terms of both volume and mass) in the WW models. We do, however, separate outward and inward MHD fluxes based on the sign of $v_z$.
In the three models shown, $\mathcal{F}_\mathrm{M}$, $\mathcal{F}_\mathrm{p,MHD}$, and $\mathcal{F}_\mathrm{E,MHD}$ decrease after $\sim 10$~Myr, as the fast-moving hot gas initially present in the simulation domain escapes the box. Overall, the fluxes at $z=3$~kpc are comparable in the three models despite two orders of magnitude difference in the injection density. In all models, the outward fluxes are on average higher than the inward fluxes (see also \autoref{sec:ww_profiles}). We note that, at $z=3$~kpc, $\mathcal{F}_\mathrm{M}$ in the M1 -- WW models is lower than in the M1 -- HW models (see \autoref{fig:HWfluxes}; the ratio is $\sim 6-10$ for time-averaged mass fluxes in HW vs. WW models). This is true even for M1 -- WW -- HD, which has an order of magnitude higher mass injection rate than the M1 -- HW models; the dense inflowing gas in M1 -- WW -- HD simply builds up at low $z$ (see \autoref{fig:WWsnap}). 
 
The bottom panels of \autoref{fig:WWfluxes} display the time evolution of the CR momentum flux $\mathcal{F}_\mathrm{p,c} \equiv \langle P_\mathrm{c} \rangle$, and CR energy flux $\mathcal{F}_\mathrm{E,c} \equiv \langle {F}_\mathrm{c}(z) \rangle$ at $z=3$~kpc. The CR fluxes are separated into outward and inward fluxes based on the sign of the vertical effective CR propagation velocity $v_\mathrm{c,eff,z}\equiv F_\mathrm{c}/(4P_\mathrm{c})$. The energy flux of CRs is completely dominated by the outflowing term. \autoref{fig:WWsnap} also shows that pressure is fairly uniform at any given height.  The regions with lower gas density generally correspond to the locations which have $v_\mathrm{c,eff,z}>0$; this is the majority of the area, such that  the momentum flux of the outflowing CRs is larger than the momentum flux of the inflowing CRs at $z=3$~kpc. In all models, the CR fluxes at $z=3$~kpc remain more than one order of magnitude larger than the corresponding MHD fluxes, although they are lower than their injection values (see \autoref{tab:tab2}). 

\subsection{Transfer of momentum to the gas}\label{sec:mom_echange_WW}

Similarly to what we have done in \autoref{sec:momexchange} for the HW models, in this section we quantify the exchange of momentum flux between gas and CRs outward along the vertical direction in different models. \autoref{fig:WWfluxdiff}
displays the time-averaged vertical profiles of the ``net'' MHD momentum flux difference $\Delta_\mathrm{z}  (\mathcal{F}_\mathrm{p,MHD} (z) - \mathcal{W}(z))$, and CR momentum flux difference $\Delta_\mathrm{z}  \mathcal{F}_\mathrm{p,c} (z)$ (see \autoref{eq:vertmomeq4}) in the three M1 -- WW models. The differences are calculated from an initial height $z_\mathrm{i}=500$~pc, and the temporal averages are over $t=50-150$~Myr. We discard the first 50 Myr from the analysis to avoid transients affected by the initial conditions, where a fast-moving hot wind is present. In all panels, $\langle \dot{p}_\mathrm{z} \rangle$ is negligible, meaning that the systems are in statistical steady state.

The main evidence of \autoref{fig:WWfluxdiff} is that unlike the situation shown for hot winds in \autoref{fig:HWprofiles}, $\Delta_\mathrm{z} \mathcal{F}_\mathrm{p,c}$ is always negative and decreasing towards larger $z$ in the WW models, confirming that CRs effectively transfer momentum to the gas as the latter flows towards higher altitudes. 
Over the height range we consider here, the gas gains more than $200 \%$ of the initial (i.e. $z  = 500$~pc) MHD momentum flux 
(corresponding to about $50 \%$ of the $z  = 500$~pc CR momentum flux) in M1 -- WW -- HD and M1 -- WW -- ID, while the gas gains about $ 100 \%$ of the initial MHD momentum flux (corresponding to less than $20 \%$ of the $z  = 500$~pc CR momentum flux) in M1 -- WW -- LD. In the next section, we use additional information from vertical profiles to analyse the main differences, in terms of CR and MHD properties, among the three models.  

\begin{figure*}
\centering
\includegraphics[width=\textwidth]{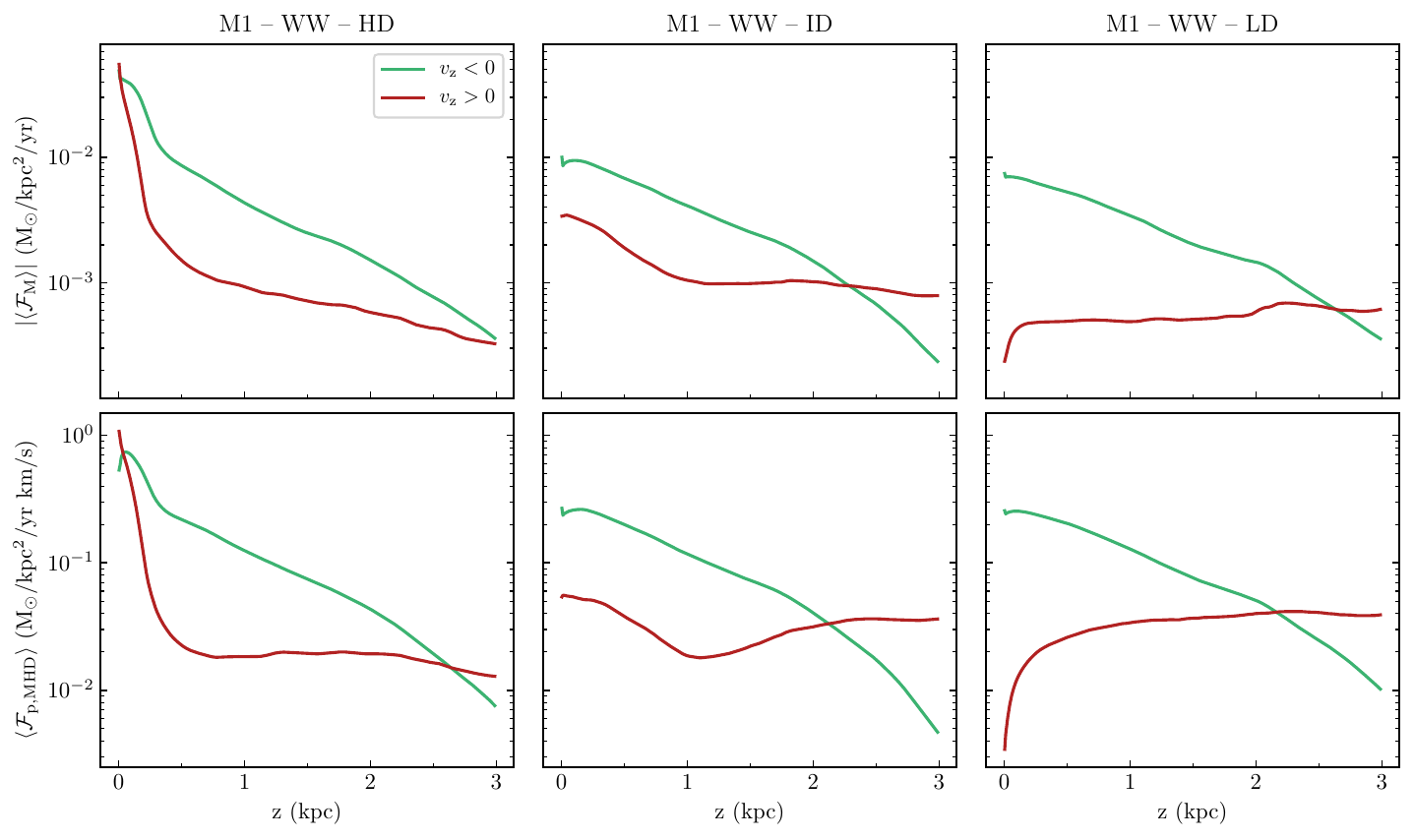}
\caption{Temporally averaged mass flux $\mathcal {F}_\mathrm{M}$ (\textit{upper panels}) and MHD momentum flux $\mathcal {F}_\mathrm{p, \, MHD}$ (\textit{bottom panels}) as a function of $z$ for the warm wind simulations M1 -- WW -- HD (\textit{left panels}), M1 -- WW -- ID (\textit{middle panels}), and M1 -- WW -- LD (\textit{right panels}); average is over $t=50-150$~Myr. The green and red lines represent the outward and inward mass fluxes, respectively.
}
\label{fig:WWMflux}
\end{figure*} 

\subsection{Vertical profiles}
\label{sec:ww_profiles}

In \autoref{fig:WWMflux}, we analyse the time-averaged vertical profiles of the outward and inward mass and MHD momentum fluxes from the three M1 -- WW models. Between $\sim 0.5-2.5$~kpc, the inward mass and momentum fluxes in all models are fairly similar; this reflects population of inflowing gas present in the initial conditions.  
In the model with higher gas density (M1 -- WW -- HD), there is more than an order of magnitude decrease in $\vert \mathcal{F}_\mathrm{M} (v_\mathrm{z}>0) \vert $ within the first $\sim 0.5$~kpc, due to rapid dropout of gas (which is injected with a velocity of only $15\,\kms$). Beyond this point, $\vert \mathcal{F}_\mathrm{M} (v_\mathrm{z}>0) \vert $ decreases more slowly with $z$, since CRs effectively transfer momentum to the gas (see \autoref{fig:WWfluxdiff}). 
In M1 -- WW -- ID, $\vert \mathcal{F}_\mathrm{M} (v_\mathrm{z}>0) \vert $ decreases by a factor of a few within 
$z\sim 1$~kpc, after which it remains flat. 
In M1 -- WW -- LD, the outward mass flux $\vert \mathcal{F}_\mathrm{M} (v_\mathrm{z}>0) \vert $  initially increases at small $z$, and then plateaus at a value slightly lower than in M1 -- WW -- ID. 
The vertical profiles of the MHD momentum fluxes are qualitatively similar to the mass flux profiles.
After a steep decrease at $z<0.5$~kpc in model M1 -- WW -- HD and a steep increase in model M1 -- WW -- LD, the outward MHD momentum flux 
$\mathcal{F}_\mathrm{p, \, MHD} (v_\mathrm{z}>0)$  remains nearly constant with $z$, at similar levels for all models.  

\begin{figure*}
\centering
\includegraphics[width=\textwidth]{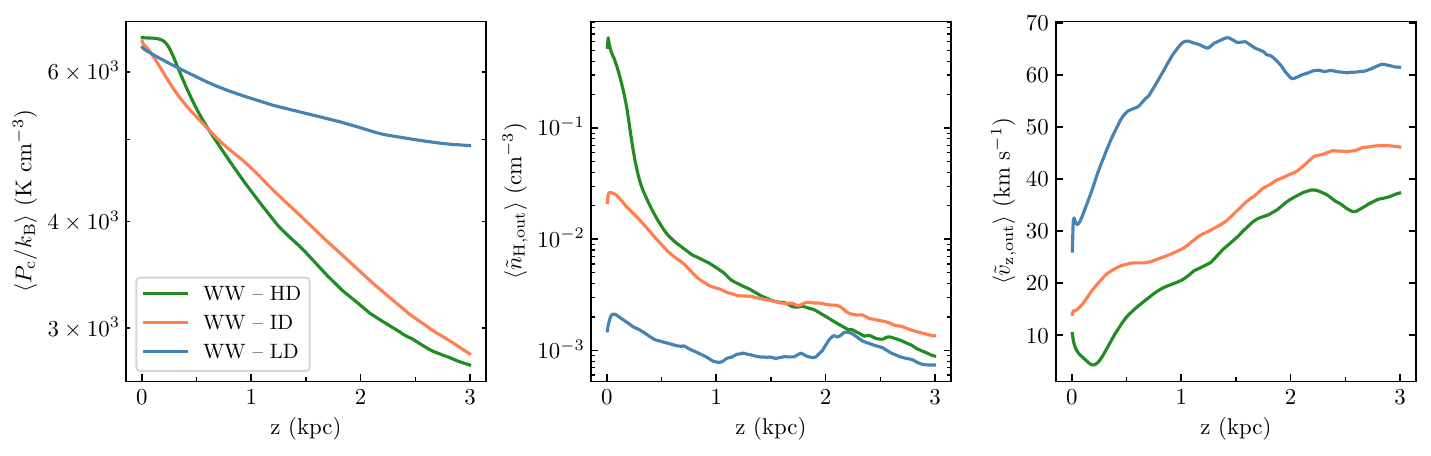}
\caption{Temporally averaged vertical profiles of CR pressure $P_\mathrm{c}$ (\textit{left panel}), typical hydrogen number density $\tilde{n}_\mathrm{H, out}$ (\textit{middle panel}), and typical vertical velocity $\tilde{v}_\mathrm{z, out}$ (\textit{right panel}) of the outflowing gas. Different colors represent different warm wind models: green for M1 -- WW -- HD, coral for M1 -- WW -- ID, blue for M1 -- WW -- LD.}
\label{fig:WWzprof1}
\end{figure*}

To understand the warm wind models in more detail, in the left panel of \autoref{fig:WWzprof1}, we plot the temporally and horizontally averaged vertical profiles of CR pressures. The CR pressure profile is less steep in M1 -- WW -- LD than in the other two models, consistent with the smaller fractional exchange in momentum flux between gas and CRs discussed in \autoref{sec:mom_echange_WW} (20\% vs. 50\% of the CR flux at $z=0.5$~kpc, where $\mathcal{F}_\mathrm{p,c} \equiv \langle P_\mathrm{c} \rangle$). 
Models M1 -- WW -- HD and M1 -- WW -- ID exhibit similar CR pressure profiles, except for the region at $z \lesssim 300$~pc, where the profile in model M1 -- WW -- HD is relatively flat due to the effective CR diffusion at high density ($n_\mathrm{H} > 0.1$~cm$^{-3}$ at $z \lesssim 300$~pc; see middle panel of \autoref{fig:WWzprof1}). At higher $z$, CR diffusion becomes negligible as the gas density rapidly drops.

In the middle panels of \autoref{fig:WWzprof1}, we display the temporally averaged typical density of the outflowing gas $\tilde{n}_\mathrm{H,out}$ as a function of $z$. In all models, the value of $\tilde{n}_\mathrm{H,out}$ at $z=0$ is larger than the density of the injected gas $n_\mathrm{H,0}$. Due to the inflow boundary conditions employed at the bottom of the vertical axis, gas flowing inward cannot leave the simulation box, but rather accumulates near the boundary, increasing its density. In M1 -- WW -- HD, $\tilde{n}_\mathrm{H,out}$ has to drop by more than one order of magnitude, to $\sim 10^{-2} \, \pcc$, before CR pressure gradient forces begin to have an effect.
This is confirmed by the right panel of \autoref{fig:WWzprof1}, showing that the outward vertical velocity $\tilde{v}_\mathrm{z,out}$ initially decreases until $z\sim 0.3$ kpc. 
At higher $z$, CR forces effectively accelerate the gas, as demonstrated by the vertical velocity increasing with $z$. In the lower-density models, much less gas builds up, and the flow accelerates outward starting near $z=0$. For $z \gtrsim 0.5$ kpc, the profiles of CR pressure and gas density in  M1 -- WW -- HD and M1 -- WW -- ID become roughly similar. This results in similar gas acceleration, as shown in the right panel of \autoref{fig:WWzprof1}, where the offset in gas velocity is due to the different velocity near the inflow (at $z \lesssim 500$ pc). The mean outward velocities at $z = 3$~kpc are $\simeq 38$ and $45\,\kms$ in M1 -- WW -- HD and M1 -- WW -- ID, respectively. The outward velocities reach $v_\mathrm{z,out} \gtrsim 100\,\kms$ (well above the averages in \autoref{fig:WWzprof1}) in low-density ($n_\mathrm{H}\lesssim 10^{-3}$~cm$^{-3}$), high-temperature ($T \gg 10^{5}$~K) regions (see \autoref{fig:WWsnap}).

In M1 -- WW -- LD, the injected gas has sufficiently low density to be very efficiently accelerated by CR forces. This is clearly visible in the right panel of \autoref{fig:WWzprof1}, showing that the outward velocity rapidly increases with $z$ in the region at $z \lesssim 1$~kpc. At high altitudes, the outward gas velocity becomes relatively flat, meaning that CR forces merely balance gravity. 
Due to the effective acceleration, the outward mass flux does not drop with $z$ as in the other models (see \autoref{fig:WWMflux}), leading to a smaller fractional variation of $\tilde{n}_\mathrm{H,out}$. As we shall see in \autoref{sec:1Dmodel}, streaming dominates the transport of CRs at low altitudes in the WW models. In this regime, $P_\mathrm{c} \propto v_\mathrm{A,i}^{-4/3} \propto {n_\mathrm{i}}^{2/3}$. With a smaller fractional drop in gas density, the drop in CR pressure is smaller in the LD model.

In summary, the transfer of momentum flux from the CR population to the gas is overall greater in the models with high/intermediate density at the base of the extra-planar region. 
For these models, the larger fractional drop in gas density leads to a larger drop in CR pressure, and, as a result, a larger momentum flux transfer.
However, the differences in the gas flows at high altitudes ($z\sim2-3$~kpc) are in fact not so large among the models: the density and velocity of the outflowing gas are within a factor $2$ of each other. 
Outward mass, momentum, and energy fluxes for the three models are therefore also similar (see \autoref{fig:WWfluxes} and \autoref{fig:WWMflux}).
This is true despite two orders of magnitude difference in the injection density.  The implication is that mass loading of CR-driven galactic winds adjusts to match the carrying capacity allowed by the momentum flux available for transfer from the CR distribution  (see \citealt{Mao&Ostriker18} and further analysis in \autoref{sec:1Dmodel}).

\subsubsection{Different cosmic-ray pressures at the wind base} 
\label{sec:WWM1vsWWM2}

So far, we have analysed the dynamical effect of CRs in models with different density of the injected gas. Here, we compare the results from models M1 -- WW -- ID and M2 -- WW -- ID. These two models have the same injected gas density, and nearly the same injected velocity and momentum flux.  The injected CR pressure $P_\mathrm{c,0}$ and CR flux $F_\mathrm{c,z,0}$  are however lower by a factor two in  M2 -- WW -- ID (see \autoref{tab:tab2}).

\begin{figure*}
\centering
\includegraphics[width=\textwidth]{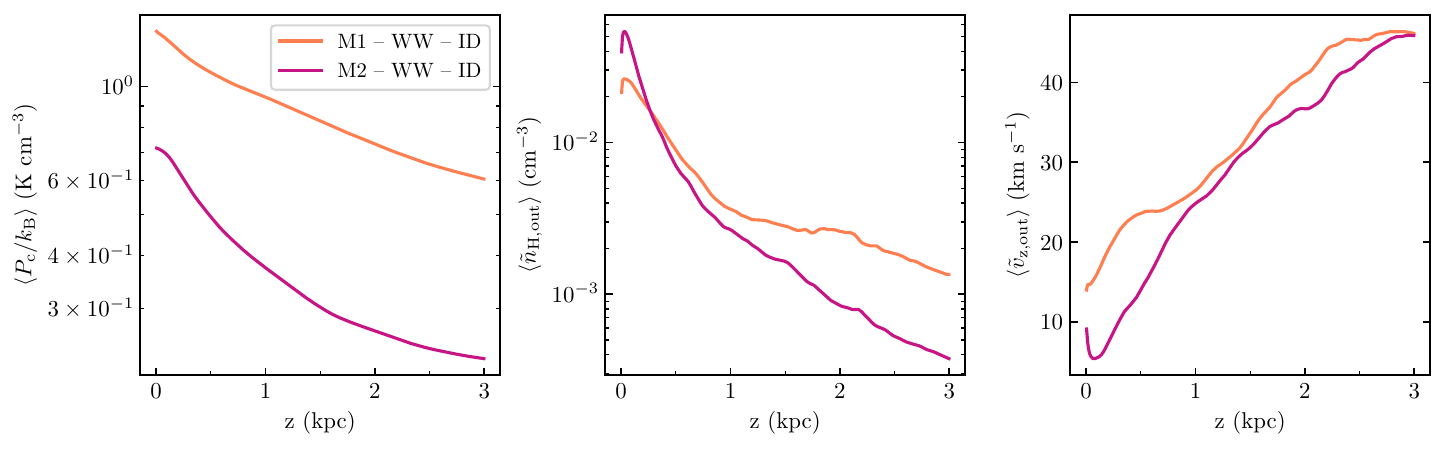}
\caption{Temporally averaged vertical profiles of CR pressure $P_\mathrm{c}$ (\textit{left panel}), typical hydrogen number density $\tilde{n}_\mathrm{H, out}$ (\textit{middle panel}), and typical vertical velocity $\tilde{v}_\mathrm{z, out}$ (\textit{right panel}) of the outflowing gas. Different colors represent different 
warm wind 
models: coral for M1 -- WW -- ID (with higher CR pressure and flux), magenta for M2 -- WW -- LD (with lower CR pressure and flux). 
}
\label{fig:WWzprof2}
\end{figure*} 

The mean vertical profiles of CR pressure, outflowing gas density, and outward gas velocity, are displayed in \autoref{fig:WWzprof2}.
In M2 -- WW -- ID, CR forces are ineffective in accelerating the gas near the injection region. As a consequence, the outward vertical velocity decreases until $z\sim100$~pc, and the gas accumulates near the boundary; the resulting gas density is a factor of 2 higher than in model M1 -- WW -- ID.
Away from the injection region, CRs are overall able to accelerate the gas. Even though the CR pressure gradients in the two models are comparable for $z\lesssim 1$~kpc, the gas density drops faster in M2 -- WW -- ID due to the smaller initial outward velocity.  
The gas acceleration produced by M2 -- WW -- ID is initially higher than the acceleration produced by M1 -- WW -- ID, while becoming comparable for $z>1$~kpc. 
Beyond $\sim 1$~kpc, the velocity profiles are similar, although the density and therefore the mass flux remains higher in M1 -- WW -- ID than in M2 -- WW -- ID.  
As we shall see in the next section, the extent to which CRs accelerate the ambient gas depends on the so-called CR sound speed, $C_\mathrm{c}=\sqrt{d{P_\mathrm{c}}/d\rho}$.

\subsection{Comparison with analytic predictions}
\label{sec:1Dmodel}

In this section, we compare the results from our warm wind simulations with the predictions of the one-dimensional (1D) steady-state model developed by \citet{Mao&Ostriker18} to describe properties of warm galactic winds driven by CRs.  We focus on 
model M1 -- WW -- ID since the density of the injected warm gas in this model is most consistent with the typical gas density in SN-driven fountain flows at the base of the extra-planar region for solar neighborhood conditions \citep[$n_\mathrm{H} \sim 0.01-0.1$~cm$^{-3}$,][]{Kim&Ostriker18}.

\citet{Mao&Ostriker18} treat the gas as a single-phase isothermal medium, and assume that velocity, magnetic field and pressure gradient streamlines follow the large-scale gravitational potential gradient. The former condition is approximately satisfied in our WW simulations, where most of the gas has a temperature of $10^{4-5}$~K.   The latter condition is also a good approximation of the mean trend in our WW simulations, where gravity is in the vertical direction, as are the mean pressure gradients and the mean magnetic field. Although our simulations have outflows and inflows simultaneously, both are primarily in the vertical direction. The \citet{Mao&Ostriker18} model includes CR advection and streaming only, under the assumption that CR diffusion is negligible in the extra-planar region. This assumption is verified in our simulations.

\begin{figure*}
\centering
\includegraphics[width=\textwidth]{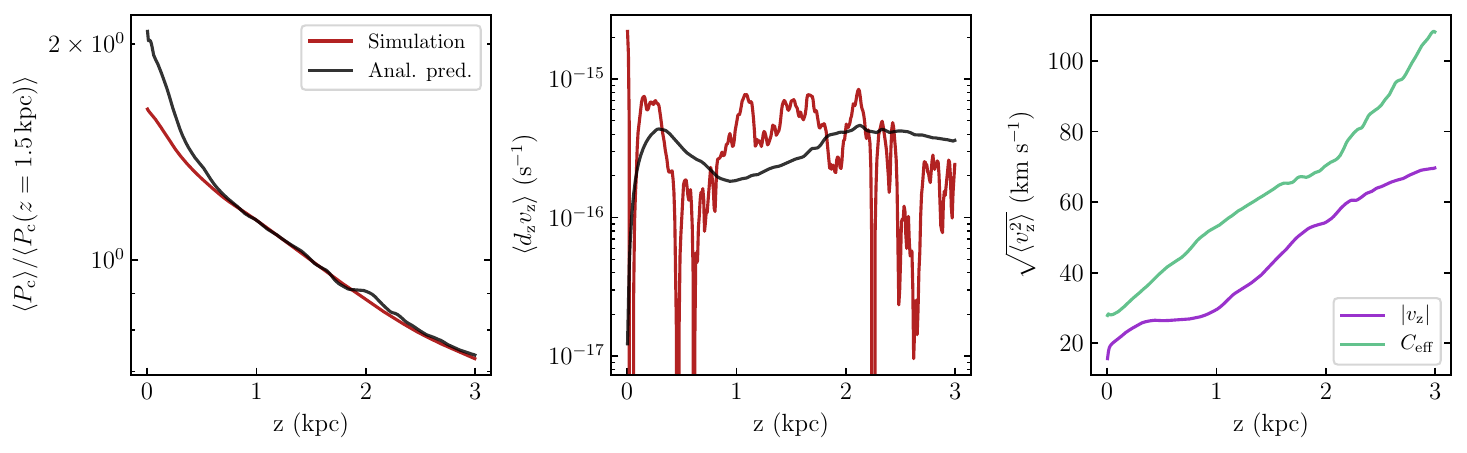}
\caption{Comparisons of warm wind simulation results with the predictions of the one-dimensional (1D) steady-state model for CR-driven winds by \citet{Mao&Ostriker18}. The \textit{left panel} shows the vertical profile of CR pressure $P_\mathrm{c}$ from the simulation M1 -- WW -- ID (red line) compared with the profile predicted by the 1D model (black line). The \textit{middle panel} shows the gas velocity variation along the $z$ axis $d_\mathrm{z} v_\mathrm{z}$ produced by the simulation M1 -- WW -- ID (red line) in comparison to the 1D model prediction (black line). The \textit{third panel} shows the vertical profiles of the root mean square gas velocity
$\langle v_\mathrm{z}^2\rangle^{1/2}$ (purple line) and the effective sound speed  $C_\mathrm{eff}^2$ (light green line). The 1D model predicts that, in the Cartesian geometry of the present simulations, $C_\mathrm{eff}^2$ has to be larger than $v_\mathrm{z}^2$ in order to have a steady-state outflow.}
\label{fig:WW1Dcomp}
\end{figure*} 

\begin{figure}
\centering
\includegraphics[width=0.47 \textwidth]{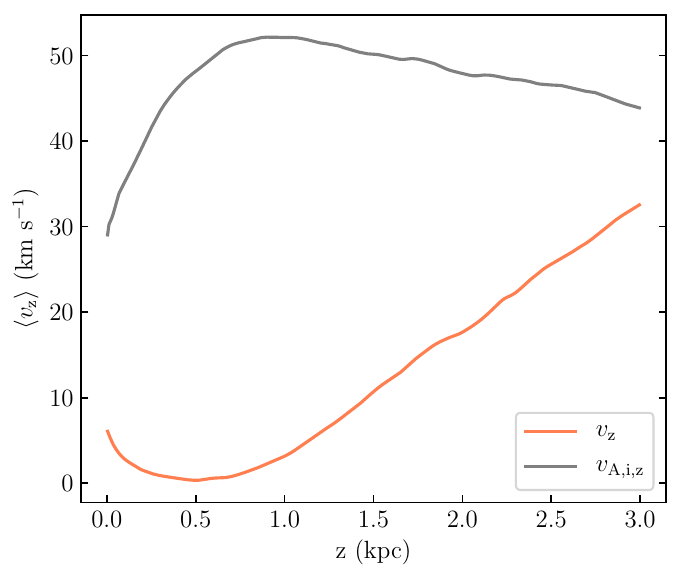}
\caption{Horizontally and temporally averaged profiles of the vertical components of the gas velocity $v_\mathrm{z}$ (orange line) and ion Alfv\`{e}n speed $v_\mathrm{A,i,z}$ (gray line) in M1 -- WW -- ID.}
\label{fig:WWvel}
\end{figure}

From the CR energy conservation law, \citet{Mao&Ostriker18} derive the steady-state expression for the CR pressure along each streamline:
\begin{equation}
P_\mathrm{c} = P_\mathrm{c,0} \left( \frac{v+v_\mathrm{A,i} }{v_0+v_\mathrm{A,i,0}} \frac{A}{A_0} \right)^{-\gamma_\mathrm{c}}  \, .
\label{eq:Mao_CRpress}
\end{equation}
Here, the ``0'' subscript denotes values at the streamline footpoint, $\gamma_\mathrm{c} = 4/3$ is the CR adiabatic index, and $A$ is the wind cross-sectional area. In the local Cartesian box of our models, $A$ does not vary with $z$, i.e.\ ${A (z)}/{A_0} = 1$ for the flow.\footnote{Strictly speaking, the area occupied by the outflowing gas increases with increasing $z$ (from 50\% at $z=0$ to 80\% at $z=3$~kpc). However, in these warm wind models, the transport of CRs is streaming-dominated ($v_\mathrm{A,i} > v$ see \autoref{fig:WWvel}), meaning that the CR flux direction primarily depends on the CR pressure gradient, rather than the velocity field direction. Since the CR pressure gradient points inward almost everywhere in the simulation domain, it is appropriate to say that the area occupied by the outflowing CR flux is constant.} 
Since in our simulations gas and CR motions are mostly along the $z$-axis, i.e.\ $\langle v \rangle \approx \langle v_\mathrm{z} \rangle $ and $\langle v_\mathrm{A,i} \rangle \approx \langle v_\mathrm{A,i,z} \rangle $, we can estimate how the mean CR pressure $\langle P_\mathrm{c}\rangle$ should vary along the vertical direction according to \autoref{eq:Mao_CRpress}. In the left panel of \autoref{fig:WW1Dcomp}, we compare the time-averaged vertical profile of CR pressure in M1 -- WW -- ID with the analytic prediction, and find an excellent agreement for $z > 500$~pc. \autoref{fig:WWvel} shows that the mean ion Alfv\`{e}n speed is always larger than the mean gas advection velocity, implying that
streaming dominates over advection in the warm extra-planar gas (see also \autoref{fig:FBvel} and \autoref{sec:transport}). We note that $\langle v_\mathrm{z} \rangle$ in \autoref{fig:WWvel} is the average between the mean inflowing and outflowing velocities, and is therefore slightly lower than $\langle v_\mathrm{z, out} \rangle$ shown in \autoref{fig:WWzprof1}.

Following \citet{Mao&Ostriker18} \citep[see their eqs.~26 and 27 for more general 1D geometry, and see also][]{Breitschwerdt+91, Everett+08}, 
the steady-state momentum equation 
along vertical streamlines in a local cartesian box is:
\begin{equation}
\left ( v_\mathrm{z}^2 - c_\mathrm{s}^2 - 
C_\mathrm{c}^2
\right ) d_\mathrm{z}v_\mathrm{z} = -v_\mathrm{z}d_\mathrm{z}\Phi \, ,
\label{eq:Mao_momeq}
\end{equation}
where the third term within the brackets is  the CR contribution to the squared sound speed, 
\begin{equation}
\dfrac{d P_\mathrm{c}}{d \rho}\equiv C_\mathrm{c}^2  = \gamma_\mathrm{c} \dfrac{P_\mathrm{c}}{\rho} \dfrac{v_\mathrm{z}+ v_\mathrm{A,i,z}/2}{v_\mathrm{z}+ v_\mathrm{A,i,z}} \;.
\label{eq:Mao_crsoundspeed}
\end{equation}
\autoref{eq:Mao_crsoundspeed} can be easily demonstrated writing $v$ and $v_\mathrm{A,i}$ as a function of $\rho$ in \autoref{eq:Mao_CRpress}. From the mass conservation law and $B_\mathrm{z}=const.$ (for $\nabla \cdot \mathbf{B} = 0$), we know that 
in steady state, the gas velocity and ion Alfv\'{e}n speed along vertical streamlines can be written as
\begin{equation}
 v \equiv v_\mathrm{z} = v_0   \left ( \dfrac{\rho}{\rho_0 } \right)^{-1}\, ,
\end{equation}
and 
\begin{equation}
 v_\mathrm{A,i} \equiv v_\mathrm{A,i,z} = v_\mathrm{A,i,0}  \left ( \dfrac{\rho_\mathrm{i}}{\rho_\mathrm{i,0} } \right)^{-\frac{1}{2}} \approx v_\mathrm{A,i,0}  \left ( \dfrac{\rho}{\rho_0 } \right)^{-\frac{1}{2}}  ,
 \label{eq:vA_rho}
\end{equation}
respectively. 
In \autoref{eq:vA_rho}, $\rho_\mathrm{i}/\rho_\mathrm{i,0}$ can be approximated to $\rho/\rho_\mathrm{0}$ since gas is mostly ionized ($x_\mathrm{i} \sim 1$) at the typical temperatures reached for extraplanar gas in our simulations ($T\gtrsim 10^4$~K). In the streaming-dominated regime (as in \autoref{fig:WWvel}), $P_\mathrm{c} \propto \rho^{2/3}$ from \autoref{eq:Mao_crsoundspeed} and \autoref{eq:vA_rho}, so that the  CR sound speed increases with decreasing $\rho$, i.e.\ ${d P_\mathrm{c}}/{d \rho} \approx 2/3\, P_\mathrm{c}/\rho \propto  \rho ^{-1/3}$. By contrast, in the advection-dominated regime, the CR sound speed decreases with decreasing $\rho$, i.e.\ ${d P_\mathrm{c}}/{d \rho} \approx 4/3\, P_\mathrm{c}/\rho \propto  \rho ^{1/3}$.
When advection dominates, as is true for fast, hot gas in our HW models, the scaling behavior of the effective sound speed does not encourage wind acceleration.  

From \autoref{eq:Mao_momeq}, the ordinary differential equation describing a steady-state vertical wind is given by
\begin{equation}
d_\mathrm{z}v_\mathrm{z} = \frac{v_\mathrm{z}}{C_\mathrm{eff}^2 - v_\mathrm{z}^2} \,d_\mathrm{z}\Phi \, ,
\label{eq:Mao_acc}
\end{equation}
where $C_\mathrm{eff}$ is the effective sound speed, defined as
\begin{equation}
 C_\mathrm{eff}^2 =  c_\mathrm{s}^2 + 
C_\mathrm{c}^2\;.
 \label{eq:eff_sound_speed}
\end{equation}  
In the middle panel of \autoref{fig:WW1Dcomp}, we compare the average outward velocity variation along the $z$ axis $\langle d_\mathrm{z}v_\mathrm{out,z}\rangle$ in M1 -- WW -- ID with the predictions of the analytic model (\autoref{eq:Mao_acc}), finding an overall good agreement. The gradient $\langle d_\mathrm{z}v_\mathrm{out,z}\rangle$ is mostly positive, as the outward velocity increases with $z$ (see \autoref{fig:WWzprof1}). From \autoref{eq:Mao_acc}, for an accelerating wind with $d_\mathrm{z}v_\mathrm{z} > 0$, it must be true that $C_\mathrm{eff}^2 > v_\mathrm{z}^2$ ($d_\mathrm{z}\Phi$ is positive by definition). The third panel of \autoref{fig:WW1Dcomp} shows that this condition is satisfied in our simulation. Of  the two terms on the RHS of \autoref{eq:eff_sound_speed}, the CR term 
$C_\mathrm{c}^2=d P_\mathrm{c}/{d \rho}$ dominates over the thermal term $c_\mathrm{s}^2$ ($c_\mathrm{s}\sim 10 \, \kms$ for $T \sim 10^4$~K). Hence, in the streaming-dominated regime of M1 -- WW -- HD, $C_\mathrm{eff}^2$ increases with $z$ as $\rho$ decreases.

The condition $C_\mathrm{eff}^2 > v_\mathrm{z}^2$ is satisfied in all WW models (at $z \gtrsim 500$~pc), except for M1 -- WW -- LD. Here, the outward velocity exceeds the effective sound speed at $z \gtrsim 1$~kpc. In fact, the wind is not accelerating in this region for this model (see right panel of \autoref{fig:WWzprof1}). 
We note also that for the subsonic regime ($C_\mathrm{eff}^2 \approx C_\mathrm{c}^2 \gg v_z^2$), it is straightforward to show that the acceleration obeys $v_z/v_\mathrm{z,0} = \left[\Delta \Phi/(3 C_\mathrm{c,0}^2) + 1   \right]^3 $ for $\Delta \Phi$ the potential difference.  
Even though both density and CR pressure are higher in M1 -- WW -- ID than in M2 -- WW -- ID, the velocity and $C_c$ are similar in the two models at $z \sim 1$ kpc, which explains the similarity of the acceleration profiles in \autoref{fig:WWzprof2}.

Finally, we note from \autoref{eq:Mao_acc} that there is no steady-state solution for a wind passing through a sonic point ($C_\mathrm{eff}^2 = v_\mathrm{z}^2$) in a local cartesian box. This would be possible only if the cross-sectional area $A$ increases as the flow moves outward. With the opening of the streamlines,  
a term $C_\mathrm{eff}^2 d_z \ln A$ is subtracted from the gradient of the gravitational potential so that the numerator of \autoref{eq:Mao_acc} passes through zero and  
winds go through a critical point to reach an asymptotic velocity \citep{Mao&Ostriker18}. 

\subsection{Cosmic-ray heating}
\label{sec:heating}

\begin{figure*}
\centering
\includegraphics[width=\textwidth]{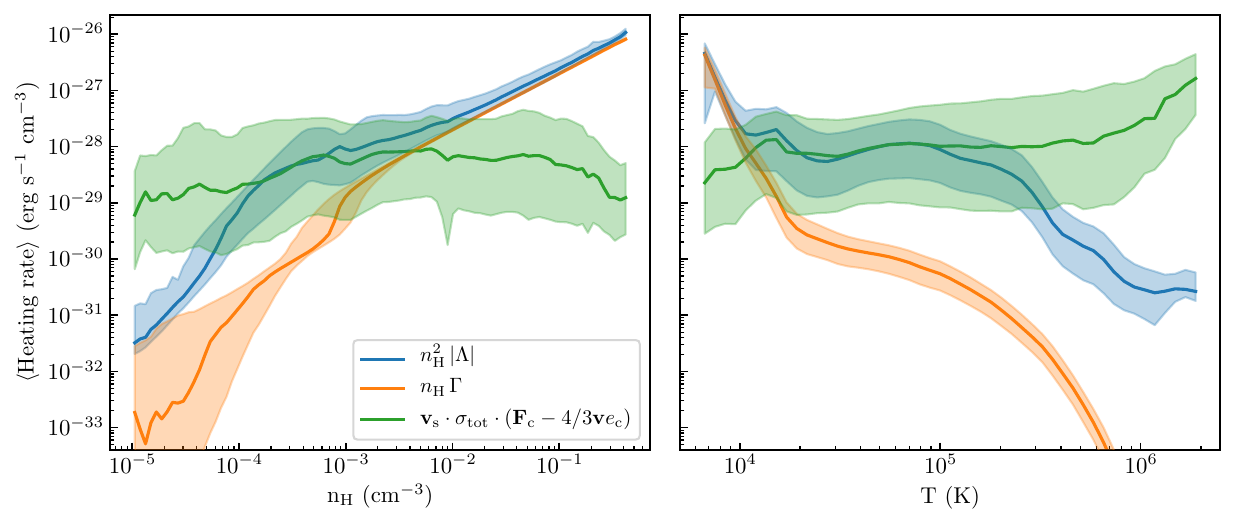}
\caption{Medians of the rates of radiative cooling $n_\mathrm{H}^2 \vert \Lambda \vert$, photoelectric heating  $n_\mathrm{H} \Gamma$, and CR streaming heating $\mathbf{v}_\mathrm{s} \cdot  \tensor{\mathrm{\sigma}}_\mathrm{tot} \cdot   ( \mathbf{F_\mathrm{c}} - 4/3 \mathbf{v} e_\mathrm{c})$ as a function of hydrogen density $n_\mathrm{H}$ (\textit{left panel}) and gas temperature $T$ (\textit{right panel}) in M1 -- WW -- ID. For both panels, the shaded areas cover the 16th and 84th percentiles of fluctuations.}
\label{fig:WWheating}
\end{figure*} 

In addition to their direct dynamical effects, energy loss from the CR bulk flow can also be an important source of heating for the ambient gas \citep[e.g.,][]{Jacob+17, Kempski+20}.\footnote{This CR heating mechanism from a collective phenomenon is entirely distinct from CR heating due to ionization.  The latter contributes at $T\lesssim 10^4$K and arises mostly from CRs at sub-GeV energies, and is not included in the current simulations.} In the self-confinement regime, damping of streaming-excited Alfv\'{e}n waves mediates transfer of energy from the CR population to the gas at a rate $\mathbf{v_\mathrm{s}} \cdot  \tensor{\mathrm{\sigma}}_\mathrm{tot} \cdot ( \mathbf{F_\mathrm{c}} - 4/3 \mathbf{v} e_\mathrm{c})$ (see RHS of \autoref{eq:eneq}). If we subtract the sum of the work equation (obtained by the dot 
product of $\mathbf{v}$ and \autoref{eq:momeq}) and the magnetic energy equation (obtained by the 
dot product of $\mathbf{B}$ and \autoref{eq:Beq}) from the total energy equation (\autoref{eq:eneq}), we can see that such energy transfer leads to an increase of the thermal energy of the gas: 
\begin{equation}
\begin{split}
& \frac{\partial e_\mathrm{t}}{\partial t} + \nabla \cdot \left [ \left(e_\mathrm{t} + P_\mathrm{t}  \right) \mathbf{v}\right] - \mathbf{v} \cdot \nabla P_\mathrm{t} = \\ 
 -n_\mathrm{H}&(n_\mathrm{H} \Lambda - \Gamma) + \mathbf{v_\mathrm{s}} \cdot 
\tensor{\mathrm{\sigma}}_\mathrm{tot} \cdot   \left (\mathbf{F_\mathrm{c}} - \dfrac{4}{3} \mathbf{v}  e_\mathrm{c} \right)  \;,
\end{split}
\label{eq:theneq}    
\end{equation}
with $e_\mathrm{t} = P_\mathrm{t}/(\gamma - 1)$ the thermal energy. We note that the last term is always positive as streaming always drains energy from CRs (see definition of $\mathbf{v}_\mathrm{s}$ in \autoref{eq:vs}); with $\tensor{\mathrm{\sigma}}_\mathrm{tot} \cdot   ( \mathbf{F_\mathrm{c}} - 4/3 \mathbf{v} e_\mathrm{c}) \rightarrow - \nabla P_c$ in steady state, the contribution to the  heating rate is $|\mathbf{v}_\mathrm{A,i} \cdot \nabla P_\mathrm{c} |$.

To evaluate the extent to which CRs affect the thermal state of the gas, we quantify the three source terms on the RHS of \autoref{eq:theneq} in our fiducial model M1 -- WW -- ID.  \autoref{fig:WWheating} shows the magnitude of the average rates of radiative cooling $n_\mathrm{H}^2 \Lambda$, photoelectic heating $n_\mathrm{H} \Gamma$, and CR heating $\mathbf{v}_\mathrm{s} \cdot  \tensor{\mathrm{\sigma}}_\mathrm{tot} \cdot   ( \mathbf{F_\mathrm{c}} - 4/3 \mathbf{v} e_\mathrm{c})$ as  a function of $n_\mathrm{H}$ and $T$. For $n_\mathrm{H} \gtrsim 3-4 \times 10^{-3}$~cm$^{-3}$ and $T \lesssim 10^4$~K, most of the gas is in thermal equilibrium with photoelectric heating balanced by cooling, $n_\mathrm{H} \Gamma \simeq n_\mathrm{H}^2 \Lambda$, while CR heating is negligible. At higher temperature and lower density, the rate of photoelectric heating quickly decreases, while the rate of CR heating only slowly decreases with density, and increases at the highest temperatures.  
In the density range $10^{-4} <n_\mathrm{H} <10^{-3}$~cm$^{-3}$ and temperature range $2\times10^4 <T<10^5$~K, most of the gas is in thermal equilibrium with CR heating balanced by cooling, $\mathbf{v}_\mathrm{s} \cdot  \tensor{\mathrm{\sigma}}_\mathrm{tot} \cdot   ( \mathbf{F_\mathrm{c}} - 4/3 \mathbf{v} e_\mathrm{c}) \simeq n_\mathrm{H}^2 \Lambda  $. At higher temperature and lower density, gas becomes thermally unstable with the rate of CR heating overall larger than the rate of radiative cooling. CRs heat the gas up to temperature of order $10^6$~K.

\section{Discussion}
\label{sec:discussion}

\subsection{Cosmic-ray pressure in the disk}

In \citetalias{Armillotta+21} and \citetalias{ Armillotta+22}, we found that for the TIGRESS model representing a solar-neighborhood environment, the average CR pressure near the midplane exceeded the other ISM pressures by a factor of $2-3$. This can be compared with the local Milky Way, where the estimated pressure of CRs with $E_\mathrm{kin} \geq 1$~GeV is closer to equipartition with the other pressures.
We speculated that this disagreement might be due to the absence of CR backreaction on the gas in the purely-postprocessing simulations of \citetalias{Armillotta+21}. Our expectation was that, with the rearrangement of the velocity and magnetic field topology in simulations with time-dependent MHD and CR physics, CRs trapped in the ISM would be able to propagate out of the disk, leading to a CR pressure decrease near the midplane. 
In \autoref{sec:fullboxsim} of the present paper, we show that rearrangement of the velocity and magnetic field topology indeed enables CRs trapped in the dense gas to propagate out of it. The decrease of pressure in the warm gas  is compensated by the increase of pressure in the hot gas, leading to an overall more uniform pressure between hot and warm/cold gas (\autoref{fig:FBsnapMHD}, \autoref{fig:Pc_nh_PDF}).  There is therefore only a marginal decrease in the mean midplane CR pressure with ``live MHD'' compared to pure post-processing CR transport (see \autoref{fig:FBpressprof}). 
This suggests that the overall CR confinement in the disk is not controlled by local magnetic geometry, but by the combination of advection in hot gas and Alfv\'enic streaming in warm gas, as limited by scattering in low density, ionized gas 
(\autoref{fig:FBvel}).  We find the scattering rate does not significantly change when ``live MHD'' is turned on (\autoref{fig:FBsigma}). 

Based on further tests, we have found that the higher-than-expected CR pressure reported from the postprocessing simulations of \citetalias{Armillotta+21} and \citetalias{Armillotta+22}
was due to spurious adiabatic work exerted from the thermal gas on the CR fluid (see \autoref{sec:cr_pressure}). The effect is particularly strong at interfaces between cold/warm and hot gas, where CR pressure gradients are high and oriented in the same directions as the velocity vectors, which results in significant gains of CR energy density ($\approx \mathbf{v} \cdot \nabla P_\mathrm{c}$). We find that with the rearrangement of the velocity field topology in self-consistent ``live MHD'' simulations, the adiabatic work diminishes in magnitude (CR pressure gradients are smaller) and becomes mostly negative (at interfaces the velocity vectors are mostly in the direction of decreasing CR pressure). 
We conclude that since adiabatic work can significantly boost the effective CR energy injection in postprocessing simulations, it is best to zero out this term, and we do so for the postprocssing simulations of 
\autoref{sec:fullboxsim}.
We find that, in steady state, the average pressure of CRs near the midplane is in equipartition with the other pressures, and in good agreement with the observed value. 

In the ISM, approximate equipartition is attained by balancing energy gains from star formation feedback and energy losses due to dissipative processes \citep{Ostriker2010, Ostriker2011, KimOstriker2015, Ostriker+22}: balancing far-UV photoelectric heating and cooling for thermal pressure, balancing momentum flux injection from supernovae with turbulent dissipation for kinetic pressure, and applying turbulent driving in combination with shear to sustain the pressure in the magnetic field. Likewise, the pressure of CRs is determined by the balance between CR energy injection from supernovae and propagation of CRs out of the disk \citep{Armillotta+21, Armillotta+22}. Other energy sink terms play a minor role in setting the midplane CR pressure: in the post-processing runs of \autoref{sec:fullboxsim}, the average rates of collisional and streaming losses relative to the rate of CR energy injection are 0.1 and 0.2, respectively.\footnote{These numbers are obtained by integrating the energy sink terms over the
simulation domain within $\vert z \vert = H_\mathrm{c,eff}$, with $H_\mathrm{c,eff}$ an effective CR scale height, and taking averages over multiple snapshots.}

In the case of negligible losses, the midplane CR pressure $P_\mathrm{c}(0)$ can be written as a function of the vertical flux of CR energy above the supernova input layer $F_\mathrm{c,inj}$ \citep[see also][]{Armillotta+21, Armillotta+22}: 
\begin{equation}
P_\mathrm{c}(0) \simeq \dfrac{H_\mathrm{c,eff}}{\kappa_\mathrm{eff}} F_\mathrm{c,inj} = \dfrac{H_\mathrm{c,eff}}{\kappa_\mathrm{eff}} \dfrac{ 0.5 \epsilon_\mathrm{c} E_\mathrm{SN} \Sigma_\mathrm{SFR}}{m_\star}\,,
\label{eq:P-F}
\end{equation}
with $\Sigma_\mathrm{SFR}$ the SFR surface density, $m_\star$ the total mass of new stars per supernova ($95.5 M_\odot$ in \citealt{Kim&Ostriker17}, from a
Kroupa IMF), $H_\mathrm{c,eff}= \langle |d\ln P_\mathrm{c}/dz| \rangle^{-1}$ an effective CR scale height (measured in the simulation through a linear fit of ${\rm ln} P_\mathrm{c}$ vs.\ $z$ within 1.5 kpc), and $\kappa_\mathrm{eff} \equiv \sigma_\mathrm{eff}^{-1} $ an effective diffusion coefficient. 
We note that $\kappa_\mathrm{eff}$ may be understood as a measurement of the efficiency of CR propagation, including not only CR diffusion, but also advection and streaming. For $H_\mathrm{c,eff}\simeq 1.3$~kpc as measured in our post-processing simulations, the midplane pressure-flux relation is satisfied for $\kappa_\mathrm{eff} \simeq 3.6 \times 10^{28}$~cm$^2$~s$^{-1}$. In the self-consistent simulations, where the distribution of CR pressure is smoother, $H_\mathrm{c,eff}$ slightly increases, and so does $\kappa_\mathrm{eff}$. After $t = 3$~Myr, $H_\mathrm{c,eff}\simeq 1.5$~kpc, while $\kappa_\mathrm{eff} \simeq 5.6 \times 10^{28}$~cm$^2$~s$^{-1}$.

We emphasize that the value of $\kappa_\mathrm{eff}$ is an order of magnitude larger than the value of  $\sigma_\parallel^{-1}$ in the low-density ionized gas that fills space outside of the midplane  (see \autoref{fig:FBsnapMHD} and \autoref{fig:FBsigma}), showing that advection and streaming (rather than pure diffusion) are essential to CR transport. 
With $F_c \equiv  (4/3)v_\mathrm{c,eff} e_\mathrm{c}= 4v_\mathrm{c,eff} P_\mathrm{c}$, the effective transport speed is $v_\mathrm{c,eff} = (1/4) \kappa_\mathrm{eff}/H_\mathrm{c,eff}$, which for the post-processing or self-consistent simulations is $v_\mathrm{c,eff}=$ 22 or 30 $\kms$, respectively.  This range is consistent with the (vertical) ion Alfv\'en speed in gas at $T\lesssim 10^4$~K, as shown in \autoref{fig:FBvel}.  Indeed, 
we conclude in \autoref{sec:transport} 
that the primary overall limit on  transport of GeV CRs in the multiphase ISM is the ion Alfv\'en speed
in warm gas at $n_\mathrm{H}\sim 0.01 - 0.1 \pcc$, since this regime is where the maximum over the diffusion, advection, and streaming speeds has the smallest value.  

In addition to the midplane CR pressure, another observable quantity that can be predicted by our simulations is the CR grammage, which is a measure of the column of gas traversed by CRs during their propagation. In our simulations, the grammage can be computed as 
\begin{equation}
X = \mu_\mathrm{H} m_\mathrm{p} v_\mathrm{p} \frac{\int n_\mathrm{H} e_\mathrm{c} dx^3}{\dot{E}_\mathrm{inj}}, 
\label{eq:grammage}
\end{equation} 
with $\mu_\mathrm{H} = 1.4$, and $\dot{E}_\mathrm{inj}$ the total CR energy injected per unit time (see Section~3.2.2 and Equation~27 in \citetalias{Armillotta+21}). In good agreement with observations ($X \sim 10$~g~cm$^{-2}$ for GeV energy, \citealt{Amato&Blasi18}), we find that the average CR grammage is $X \simeq 10.8$~g~cm$^{-2}$ in the postprocessing simulations, and $X \simeq 7.2$~g~cm$^{-2}$ in the simulations with MHD at $t=3$~Myr.
Provided that the CR scale height is larger than that of the gas, it is straightforward to show that \autoref{eq:grammage} is equivalent to $X \approx (3/8)\Sigma_\mathrm{gas} v_\mathrm{p}/v_\mathrm{c,eff}$ for $\Sigma_\mathrm{gas}$ the total gas surface density.  With an average $\Sigma_\mathrm{gas}=9.5 \,\mo$~pc$^{-2}$ in the TIGRESS solar neighborhood simulation, the resulting estimated values of the grammage are in good agreement with computed values when adopting  $v_\mathrm{c,eff}$ listed above.     

Based on the success of our simulations in reproducing key observed properties in the local ISM, we conclude that the self-confinement paradigm is able to explain transport of GeV CRs.  Provided that the underlying MHD simulation has a realistic representation of the multiphase, star-forming ISM, it does not appear necessary to include sources of small-scale MHD waves other than self-excitation by CRs, or more generally to adopt an ad hoc scattering coefficient.  When combined with realistic flow and Alfv\'en velocities in hot and warm ionized gas, scattering set by the balance between growth from CR streaming instability and nonlinear Landau damping produces realistic transport.  
We emphasize, however, that the quantitative results presented here apply only to CRs with energy $E_\mathrm{kin} \sim 1$~GeV.  Because the scattering coefficient decreases with increasing CR particle momentum (through $n_1$, see \autoref{eq:n1}), we expect the CR confinement time within the disk and, as a consequence, the CR grammage (and measured B/C ratio) to decrease with increasing momentum, in agreement with the observations \citep[e.g.,][]{Ptuskin+09}. It will be very interesting to extend the current models to higher CR energy in order to quantify the scaling behavior of the grammage, and to identify when self-excited waves fail to produce results consistent with observations.

To date, the only other numerical studies that have tested CR transport with a variable scattering model based on self-confinement are those of \citet{Hopkins2021, Hopkins+22} employing cosmological zoom-in FIRE simulations \citep{Hopkins+18} with CRs, adopting the same CR injection efficiency $\epsilon_\mathrm{c}=0.1$ that we do. In contrast to our conclusions, they suggest that the standard self-confinement model is incompatible with observations.  In their simulations, CRs are excessively confined, with grammage, residence time, and CR pressure up to an order of magnitude higher than observations. They argue that the failure is the consequence of self-confinement instability: if $P_\mathrm{c}$ rises, $\sigma_\parallel$ increases, which slows down CR diffusion, further increasing $P_\mathrm{c}$ and $\sigma_\parallel$ in a runaway process that stops when the effective propagation speed decreases to the local ion Alfv\`{e}n speed. While we agree with their argument that streaming dominates diffusion in low-density regions, 
and we find that streaming is indeed the main limit in some regimes, 
this is not the only factor controlling CR transport.  
In our simulations, advection is the dominant transport mechanism in hot, rarefied gas at $T\gtrsim 2\times 10^4$~K and $n_\mathrm{H}\lesssim 10^{-2}\,\pcc$, while waves are strongly damped by ion-neutral collisions so that diffusion is dominant at $n_\mathrm{H} \gtrsim 0.5 \, \pcc$ and $T \lesssim 5 \times10^3$~K (see \autoref{fig:FBvel}). 
Thus, in supernova remnants where CRs are injected, and in superbubbles and chimneys that channel hot gas from the disk to the halo, gas velocities are so high that CRs are quickly advected away from the midplane. Furthermore, the high diffusion in the predominantly-neutral gas prevents CRs from being trapped there.  As a result, the typical CR pressure we find is compatible with ISM observations in the Milky Way.  

One possible reason for the difference between our results and those of 
\citet{Hopkins2021, Hopkins+22}
is that the mass resolution adopted in cosmological zoom-in simulations ($\Delta M \sim 10^{3}-10^{4}\,M_\odot$) is insufficient to properly resolve the hot phase of the ISM. In our local-disk simulations, the mass resolution at the typical densities ($n_\mathrm{H} \sim 10^{-4}-10^{-3}$~cm$^{-3}$) of the hot ISM is $\Delta M = \rho \Delta x^3 \sim 10^{-3}-10^{-2}\,M_\odot$, far lower than the value $7000 \, M_\odot$ 
employed by \citet{Hopkins2021} in their Milky Way-like model. Since the Sedov stage of SNR evolution ends when remnant mass is $\sim 1000 \, M_\odot$ \citep{KimOstriker2015a}, it is not possible to capture the Sedov stage of evolution of supernova remnants with mass elements $7000 \, M_\odot$, and therefore supernova feedback in FIRE is realized via momentum injection  \citep{Hopkins2018a,Hopkins+18} without resolving the hot ISM.   
Interestingly, \citet{Hopkins2021} remark that the densities of their superbubble regions are $\sim 0.01 \,\pcc$ (one or two orders of magnitude denser than in our simulations), and that these regions often have very low effective diffusion.  Though they do not report the advection velocities within superbubbles, it is possible that streaming limits CR transport there, unlike the case for our higher resolution models. 
\citet{Hopkins+22} explore a variety of alternatives to standard self-confinement, but our conclusion is that at least for GeV CRs, it appears that the standard model is satisfactory.  Crucially, this requires that the underlying MHD simulation of the multiphase ISM is sufficiently accurate, which demands much higher resolution than is possible in any present cosmological simulations (or zoom simulations of massive galaxies) to follow the creation and propagation of hot gas.  Potentially, high resolution MHD + CR simulations, using TIGRESS or similar frameworks, can be employed to calibrate effective CR scattering rates for use in lower resolution cosmological simulations.  

\subsection{Cosmic-ray driven outflows}
\label{sec:otherworks}

As reviewed in \autoref{sec:introduction}, the last decade has seen an increasing number of studies, both analytic and numerical, exploring the role of CRs in accelerating large-scale galactic outflows. However, the predictions of these studies are highly sensitive to the physics of CR transport included in the model. 

A key distinction is between the case when CR transport is dominated by diffusion, and when it is dominated by streaming.  \citet{Quataert+21a, Quataert+21b} have used a combination of analytic estimates and time-dependent numerical simulations to study the properties of galactic winds driven in each of these limits, considering spherically symmetric geometry as would be appropriate for a wind emerging from the center of a galaxy.  
Their diffusive solutions primarily focus cases with $\kappa \equiv 1/\sigma_\parallel > r_\mathrm{0} c_\mathrm{i}$,   where $ r_\mathrm{0}$ is the base radius of the wind and $c_\mathrm{i}$ is the isothermal gas sound speed, and they argue that when this inequality holds, diffusive CR transport yields mass-loss rates and wind powers higher than streaming transport. When $\kappa \equiv 1/\sigma_\parallel < r_\mathrm{0} c_\mathrm{i}$, they instead find much reduced wind fluxes for the diffusive case.  For $r_0$ of a few kpc in a galactic disk,  and $c_i\sim 10 \, \kms$ for warm ionized gas, $r_0 c_i \sim 10^{28} {\rm cm^2\ s^{-1}}$.  This can be compared to 
the results shown in \autoref{fig:FBsigma} for $\sigma_\parallel$, based on our self-consistent computation of the scattering coefficient assuming that waves are excited by the streaming instability.  There, 
for typical densities $n_\mathrm{H} \simeq 10^{-3}-10^{-2}$~cm$^{-3}$ (see \autoref{fig:WWzprof1} from our warm wind simulations), $\sigma_\parallel^{-1} \sim \rm{few} \times 10^{-27}\,{\rm cm^2\ s^{-1}}$. This, together with our previous results in \citet{Armillotta+22} showing that $\sigma_\parallel^{-1}$ is even smaller in inner-galaxy environments, suggests that the large-diffusion regime $\kappa > r_\mathrm{0} c_\mathrm{i}$ would not be applicable. Although quantitative results would presumably vary somewhat with environment, \autoref{fig:FBvel} shows that in fact only relatively
dense and cool gas is diffusion dominated, with hot gas advection-dominated and warm, intermediate-density gas streaming-dominated.  Note that these statements refer to the $\sim$ GeV protons that comprise the bulk of the CR energy density; high energy CRs have lower scattering rates and would therefore be diffusion dominated down to lower density (i.e. the blue curves in \autoref{fig:FBvel} would shift up), but they could not power significant winds.  

In our warm wind simulations, diffusion is minimal, and the primary limit on CR transport is initially the Alfv\'en speed, with advection increasing as gas accelerates outward.  We show in \autoref{sec:1Dmodel} that the flow profiles from the simulations (e.g.\ profiles of CR pressure  and acceleration rate) are in very good agreement with the predictions of the 1D analytic model of \citet{Mao&Ostriker18}, which includes CR advection and streaming only, and assumes an isothermal flow. \citet{Mao&Ostriker18} proposed (see their Eq. 46) that the mass-loss rate per unit area in a CR-driven warm wind is expected to scale as 
\begin{equation}\label{eq:MO_scaling}
\dot \Sigma_\mathrm{wind} \sim  P_{c,0} V_H^{-5/3} u_0^{2/3}
\end{equation}
for $P_{c,0}$ the CR pressure at the base of the wind, $V_H$ the characteristic halo velocity, and $u_0$ the velocity of warm gas in the fountain flow in the region where the wind originates. The limited spatial domain in our present simulations precludes testing whether this predicted scaling holds, since the CR-driven wind does not pass through a critical point. The acceleration of warm gas by CRs within our domain is, however, comparable to that achieved by transfer of momentum from hot gas to warm cloudlets when a powerful hot wind is present (compare velocities profiles in \autoref{fig:HWprofiles}, \autoref{fig:WWzprof1}, \autoref{fig:WWzprof2}).  Also,
our warm wind simulations show that the fluxes carried by the CR-accelerated wind are  very insensitive to the density of the gas flowing into the extraplanar region: 
\autoref{fig:WWMflux} shows only a factor of a few variation in the wind fluxes among models where the inflow density varies by two orders of magnitude. This suggests that the mass flux of a CR-driven warm wind may be set by the velocity $u_0$ of warm gas in the fountain region, consistent with \autoref{eq:MO_scaling}, rather than the mean density $\rho_0$ of the fountain.  An alternative way of writing the predicted CR-driven wind mass is 
\begin{equation}\label{eq:alt_flux}
\dot \Sigma_\mathrm{wind} \sim  P_{c,0}^3 V_H^{-5} \rho_0^{-2}    \end{equation}
\citep[see Eq. 44 of][]{Mao&Ostriker18}; this is mathematically completely equivalent to \autoref{eq:MO_scaling},
but the mass fluxes from \autoref{eq:alt_flux} are lower by a factor $\sim  (P_{c,0}/\rho_0)^2 V_H^{-10/3} u_0^{-2/3}\ll 1$ 
compared to \autoref{eq:MO_scaling} if one adopts typical fountain densities for $\rho_0$.  

While several recent studies have focused on the role of CR pressure in launching warm outflows, \citet{Everett+08} presented a semi-analytic 1D model of hot outflows driven by both thermal and CR pressure, aimed at explaining the diffuse soft X-ray emission observed at the Galactic Center. \citet{Everett+08} concluded that CR pressure is crucial to driving a hot wind from the Milky Way's Center, and that cases with CR pressure comparable to thermal gas pressure produce the best fit to the observations. 
Our simulation that is most similar to those of \citet{Everett+08} is M1 -- HW -- HPCR, in which most of the volume is occupied by hot, fast-moving gas,  and CR and gas pressures are comparable (see \autoref{fig:HWprofiles}).  However, our conclusion 
is that CRs do not contribute to the hot wind acceleration, but rather slow down the gas. 

We identify two main differences between our hot wind simulations and the models of \citet{Everett+08}.   First, they treat the gas as a single-phase medium, while gas is multiphase in our simulation. Even though subdominant in terms of volume, the increasing fraction of warm, slow-moving gas cloudlets at high altitude will help to slow down CRs and in doing so help to create a CR gradient pointing upward. Furthermore, while CR transport is advection-dominated in our simulation (since advection dominates in the volume-filling hot gas; see \autoref{fig:FBvel}), it is streaming-dominated in \citeauthor{Everett+08}'s best model (see their Fig.~8 for $z<3$~kpc). This is because the (imposed) magnetic field at the wind base in their model is two orders of magnitude higher than the average value for the hot wind of our simulation (see \autoref{tab:tab1}). Although we reach different conclusions regarding the role of CRs in hot winds, we agree with \citeauthor{Everett+08}~that  streaming transport is crucial to accelerating gas. Indeed, in \autoref{sec:warmwind}, we demonstrate that CRs can efficiently accelerate flows of gas in the warm medium, where CR transport is streaming-dominated.

Several other recent studies \citep[e.g.,][]{Girichidis+16, Simpson+16, Farber+18, Girichidis+18, Rathjen2021} have investigated CR driving of outflows using simulations of galactic disk patches, similar in geometry and numerical resolution to our models. 
Similar to our conclusions, they find that CRs are able to drive outflows of warm gas that reach moderate velocity ($v\lesssim 100\,\kms$) within a few kpc, and which are also somewhat denser than hot winds driven by SNe. Of course, densities in both hot SN-driven winds and warm CR-driven winds vary (in space and time) depending on the input energy and momentum fluxes driving them, and hot winds also help in accelerating warm embedded clouds. Comparing our M1 -- HW models with our M1 -- WW models, which have comparable hot-gas and CR pressure injected (see \autoref{tab:tab2}), the density of warm gas driven by CRs is a factor of a few higher in the M1 -- WW models than the density of hot gas in the M1 -- HW models (compare center panel of \autoref{fig:WWzprof1} with bottom panels of \autoref{fig:HWprofiles}). However, the density and mass flux of warm gas driven by interaction with the hot gas in the M1 -- HW models is higher than the density and mass flux of warm gas driven by CRs in the M1 -- WW models (mass fluxes of warm gas for the HW and WW cases are shown in top-left  panels of \autoref{fig:HWfluxes} and \autoref{fig:WWfluxes}, respectively). We point out that this difference in density and, most importantly, in pressure (given that the temperature is $\sim 10^4$~K in both cases) could be used to distinguish between warm outflows generated by interaction with hot gas and those driven by CRs in observational studies. 

An important difference between our warm wind simulations and the above disk-patch simulations is that the latter do not include CR streaming, only diffusion. We find that streaming dominates CR transport in our WW simulations (and in warm ionized gas generally), and, in doing so, it controls the dynamical and thermal evolution of the outflow. The absence of gas heating via CR streaming might explain why, in the above simulations, most of the outflowing gas has a temperature of $10^4$~K \citep[see e.g.,][]{Simpson+16, Farber+18, Girichidis+18}, while, in our WW simulations, a significant fraction of gas is heated up to temperature $T > 10^5$~K. 

In work to date, CR streaming has been included only in larger-scale simulations of isolated galaxies or cosmological zoom-ins only \citep[e.g.,][]{Ruszkowski+17, Chan+19, Hopkins+19, Hopkins2021, Thosmas+23}, which have much lower resolution than our simulations. While most of these works acknowledge the importance of CR streaming in accelerating galactic outflows, they do not discuss the impact of CR-streaming heating on the thermal state of the gas. \citet{Ji+20} analyse the simulation of a Milky Way-like galaxy performed by \citet{Hopkins+19}, finding that heating from CRs via streaming losses is  much weaker than radiative cooling. Unlike us, they do not compare heating and cooling rates in different gas phases, but rather they compute their average values as a function of the distance from the galactic plane. Similarly, in \autoref{sec:heating}, we find that on average radiative cooling dominates over heating from CRs (see \autoref{fig:WWheating}).
However, gas cooling rates become much lower than CR heating rates in the low-density gas. 

The right panel of \autoref{fig:WWheating} shows that CR heating exceeds cooling gas when $T\gtrsim 10^5$ K, which is consistent with the prediction for the onset of thermal instability in \citet[][see their Eq.\ 47]{Kempski+20}.  Their prediction is that under conditions where the CR pressure is large compared to the gas pressure (true in the extraplanar region -- see e.g.\ \autoref{fig:FBpressprof} or the $\mathcal{F}_\mathrm{p}$ panels of \autoref{fig:WWfluxes}), 
the onset of thermal instability is where $d\Lambda/dT \sim 0$.  For our cooling function \citep[Fig.\ 1 of][]{Kim&Ostriker17}, there is a  maximum at $T \sim 10^5$~K.  We note however that thermal instability only develops in selected lower-density regions within our warm winds (see \autoref{fig:WWsnap}), rather than throughout the flow.  It is unknown whether on larger scales this could ultimately lead to a thermally-driven wind, as has been suggested by \citet{modak2023}; within our simulation domain, the CR pressure remains quite large compared to the thermal gas pressure.

\section{Conclusions and prospects}
\label{sec:conclusions}

The primary focus of this paper is on the dynamical interaction of CRs with multiphase gas in extraplanar regions, in order to understand when and how CRs are able to drive galactic winds. Our study makes use of the TIGRESS MHD simulations of portions of star-forming galactic disks \citep{Kim&Ostriker17,Kim+20} for initial conditions of the gas and magnetic fields, and also to set CR injection rates based on SN rates in the simulations.  

As a prelude to our investigation of wind driving, we follow up on \citetalias{Armillotta+21} and \citetalias{Armillotta+22} to analyse how the MHD backreaction modifies the transport and distribution of CRs, thereby testing the conclusions of our previous studies.
In the postprocessing CR simulations of \citetalias{Armillotta+21} and \citetalias{Armillotta+22} (with frozen MHD variables), the prevailing orientations of the velocity and magnetic field lines
cause CRs to be confined within the warm+cold, dense gas. However, once ``live MHD'' is turned on, the excess of CR pressure in these regions forces dense gas to expand into the hot gas. The resulting rearrangement of the velocity and magnetic field topology enables CRs to propagate out of the dense gas. 
The net result is a more uniform CR distribution between the hot and warm+cold gas regions when MHD is ``live'' (compare \autoref{fig:FBsnapIni}, \autoref{fig:FBsnapMHD}). 
Even though the ``live MHD'' simulations produce a smoother CR distribution compared to the postprocessing simulations, 
our other previous conclusions regarding CR transport remain valid. In particular, we confirm that the scattering coefficient varies over more than four orders of magnitude depending on the gas properties (e.g., gas density and ionization), and that the transport of CRs is different in different thermal phases of the gas (\autoref{fig:FBsigma}, \autoref{fig:FBvel}). In ionized gas at $T>10^4$ K, the scattering coefficient is $\sigma_\parallel \sim 10^{-28}-10^{-27} {\rm cm^{-2}\ s}$, whereas in mostly-neutral gas at $T<10^4$ K, $\sigma_\parallel <10^{-31} {\rm cm^{-2}\ s}$. As a result of the high scattering rate, CR transport is primarily by advection in the high-velocity, low-density hot gas, and by streaming at the Alfv\'en speed in intermediate-density warm ionized gas. Given the extremely low scattering rate in neutral warm and cold gas (which is the majority of mass in the ISM), diffusion dominates CR transport in these regions. The CR pressure near the midplane is comparable to the kinetic, magnetic, and thermal pressure in the gas, but the scale height of CRs is far larger than that of the warm+cold gas (\autoref{fig:FBpressprof}).  

In our studies of CR-driven winds, for the initial conditions we extract a domain from the extra-planar region ($z > 500$~pc) of the TIGRESS solar neighborhood MHD simulation outputs, considering both an epoch of strong and weaker outflow (see \autoref{tab:tab1} for conditions of each case). We conduct simulations (with duration $\sim 100$ Myr) in which we inject both gas and CRs with prescribed boundary conditions at the bottom of the simulation box. 
In the hot wind models (entries 1--6 in \autoref{tab:tab2}), the inflow boundary conditions for the MHD variables are set to their initial mean values from TIGRESS in the hot outflowing gas at $z = 500$~pc, while exploring different conditions of CR pressure. In the warm wind models (entries 7--12 in \autoref{tab:tab2}), the injected gas is warm ($T = 10^4$~K) and slow-moving; the gas velocity and magnetic field inflow boundary conditions are set to the mean values within the warm outflow from TIGRESS at $z = 500$~pc, while exploring different values of injected gas density and CR pressure.

We find that CRs in fast-moving, predominantly-hot winds do not aid in the outflow acceleration, but instead slow down the gas. In both of our M1 -- HW models with fast-moving hot gas outflows, the CR pressure evolves to increase outward because the advection velocity decreases outward, so that momentum is transferred from gas to CRs (see \autoref{fig:HWfluxdiff}).  When a powerful hot wind is present, its interaction with embedded denser/cooler cloudlets shreds and accelerates them, effectively transferring momentum to create large outward mass fluxes of warm gas (\autoref{fig:HWfluxes}).

By contrast, our warm wind models demonstrate that CRs can effectively transfer momentum to the surrounding gas as the latter moves towards higher altitudes, with the MHD momentum flux increasing by $\sim 200\% $ and warm gas accelerating to $\sim 50-60 \ \kms$ within 3 kpc (\autoref{fig:WWfluxdiff}, \autoref{fig:WWzprof1}).
The reason that CR-driven acceleration is effective for the warm wind case is that streaming at the Alfv\'en speed controls CR transport for the conditions of warm gas in the extra-planar region (\autoref{fig:WWvel}).  When streaming dominates, the CR pressure decreases as $P_\mathrm{c} \propto \rho^{2/3}$, and therefore the effective sound speed increases upward as $C_\mathrm{eff}^2\approx dP_\mathrm{c}/d\rho \propto \rho^{-1/3}$).
This condition ensures that the CR sound speed remains larger than the gas vertical velocity, which is an essential requirement for the gas to be accelerated (see \autoref{eq:Mao_acc}).  
In our WW simulations, mass, momentum, and energy fluxes of the gas flowing out of the box are nearly independent of the injected gas density, while the fluxes increase with the level of CR pressure at the base of the extra-planar region. We also find that CRs affect 
the thermal state of the outflow: in the low-density ($n_\mathrm{H} < 10^{-4}$~cm$^{-3}$), higher-temperature ($T > 10^{5}$~K) regime, gas becomes thermally unstable with the rate of heating due to CR streaming losses larger than the rate of radiative cooling.

In conclusion, our findings indicate that under realistic scattering rates that are self-consistently determined, CRs can effectively drive flows of warm gas away from the galactic disk. 
This process is likely to be important in locations where there has not been a recent superbubble breakout powering a hot wind.  
Our results suggest that CRs remain a promising candidate to help remove gas from galaxies and limit star formation over cosmic timescales.  

It is important to emphasize, however, that while the simplified simulations presented in this study provide valuable insights into the conditions under which CRs are able to accelerate the surrounding gas, there is still much work to be done.  In particular, to fully understand the initiation of multiphase outflows and the relative importance of the CR pressure force vs. transfer of momentum from hot gas in accelerating cooler and denser outflows, it is necessary to self-consistently include CR energy injection together with thermal/kinetic energy injection as part of the supernova feedback process.  Incorporation of CRs within local star-forming ISM simulations such as TIGRESS offers a practical way to explore parameter space with sufficient numerical resolution to follow all of the ISM phases and their interaction with CRs.  For comprehensive understanding of the complete outflow formation and evolution process, including acceleration to velocities exceeding the escape speed as streamlines open up, it will also be necessary to conduct simulations of entire galactic disks. For simulations of dwarf galaxies, it will be possible to employ
sufficiently high resolution to directly follow star formation and
feedback processes including CR physics; for simulations of massive galaxies at lower resolution, a wind launching model may be calibrated based on 
resolved local-disk simulations \citep[as in][]{Kim+20b} but  including CRs.

\section*{Acknowledgements}

We thank the anonymous referee for helpful comments and suggestions. This work was supported in part by grant 510940 from the Simons Foundation to E.~C.\ Ostriker. The work of C.-G.K.\ was supported in part by NASA ATP grant No.\ 80NSSC22K0717. Computational resources were
provided by the Princeton Institute for Computational Science and Engineering
(PICSciE) and the Office of Information Technology's High Performance Computing
Center at Princeton University. The Center for Computational Astrophysics at the Flatiron Institute is supported by the Simons Foundation.

\bibliography{bib}{}

\begin{thebibliography}{}
\expandafter\ifx\csname natexlab\endcsname\relax\def\natexlab#1{#1}\fi
\providecommand{\url}[1]{\href{#1}{#1}}
\providecommand{\dodoi}[1]{doi:~\href{http://doi.org/#1}{\nolinkurl{#1}}}
\providecommand{\doeprint}[1]{\href{http://ascl.net/#1}{\nolinkurl{http://ascl.net/#1}}}
\providecommand{\doarXiv}[1]{\href{https://arxiv.org/abs/#1}{\nolinkurl{https://arxiv.org/abs/#1}}}

\bibitem[{{Ackermann} {et~al.}(2014){Ackermann}, {Ajello}, {Albert}, {Baldini},
  {Ballet}, {Barbiellini}, {Bastieri}, {Bellazzini}, {Bissaldi}, {Blandford},
  {Bloom}, {Bottacini}, {Brandt}, {Bregeon}, {Bruel}, {Buehler}, {Buson},
  {Caliandro}, {Cameron}, {Caragiulo}, {Caraveo}, {Cavazzuti}, {Charles},
  {Chekhtman}, {Cheung}, {Chiang}, {Chiaro}, {Ciprini}, {Claus},
  {Cohen-Tanugi}, {Conrad}, {Corbel}, {D'Ammando}, {de Angelis}, {den Hartog},
  {de Palma}, {Dermer}, {Desiante}, {Digel}, {Di Venere}, {do Couto e Silva},
  {Donato}, {Drell}, {Drlica-Wagner}, {Favuzzi}, {Ferrara}, {Focke},
  {Franckowiak}, {Fuhrmann}, {Fukazawa}, {Fusco}, {Gargano}, {Gasparrini},
  {Germani}, {Giglietto}, {Giordano}, {Giroletti}, {Glanzman}, {Godfrey},
  {Grenier}, {Grove}, {Guiriec}, {Hadasch}, {Harding}, {Hayashida}, {Hays},
  {Hewitt}, {Hill}, {Hou}, {Jean}, {Jogler}, {J{\'o}hannesson}, {Johnson},
  {Johnson}, {Kerr}, {Kn{\"o}dlseder}, {Kuss}, {Larsson}, {Latronico},
  {Lemoine-Goumard}, {Longo}, {Loparco}, {Lott}, {Lovellette}, {Lubrano},
  {Manfreda}, {Martin}, {Massaro}, {Mayer}, {Mazziotta}, {McEnery},
  {Michelson}, {Mitthumsiri}, {Mizuno}, {Monzani}, {Morselli}, {Moskalenko},
  {Murgia}, {Nemmen}, {Nuss}, {Ohsugi}, {Omodei}, {Orienti}, {Orlando},
  {Ormes}, {Paneque}, {Panetta}, {Perkins}, {Pesce-Rollins}, {Piron}, {Pivato},
  {Porter}, {Rain{\`o}}, {Rando}, {Razzano}, {Razzaque}, {Reimer}, {Reimer},
  {Reposeur}, {Saz Parkinson}, {Schaal}, {Schulz}, {Sgr{\`o}}, {Siskind},
  {Spandre}, {Spinelli}, {Stawarz}, {Suson}, {Takahashi}, {Tanaka}, {Thayer},
  {Thayer}, {Thompson}, {Tibaldo}, {Tinivella}, {Torres}, {Tosti}, {Troja},
  {Uchiyama}, {Vianello}, {Winer}, {Wolff}, {Wood}, {Wood}, {Wood},
  {Charbonnel}, {Corbet}, {De Gennaro Aquino}, {Edlin}, {Mason}, {Schwarz},
  {Shore}, {Starrfield}, {Teyssier}, \& {Fermi-LAT
  Collaboration}}]{Ackermann+13}
{Ackermann}, M., {Ajello}, M., {Albert}, A., {et~al.} 2014, Science, 345, 554,
  \dodoi{10.1126/science.1253947}

\bibitem[{{Aguilar} {et~al.}(2014){Aguilar}, {Aisa}, {Alvino}, {Ambrosi},
  {Andeen}, {Arruda}, {Attig}, {Azzarello}, {Bachlechner}, {Barao}, {Barrau},
  {Barrin}, {Bartoloni}, {Basara}, {Battarbee}, {Battiston}, {Bazo}, {Becker},
  {Behlmann}, {Beischer}, {Berdugo}, {Bertucci}, {Bigongiari}, {Bindi},
  {Bizzaglia}, {Bizzarri}, {Boella}, {de Boer}, {Bollweg}, {Bonnivard},
  {Borgia}, {Borsini}, {Boschini}, {Bourquin}, {Burger}, {Cadoux}, {Cai},
  {Capell}, {Caroff}, {Casaus}, {Cascioli}, {Castellini}, {Cernuda},
  {Cervelli}, {Chae}, {Chang}, {Chen}, {Chen}, {Cheng}, {Chen}, {Cheng},
  {Chikanian}, {Chou}, {Choumilov}, {Choutko}, {Chung}, {Clark}, {Clavero},
  {Coignet}, {Consolandi}, {Contin}, {Corti}, {Coste}, {Cui}, {Dai}, {Delgado},
  {Della Torre}, {Demirk{\"o}z}, {Derome}, {Di Falco}, {Di Masso}, {Dimiccoli},
  {D{\'\i}az}, {von Doetinchem}, {Du}, {Duranti}, {D'Urso}, {Eline}, {Eppling},
  {Eronen}, {Fan}, {Farnesini}, {Feng}, {Fiandrini}, {Fiasson}, {Finch},
  {Fisher}, {Galaktionov}, {Gallucci}, {Garc{\'\i}a}, {Garc{\'\i}a-L{\'o}pez},
  {Gast}, {Gebauer}, {Gervasi}, {Ghelfi}, {Gillard}, {Giovacchini}, {Goglov},
  {Gong}, {Goy}, {Grabski}, {Grand i}, {Graziani}, {Guandalini}, {Guerri},
  {Guo}, {Habiby}, {Haino}, {Han}, {He}, {Heil}, {Hoffman}, {Hsieh}, {Huang},
  {Huh}, {Incagli}, {Ionica}, {Jang}, {Jinchi}, {Kanishev}, {Kim}, {Kim},
  {Kirn}, {Kossakowski}, {Kounina}, {Kounine}, {Koutsenko}, {Krafczyk}, {Kunz},
  {La Vacca}, {Laudi}, {Laurenti}, {Lazzizzera}, {Lebedev}, {Lee}, {Lee},
  {Leluc}, {Li}, {Li}, {Li}, {Li}, {Li}, {Li}, {Li}, {Li}, {Li}, {Lim}, {Lin},
  {Lipari}, {Lippert}, {Liu}, {Liu}, {Lomtadze}, {Lu}, {Lu}, {Luebelsmeyer},
  {Luo}, {Luo}, {Lv}, {Majka}, {Malinin}, {Ma{\~n}{\'a}}, {Mar{\'\i}n},
  {Martin}, {Mart{\'\i}nez}, {Masi}, {Maurin}, {Menchaca-Rocha}, {Meng}, {Mo},
  {Morescalchi}, {Mott}, {M{\"u}ller}, {Ni}, {Nikonov}, {Nozzoli}, {Nunes},
  {Obermeier}, {Oliva}, {Orcinha}, {Palmonari}, {Palomares}, {Paniccia},
  {Papi}, {Pedreschi}, {Pensotti}, {Pereira}, {Pilo}, {Piluso}, {Pizzolotto},
  {Plyaskin}, {Pohl}, {Poireau}, {Postaci}, {Putze}, {Quadrani}, {Qi},
  {Rancoita}, {Rapin}, {Ricol}, {Rodr{\'\i}guez}, {Rosier-Lees}, {Rozhkov},
  {Rozza}, {Sagdeev}, {Sandweiss}, {Saouter}, {Sbarra}, {Schael}, {Schmidt},
  {Schuckardt}, {von Dratzig}, {Schwering}, {Scolieri}, {Seo}, {Shan}, {Shan},
  {Shi}, {Shi}, {Shi}, {Siedenburg}, {Son}, {Spada}, {Spinella}, {Sun}, {Sun},
  {Tacconi}, {Tang}, {Tang}, {Tang}, {Tao}, {Tescaro}, {Ting}, {Ting},
  {Tomassetti}, {Torsti}, {T{\"u}rko{\v{g}}lu}, {Urban}, {Vagelli}, {Valente},
  {Vannini}, {Valtonen}, {Vaurynovich}, {Vecchi}, {Velasco}, {Vialle}, {Wang},
  {Wang}, {Wang}, {Wang}, {Wang}, {Weng}, {Whitman}, {Wienkenh{\"o}ver}, {Wu},
  {Xia}, {Xie}, {Xie}, {Xiong}, {Xin}, {Xu}, {Xu}, {Yan}, {Yang}, {Yang}, {Ye},
  {Yi}, {Yu}, {Yu}, {Zeissler}, {Zhang}, {Zhang}, {Zhang}, {Zhang}, {Zheng},
  {Zhuang}, {Zhukov}, {Zichichi}, {Zimmermann}, {Zuccon}, {Zurbach}, \& {AMS
  Collaboration}}]{Aguilar+14}
{Aguilar}, M., {Aisa}, D., {Alvino}, A., {et~al.} 2014, \prl, 113, 121102,
  \dodoi{10.1103/PhysRevLett.113.121102}

\bibitem[{{Aguilar} {et~al.}(2015){Aguilar}, {Aisa}, {Alpat}, {Alvino},
  {Ambrosi}, {Andeen}, {Arruda}, {Attig}, {Azzarello}, {Bachlechner}, {Barao},
  {Barrau}, {Barrin}, {Bartoloni}, {Basara}, {Battarbee}, {Battiston}, {Bazo},
  {Becker}, {Behlmann}, {Beischer}, {Berdugo}, {Bertucci}, {Bindi},
  {Bizzaglia}, {Bizzarri}, {Boella}, {de Boer}, {Bollweg}, {Bonnivard},
  {Borgia}, {Borsini}, {Boschini}, {Bourquin}, {Burger}, {Cadoux}, {Cai},
  {Capell}, {Caroff}, {Casaus}, {Castellini}, {Cernuda}, {Cerreta}, {Cervelli},
  {Chae}, {Chang}, {Chen}, {Chen}, {Chen}, {Chen}, {Cheng}, {Chou},
  {Choumilov}, {Choutko}, {Chung}, {Clark}, {Clavero}, {Coignet}, {Consolandi},
  {Contin}, {Corti}, {Gil}, {Coste}, {Creus}, {Crispoltoni}, {Cui}, {Dai},
  {Delgado}, {Della Torre}, {Demirk{\"o}z}, {Derome}, {Di Falco}, {Di Masso},
  {Dimiccoli}, {D{\'\i}az}, {von Doetinchem}, {Donnini}, {Duranti}, {D'Urso},
  {Egorov}, {Eline}, {Eppling}, {Eronen}, {Fan}, {Farnesini}, {Feng},
  {Fiandrini}, {Fiasson}, {Finch}, {Fisher}, {Formato}, {Galaktionov},
  {Gallucci}, {Garc{\'\i}a}, {Garc{\'\i}a-L{\'o}pez}, {Gargiulo}, {Gast},
  {Gebauer}, {Gervasi}, {Ghelfi}, {Giovacchini}, {Goglov}, {Gong}, {Goy},
  {Grabski}, {Grandi}, {Graziani}, {Guand alini}, {Guerri}, {Guo}, {Haas},
  {Habiby}, {Haino}, {Han}, {He}, {Heil}, {Hoffman}, {Hsieh}, {Huang}, {Huh},
  {Incagli}, {Ionica}, {Jang}, {Jinchi}, {Kanishev}, {Kim}, {Kim}, {Kirn},
  {Korkmaz}, {Kossakowski}, {Kounina}, {Kounine}, {Koutsenko}, {Krafczyk}, {La
  Vacca}, {Laudi}, {Laurenti}, {Lazzizzera}, {Lebedev}, {Lee}, {Lee}, {Leluc},
  {Li}, {Li}, {Li}, {Li}, {Li}, {Li}, {Li}, {Li}, {Li}, {Li}, {Lim}, {Lin},
  {Lipari}, {Lippert}, {Liu}, {Liu}, {Liu}, {Lolli}, {Lomtadze}, {Lu}, {Lu},
  {Lu}, {Luebelsmeyer}, {Luo}, {Luo}, {Lv}, {Majka}, {Ma{\~n}{\'a}},
  {Mar{\'\i}n}, {Martin}, {Mart{\'\i}nez}, {Masi}, {Maurin}, {Menchaca-Rocha},
  {Meng}, {Mo}, {Morescalchi}, {Mott}, {M{\"u}ller}, {Nelson}, {Ni}, {Nikonov},
  {Nozzoli}, {Nunes}, {Obermeier}, {Oliva}, {Orcinha}, {Palmonari},
  {Palomares}, {Paniccia}, {Papi}, {Pauluzzi}, {Pedreschi}, {Pensotti},
  {Pereira}, {Picot-Clemente}, {Pilo}, {Piluso}, {Pizzolotto}, {Plyaskin},
  {Pohl}, {Poireau}, {Putze}, {Quadrani}, {Qi}, {Qin}, {Qu}, {R{\"a}ih{\"a}},
  {Rancoita}, {Rapin}, {Ricol}, {Rodr{\'\i}guez}, {Rosier-Lees}, {Rozhkov},
  {Rozza}, {Sagdeev}, {Sandweiss}, {Saouter}, {Schael}, {Schmidt}, {von
  Dratzig}, {Schwering}, {Scolieri}, {Seo}, {Shan}, {Shan}, {Shi}, {Shi},
  {Shi}, {Siedenburg}, {Son}, {Song}, {Spada}, {Spinella}, {Sun}, {Sun},
  {Tacconi}, {Tang}, {Tang}, {Tang}, {Tao}, {Tescaro}, {Ting}, {Ting},
  {Tomassetti}, {Torsti}, {T{\"u}rko{\v{g}}lu}, {Urban}, {Vagelli}, {Valente},
  {Vannini}, {Valtonen}, {Vaurynovich}, {Vecchi}, {Velasco}, {Vialle},
  {Vitale}, {Vitillo}, {Wang}, {Wang}, {Wang}, {Wang}, {Wang}, {Wang}, {Weng},
  {Whitman}, {Wienkenh{\"o}ver}, {Willenbrock}, {Wu}, {Wu}, {Xia}, {Xie},
  {Xie}, {Xiong}, {Xu}, {Xu}, {Yan}, {Yang}, {Yang}, {Yang}, {Ye}, {Yi}, {Yu},
  {Yu}, {Zeissler}, {Zhang}, {Zhang}, {Zhang}, {Zhang}, {Zhang}, {Zhang},
  {Zhang}, {Zheng}, {Zhuang}, {Zhukov}, {Zichichi}, {Zimmermann}, {Zuccon}, \&
  {AMS Collaboration}}]{Aguilar+15}
{Aguilar}, M., {Aisa}, D., {Alpat}, B., {et~al.} 2015, \prl, 115, 211101,
  \dodoi{10.1103/PhysRevLett.115.211101}

\bibitem[{{Amato} \& {Blasi}(2018)}]{Amato&Blasi18}
{Amato}, E., \& {Blasi}, P. 2018, Advances in Space Research, 62, 2731,
  \dodoi{10.1016/j.asr.2017.04.019}

\bibitem[{{Armillotta} {et~al.}(2016){Armillotta}, {Fraternali}, \&
  {Marinacci}}]{Armillotta+16}
{Armillotta}, L., {Fraternali}, F., \& {Marinacci}, F. 2016, \mnras, 462, 4157,
  \dodoi{10.1093/mnras/stw1930}

\bibitem[{{Armillotta} {et~al.}(2021){Armillotta}, {Ostriker}, \&
  {Jiang}}]{Armillotta+21}
{Armillotta}, L., {Ostriker}, E.~C., \& {Jiang}, Y.-F. 2021, \apj, 922, 11,
  \dodoi{10.3847/1538-4357/ac1db2}

\bibitem[{{Armillotta} {et~al.}(2022){Armillotta}, {Ostriker}, \&
  {Jiang}}]{Armillotta+22}
---. 2022, \apj, 929, 170, \dodoi{10.3847/1538-4357/ac5fa9}

\bibitem[{{Bai} {et~al.}(2019){Bai}, {Ostriker}, {Plotnikov}, \&
  {Stone}}]{Bai2019}
{Bai}, X.-N., {Ostriker}, E.~C., {Plotnikov}, I., \& {Stone}, J.~M. 2019, \apj,
  876, 60, \dodoi{10.3847/1538-4357/ab1648}

\bibitem[{{Bambic} {et~al.}(2021){Bambic}, {Bai}, \& {Ostriker}}]{Bambic2021}
{Bambic}, C.~J., {Bai}, X.-N., \& {Ostriker}, E.~C. 2021, \apj, 920, 141,
  \dodoi{10.3847/1538-4357/ac0ce7}

\bibitem[{{Banda-Barrag{\'a}n} {et~al.}(2021){Banda-Barrag{\'a}n},
  {Br{\"u}ggen}, {Heesen}, {Scannapieco}, {Cottle}, {Federrath}, \&
  {Wagner}}]{Banda-Barragan+21}
{Banda-Barrag{\'a}n}, W.~E., {Br{\"u}ggen}, M., {Heesen}, V., {et~al.} 2021,
  \mnras, 506, 5658, \dodoi{10.1093/mnras/stab1884}

\bibitem[{{Beck}(2001)}]{Beck01}
{Beck}, R. 2001, \ssr, 99, 243.
\newblock \doarXiv{astro-ph/0012402}

\bibitem[{{Bell}(2004)}]{Bell04}
{Bell}, A.~R. 2004, \mnras, 353, 550, \dodoi{10.1111/j.1365-2966.2004.08097.x}

\bibitem[{{Blasi}(2013)}]{Blasi13}
{Blasi}, P. 2013, \aapr, 21, 70, \dodoi{10.1007/s00159-013-0070-7}

\bibitem[{{Blasi} {et~al.}(2012){Blasi}, {Amato}, \& {Serpico}}]{Blasi+12}
{Blasi}, P., {Amato}, E., \& {Serpico}, P.~D. 2012, \prl, 109, 061101,
  \dodoi{10.1103/PhysRevLett.109.061101}

\bibitem[{{Bolatto} {et~al.}(2013){Bolatto}, {Warren}, {Leroy}, {Walter},
  {Veilleux}, {Ostriker}, {Ott}, {Zwaan}, {Fisher}, {Weiss}, {Rosolowsky}, \&
  {Hodge}}]{Bolatto+13}
{Bolatto}, A.~D., {Warren}, S.~R., {Leroy}, A.~K., {et~al.} 2013, \nat, 499,
  450, \dodoi{10.1038/nature12351}

\bibitem[{{Boulares} \& {Cox}(1990)}]{Boulares&Cox90}
{Boulares}, A., \& {Cox}, D.~P. 1990, \apj, 365, 544, \dodoi{10.1086/169509}

\bibitem[{{Breitschwerdt} {et~al.}(1991){Breitschwerdt}, {McKenzie}, \&
  {Voelk}}]{Breitschwerdt+91}
{Breitschwerdt}, D., {McKenzie}, J.~F., \& {Voelk}, H.~J. 1991, \aap, 245, 79

\bibitem[{{Chan} {et~al.}(2021){Chan}, {Keres}, {Gurvich}, {Hopkins}, {Trapp},
  {Ji}, \& {Faucher-Giguere}}]{Chan+21}
{Chan}, T.~K., {Keres}, D., {Gurvich}, A.~B., {et~al.} 2021, arXiv e-prints,
  arXiv:2110.06231.
\newblock \doarXiv{2110.06231}

\bibitem[{{Chan} {et~al.}(2019){Chan}, {Kere{\v{s}}}, {Hopkins}, {Quataert},
  {Su}, {Hayward}, \& {Faucher-Gigu{\`e}re}}]{Chan+19}
{Chan}, T.~K., {Kere{\v{s}}}, D., {Hopkins}, P.~F., {et~al.} 2019, \mnras, 488,
  3716, \dodoi{10.1093/mnras/stz1895}

\bibitem[{{Chandran}(2000)}]{Chandran00}
{Chandran}, B. D.~G. 2000, \apj, 529, 513, \dodoi{10.1086/308232}

\bibitem[{{Chisholm} {et~al.}(2017){Chisholm}, {Tremonti}, {Leitherer}, \&
  {Chen}}]{Chisholm+17}
{Chisholm}, J., {Tremonti}, C.~A., {Leitherer}, C., \& {Chen}, Y. 2017, \mnras,
  469, 4831, \dodoi{10.1093/mnras/stx1164}

\bibitem[{{Concas} {et~al.}(2019){Concas}, {Popesso}, {Brusa}, {Mainieri}, \&
  {Thomas}}]{Concas+19}
{Concas}, A., {Popesso}, P., {Brusa}, M., {Mainieri}, V., \& {Thomas}, D. 2019,
  \aap, 622, A188, \dodoi{10.1051/0004-6361/201732152}

\bibitem[{{Cooper} {et~al.}(2009){Cooper}, {Bicknell}, {Sutherland}, \&
  {Bland-Hawthorn}}]{Cooper+09}
{Cooper}, J.~L., {Bicknell}, G.~V., {Sutherland}, R.~S., \& {Bland-Hawthorn},
  J. 2009, \apj, 703, 330, \dodoi{10.1088/0004-637X/703/1/330}

\bibitem[{{Cummings} {et~al.}(2016){Cummings}, {Stone}, {Heikkila}, {Lal},
  {Webber}, {J{\'o}hannesson}, {Moskalenko}, {Orlando}, \&
  {Porter}}]{Cummings+16}
{Cummings}, A.~C., {Stone}, E.~C., {Heikkila}, B.~C., {et~al.} 2016, \apj, 831,
  18, \dodoi{10.3847/0004-637X/831/1/18}

\bibitem[{{Di Teodoro} {et~al.}(2020){Di Teodoro}, {McClure-Griffiths},
  {Lockman}, \& {Armillotta}}]{DiTeodoro+20}
{Di Teodoro}, E.~M., {McClure-Griffiths}, N.~M., {Lockman}, F.~J., \&
  {Armillotta}, L. 2020, \nat, 584, 364, \dodoi{10.1038/s41586-020-2595-z}

\bibitem[{{Di Teodoro} {et~al.}(2019){Di Teodoro}, {McClure-Griffiths}, {De
  Breuck}, {Armillotta}, {Pingel}, {Jameson}, {Dickey}, {Rubio},
  {Stanimirovi{\'c}}, \& {Staveley-Smith}}]{DiTeodoro+19}
{Di Teodoro}, E.~M., {McClure-Griffiths}, N.~M., {De Breuck}, C., {et~al.}
  2019, \apjl, 885, L32, \dodoi{10.3847/2041-8213/ab4fe9}

\bibitem[{{Dorfi} \& {Breitschwerdt}(2012)}]{Dorfi&Breitschwerdt12}
{Dorfi}, E.~A., \& {Breitschwerdt}, D. 2012, \aap, 540, A77,
  \dodoi{10.1051/0004-6361/201118082}

\bibitem[{{Draine}(2011)}]{Draine11}
{Draine}, B.~T. 2011, {Physics of the Interstellar and Intergalactic Medium}

\bibitem[{{Everett} {et~al.}(2008){Everett}, {Zweibel}, {Benjamin}, {McCammon},
  {Rocks}, \& {Gallagher}}]{Everett+08}
{Everett}, J.~E., {Zweibel}, E.~G., {Benjamin}, R.~A., {et~al.} 2008, \apj,
  674, 258, \dodoi{10.1086/524766}

\bibitem[{{Evoli} {et~al.}(2018){Evoli}, {Blasi}, {Morlino}, \&
  {Aloisio}}]{Evoli+18}
{Evoli}, C., {Blasi}, P., {Morlino}, G., \& {Aloisio}, R. 2018, \prl, 121,
  021102, \dodoi{10.1103/PhysRevLett.121.021102}

\bibitem[{{Farber} {et~al.}(2018){Farber}, {Ruszkowski}, {Yang}, \&
  {Zweibel}}]{Farber+18}
{Farber}, R., {Ruszkowski}, M., {Yang}, H. Y.~K., \& {Zweibel}, E.~G. 2018,
  \apj, 856, 112, \dodoi{10.3847/1538-4357/aab26d}

\bibitem[{{Fielding} \& {Bryan}(2022)}]{Fielding2022}
{Fielding}, D.~B., \& {Bryan}, G.~L. 2022, \apj, 924, 82,
  \dodoi{10.3847/1538-4357/ac2f41}

\bibitem[{{Girichidis} {et~al.}(2018){Girichidis}, {Naab}, {Hanasz}, \&
  {Walch}}]{Girichidis+18}
{Girichidis}, P., {Naab}, T., {Hanasz}, M., \& {Walch}, S. 2018, \mnras, 479,
  3042, \dodoi{10.1093/mnras/sty1653}

\bibitem[{{Girichidis} {et~al.}(2022){Girichidis}, {Pfrommer}, {Pakmor}, \&
  {Springel}}]{Girichidis+22}
{Girichidis}, P., {Pfrommer}, C., {Pakmor}, R., \& {Springel}, V. 2022, \mnras,
  510, 3917, \dodoi{10.1093/mnras/stab3462}

\bibitem[{{Girichidis} {et~al.}(2016){Girichidis}, {Naab}, {Walch}, {Hanasz},
  {Mac Low}, {Ostriker}, {Gatto}, {Peters}, {W{\"u}nsch}, {Glover}, {Klessen},
  {Clark}, \& {Baczynski}}]{Girichidis+16}
{Girichidis}, P., {Naab}, T., {Walch}, S., {et~al.} 2016, \apjl, 816, L19,
  \dodoi{10.3847/2041-8205/816/2/L19}

\bibitem[{{Grenier} {et~al.}(2015){Grenier}, {Black}, \& {Strong}}]{Grenier+15}
{Grenier}, I.~A., {Black}, J.~H., \& {Strong}, A.~W. 2015, \araa, 53, 199,
  \dodoi{10.1146/annurev-astro-082214-122457}

\bibitem[{{Gronke} \& {Oh}(2020)}]{Gronke+20}
{Gronke}, M., \& {Oh}, S.~P. 2020, \mnras, 492, 1970,
  \dodoi{10.1093/mnras/stz3332}

\bibitem[{{Hanasz} {et~al.}(2021){Hanasz}, {Strong}, \&
  {Girichidis}}]{Hanasz+21}
{Hanasz}, M., {Strong}, A., \& {Girichidis}, P. 2021, arXiv e-prints,
  arXiv:2106.08426.
\newblock \doarXiv{2106.08426}

\bibitem[{{Heckman} {et~al.}(2015){Heckman}, {Alexandroff}, {Borthakur},
  {Overzier}, \& {Leitherer}}]{Heckman+15}
{Heckman}, T.~M., {Alexandroff}, R.~M., {Borthakur}, S., {Overzier}, R., \&
  {Leitherer}, C. 2015, \apj, 809, 147, \dodoi{10.1088/0004-637X/809/2/147}

\bibitem[{{Heckman} \& {Thompson}(2017)}]{Heckman+17}
{Heckman}, T.~M., \& {Thompson}, T.~A. 2017, arXiv e-prints, arXiv:1701.09062,
  \dodoi{10.48550/arXiv.1701.09062}

\bibitem[{{Hodges-Kluck} {et~al.}(2020){Hodges-Kluck}, {Yukita}, {Tanner},
  {Ptak}, {Bregman}, \& {Li}}]{Hodges-Kluck+20}
{Hodges-Kluck}, E.~J., {Yukita}, M., {Tanner}, R., {et~al.} 2020, \apj, 903,
  35, \dodoi{10.3847/1538-4357/abb884}

\bibitem[{{Hopkins} {et~al.}(2022){Hopkins}, {Squire}, {Butsky}, \&
  {Ji}}]{Hopkins+22}
{Hopkins}, P.~F., {Squire}, J., {Butsky}, I.~S., \& {Ji}, S. 2022, \mnras, 517,
  5413, \dodoi{10.1093/mnras/stac2909}

\bibitem[{{Hopkins} {et~al.}(2021){Hopkins}, {Squire}, {Chan}, {Quataert},
  {Ji}, {Kere{\v{s}}}, \& {Faucher-Gigu{\`e}re}}]{Hopkins2021}
{Hopkins}, P.~F., {Squire}, J., {Chan}, T.~K., {et~al.} 2021, \mnras, 501,
  4184, \dodoi{10.1093/mnras/staa3691}

\bibitem[{{Hopkins} {et~al.}(2018{\natexlab{a}}){Hopkins}, {Wetzel},
  {Kere{\v{s}}}, {Faucher-Gigu{\`e}re}, {Quataert}, {Boylan-Kolchin}, {Murray},
  {Hayward}, {Garrison-Kimmel}, {Hummels}, {Feldmann}, {Torrey}, {Ma},
  {Angl{\'e}s-Alc{\'a}zar}, {Su}, {Orr}, {Schmitz}, {Escala}, {Sanderson},
  {Grudi{\'c}}, {Hafen}, {Kim}, {Fitts}, {Bullock}, {Wheeler}, {Chan},
  {Elbert}, \& {Narayanan}}]{Hopkins+18}
{Hopkins}, P.~F., {Wetzel}, A., {Kere{\v{s}}}, D., {et~al.} 2018{\natexlab{a}},
  \mnras, 480, 800, \dodoi{10.1093/mnras/sty1690}

\bibitem[{{Hopkins} {et~al.}(2018{\natexlab{b}}){Hopkins}, {Wetzel},
  {Kere{\v{s}}}, {Faucher-Gigu{\`e}re}, {Quataert}, {Boylan-Kolchin}, {Murray},
  {Hayward}, \& {El-Badry}}]{Hopkins2018a}
---. 2018{\natexlab{b}}, \mnras, 477, 1578, \dodoi{10.1093/mnras/sty674}

\bibitem[{{Hopkins} {et~al.}(2020){Hopkins}, {Chan}, {Garrison-Kimmel}, {Ji},
  {Su}, {Hummels}, {Kere{\v{s}}}, {Quataert}, \&
  {Faucher-Gigu{\`e}re}}]{Hopkins+19}
{Hopkins}, P.~F., {Chan}, T.~K., {Garrison-Kimmel}, S., {et~al.} 2020, \mnras,
  492, 3465, \dodoi{10.1093/mnras/stz3321}

\bibitem[{{Jacob} \& {Pfrommer}(2017)}]{Jacob+17}
{Jacob}, S., \& {Pfrommer}, C. 2017, \mnras, 467, 1449,
  \dodoi{10.1093/mnras/stx131}

\bibitem[{{Ji} {et~al.}(2020){Ji}, {Chan}, {Hummels}, {Hopkins}, {Stern},
  {Kere{\v{s}}}, {Quataert}, {Faucher-Gigu{\`e}re}, \& {Murray}}]{Ji+20}
{Ji}, S., {Chan}, T.~K., {Hummels}, C.~B., {et~al.} 2020, \mnras, 496, 4221,
  \dodoi{10.1093/mnras/staa1849}

\bibitem[{{Jiang} \& {Oh}(2018)}]{Jiang&Oh18}
{Jiang}, Y.-F., \& {Oh}, S.~P. 2018, \apj, 854, 5,
  \dodoi{10.3847/1538-4357/aaa6ce}

\bibitem[{{Kempski} \& {Quataert}(2020)}]{Kempski+20}
{Kempski}, P., \& {Quataert}, E. 2020, \mnras, 493, 1801,
  \dodoi{10.1093/mnras/staa385}

\bibitem[{{Kim} \& {Ostriker}(2015{\natexlab{a}})}]{KimOstriker2015}
{Kim}, C.-G., \& {Ostriker}, E.~C. 2015{\natexlab{a}}, \apj, 815, 67,
  \dodoi{10.1088/0004-637X/815/1/67}

\bibitem[{{Kim} \& {Ostriker}(2015{\natexlab{b}})}]{KimOstriker2015a}
---. 2015{\natexlab{b}}, \apj, 802, 99, \dodoi{10.1088/0004-637X/802/2/99}

\bibitem[{{Kim} \& {Ostriker}(2017)}]{Kim&Ostriker17}
---. 2017, \apj, 846, 133, \dodoi{10.3847/1538-4357/aa8599}

\bibitem[{{Kim} \& {Ostriker}(2018)}]{Kim&Ostriker18}
---. 2018, \apj, 853, 173, \dodoi{10.3847/1538-4357/aaa5ff}

\bibitem[{{Kim} {et~al.}(2020{\natexlab{a}}){Kim}, {Ostriker}, {Somerville},
  {Bryan}, {Fielding}, {Forbes}, {Hayward}, {Hernquist}, \& {Pandya}}]{Kim+20}
{Kim}, C.-G., {Ostriker}, E.~C., {Somerville}, R.~S., {et~al.}
  2020{\natexlab{a}}, \apj, 900, 61, \dodoi{10.3847/1538-4357/aba962}

\bibitem[{{Kim} {et~al.}(2020{\natexlab{b}}){Kim}, {Ostriker}, {Fielding},
  {Smith}, {Bryan}, {Somerville}, {Forbes}, {Genel}, \& {Hernquist}}]{Kim+20b}
{Kim}, C.-G., {Ostriker}, E.~C., {Fielding}, D.~B., {et~al.}
  2020{\natexlab{b}}, \apjl, 903, L34, \dodoi{10.3847/2041-8213/abc252}

\bibitem[{{Koyama} \& {Inutsuka}(2002)}]{Koyama&Inutsuka02}
{Koyama}, H., \& {Inutsuka}, S.-i. 2002, \apjl, 564, L97,
  \dodoi{10.1086/338978}

\bibitem[{{Kulsrud} \& {Pearce}(1969)}]{Kulsrud&Pearce69}
{Kulsrud}, R., \& {Pearce}, W.~P. 1969, \apj, 156, 445, \dodoi{10.1086/149981}

\bibitem[{{Kulsrud}(2005)}]{Kulsrud05}
{Kulsrud}, R.~M. 2005, {Plasma physics for astrophysics}

\bibitem[{{Kulsrud} \& {Cesarsky}(1971)}]{Kulsrud&Cesarsky1971}
{Kulsrud}, R.~M., \& {Cesarsky}, C.~J. 1971, \aplett, 8, 189

\bibitem[{{Leroy} {et~al.}(2015){Leroy}, {Walter}, {Martini}, {Roussel},
  {Sandstrom}, {Ott}, {Weiss}, {Bolatto}, {Schuster}, \&
  {Dessauges-Zavadsky}}]{Leroy+15}
{Leroy}, A.~K., {Walter}, F., {Martini}, P., {et~al.} 2015, \apj, 814, 83,
  \dodoi{10.1088/0004-637X/814/2/83}

\bibitem[{{Mao} \& {Ostriker}(2018)}]{Mao&Ostriker18}
{Mao}, S.~A., \& {Ostriker}, E.~C. 2018, \apj, 854, 89,
  \dodoi{10.3847/1538-4357/aaa88e}

\bibitem[{{Marasco} {et~al.}(2023){Marasco}, {Belfiore}, {Cresci}, {Lelli},
  {Venturi}, {Hunt}, {Concas}, {Marconi}, {Mannucci}, {Mingozzi}, {McLeod},
  {Kumari}, {Carniani}, {Vanzi}, \& {Ginolfi}}]{Marasco+23}
{Marasco}, A., {Belfiore}, F., {Cresci}, G., {et~al.} 2023, \aap, 670, A92,
  \dodoi{10.1051/0004-6361/202244895}

\bibitem[{{Martin}(2005)}]{Martin+05}
{Martin}, C.~L. 2005, \apj, 621, 227, \dodoi{10.1086/427277}

\bibitem[{{Martini} {et~al.}(2018){Martini}, {Leroy}, {Mangum}, {Bolatto},
  {Keating}, {Sandstrom}, \& {Walter}}]{Martini+18}
{Martini}, P., {Leroy}, A.~K., {Mangum}, J.~G., {et~al.} 2018, \apj, 856, 61,
  \dodoi{10.3847/1538-4357/aab08e}

\bibitem[{{McQuinn} \& {Werk}(2018)}]{McQuinn+18}
{McQuinn}, M., \& {Werk}, J.~K. 2018, \apj, 852, 33,
  \dodoi{10.3847/1538-4357/aa9d3f}

\bibitem[{{Modak} {et~al.}(2023){Modak}, {Quataert}, {Jiang}, \&
  {Thompson}}]{modak2023}
{Modak}, S., {Quataert}, E., {Jiang}, Y.-F., \& {Thompson}, T.~A. 2023, \mnras,
  524, 6374, \dodoi{10.1093/mnras/stad2257}

\bibitem[{{Morlino} \& {Caprioli}(2012)}]{Morlino&Caprioli12}
{Morlino}, G., \& {Caprioli}, D. 2012, \aap, 538, A81,
  \dodoi{10.1051/0004-6361/201117855}

\bibitem[{{Noon} {et~al.}(2023){Noon}, {Krumholz}, {Di Teodoro},
  {McClure-Griffiths}, {Lockman}, \& {Armillotta}}]{Noon+23}
{Noon}, K.~A., {Krumholz}, M.~R., {Di Teodoro}, E.~M., {et~al.} 2023, \mnras,
  524, 1258, \dodoi{10.1093/mnras/stad1890}

\bibitem[{{Ostriker} \& {Kim}(2022)}]{Ostriker+22}
{Ostriker}, E.~C., \& {Kim}, C.-G. 2022, \apj, 936, 137,
  \dodoi{10.3847/1538-4357/ac7de2}

\bibitem[{{Ostriker} {et~al.}(2010){Ostriker}, {McKee}, \&
  {Leroy}}]{Ostriker2010}
{Ostriker}, E.~C., {McKee}, C.~F., \& {Leroy}, A.~K. 2010, \apj, 721, 975,
  \dodoi{10.1088/0004-637X/721/2/975}

\bibitem[{{Ostriker} \& {Shetty}(2011)}]{Ostriker2011}
{Ostriker}, E.~C., \& {Shetty}, R. 2011, \apj, 731, 41,
  \dodoi{10.1088/0004-637X/731/1/41}

\bibitem[{{Padovani} {et~al.}(2020){Padovani}, {Ivlev}, {Galli}, {Offner},
  {Indriolo}, {Rodgers-Lee}, {Marcowith}, {Girichidis}, {Bykov}, \&
  {Kruijssen}}]{Padovani+20}
{Padovani}, M., {Ivlev}, A.~V., {Galli}, D., {et~al.} 2020, \ssr, 216, 29,
  \dodoi{10.1007/s11214-020-00654-1}

\bibitem[{{Pakmor} {et~al.}(2016){Pakmor}, {Pfrommer}, {Simpson}, \&
  {Springel}}]{Pakmor+16}
{Pakmor}, R., {Pfrommer}, C., {Simpson}, C.~M., \& {Springel}, V. 2016, \apjl,
  824, L30, \dodoi{10.3847/2041-8205/824/2/L30}

\bibitem[{{Peschken} {et~al.}(2023){Peschken}, {Hanasz}, {Naab},
  {W{\'o}lta{\'n}ski}, \& {Gawryszczak}}]{Peschken+23}
{Peschken}, N., {Hanasz}, M., {Naab}, T., {W{\'o}lta{\'n}ski}, D., \&
  {Gawryszczak}, A. 2023, \mnras, 522, 5529, \dodoi{10.1093/mnras/stad1358}

\bibitem[{{Piontek} \& {Ostriker}(2007)}]{PiontekOstriker2007}
{Piontek}, R.~A., \& {Ostriker}, E.~C. 2007, \apj, 663, 183,
  \dodoi{10.1086/518103}

\bibitem[{{Plotnikov} {et~al.}(2021){Plotnikov}, {Ostriker}, \&
  {Bai}}]{Plotnikov2021}
{Plotnikov}, I., {Ostriker}, E.~C., \& {Bai}, X.-N. 2021, \apj, 914, 3,
  \dodoi{10.3847/1538-4357/abf7b3}

\bibitem[{{Ptuskin} {et~al.}(2009){Ptuskin}, {Strelnikova}, \&
  {Sveshnikova}}]{Ptuskin+09}
{Ptuskin}, V.~S., {Strelnikova}, O.~N., \& {Sveshnikova}, L.~G. 2009,
  Astroparticle Physics, 31, 284, \dodoi{10.1016/j.astropartphys.2009.02.004}

\bibitem[{{Quataert} {et~al.}(2021{\natexlab{a}}){Quataert}, {Thompson}, \&
  {Jiang}}]{Quataert+21a}
{Quataert}, E., {Thompson}, T.~A., \& {Jiang}, Y.-F. 2021{\natexlab{a}},
  \mnras, \dodoi{10.1093/mnras/stab3273}

\bibitem[{{Quataert} {et~al.}(2021{\natexlab{b}}){Quataert}, {Thompson}, \&
  {Jiang}}]{Quataert+21b}
---. 2021{\natexlab{b}}, \mnras, \dodoi{10.1093/mnras/stab3273}

\bibitem[{{Rathjen} {et~al.}(2021){Rathjen}, {Naab}, {Girichidis}, {Walch},
  {W{\"u}nsch}, {Dinnbier}, {Seifried}, {Klessen}, \& {Glover}}]{Rathjen2021}
{Rathjen}, T.-E., {Naab}, T., {Girichidis}, P., {et~al.} 2021, arXiv e-prints,
  arXiv:2103.14128.
\newblock \doarXiv{2103.14128}

\bibitem[{{Recchia}(2021)}]{Recchia21}
{Recchia}, S. 2021, arXiv e-prints, arXiv:2101.02052.
\newblock \doarXiv{2101.02052}

\bibitem[{{Recchia} {et~al.}(2016){Recchia}, {Blasi}, \&
  {Morlino}}]{Recchia+16}
{Recchia}, S., {Blasi}, P., \& {Morlino}, G. 2016, \mnras, 462, 4227,
  \dodoi{10.1093/mnras/stw1966}

\bibitem[{{Rupke} {et~al.}(2005){Rupke}, {Veilleux}, \& {Sanders}}]{Rupke+05}
{Rupke}, D.~S., {Veilleux}, S., \& {Sanders}, D.~B. 2005, \apjs, 160, 115,
  \dodoi{10.1086/432889}

\bibitem[{{Ruszkowski} {et~al.}(2017){Ruszkowski}, {Yang}, \&
  {Zweibel}}]{Ruszkowski+17}
{Ruszkowski}, M., {Yang}, H. Y.~K., \& {Zweibel}, E. 2017, \apj, 834, 208,
  \dodoi{10.3847/1538-4357/834/2/208}

\bibitem[{{Salem} \& {Bryan}(2014)}]{Salem&Bryan13}
{Salem}, M., \& {Bryan}, G.~L. 2014, \mnras, 437, 3312,
  \dodoi{10.1093/mnras/stt2121}

\bibitem[{{Scannapieco} \& {Br{\"u}ggen}(2015)}]{Scannapieco+15}
{Scannapieco}, E., \& {Br{\"u}ggen}, M. 2015, \apj, 805, 158,
  \dodoi{10.1088/0004-637X/805/2/158}

\bibitem[{{Schneider} {et~al.}(2020){Schneider}, {Ostriker}, {Robertson}, \&
  {Thompson}}]{Schneider+20}
{Schneider}, E.~E., {Ostriker}, E.~C., {Robertson}, B.~E., \& {Thompson}, T.~A.
  2020, \apj, 895, 43, \dodoi{10.3847/1538-4357/ab8ae8}

\bibitem[{{Schneider} \& {Robertson}(2017)}]{Schneider+17}
{Schneider}, E.~E., \& {Robertson}, B.~E. 2017, \apj, 834, 144,
  \dodoi{10.3847/1538-4357/834/2/144}

\bibitem[{{Simpson} {et~al.}(2016){Simpson}, {Pakmor}, {Marinacci}, {Pfrommer},
  {Springel}, {Glover}, {Clark}, \& {Smith}}]{Simpson+16}
{Simpson}, C.~M., {Pakmor}, R., {Marinacci}, F., {et~al.} 2016, \apjl, 827,
  L29, \dodoi{10.3847/2041-8205/827/2/L29}

\bibitem[{{Skinner} \& {Ostriker}(2013)}]{SkinnerOstriker2013}
{Skinner}, M.~A., \& {Ostriker}, E.~C. 2013, \apjs, 206, 21,
  \dodoi{10.1088/0067-0049/206/2/21}

\bibitem[{{Sparre} {et~al.}(2020){Sparre}, {Pfrommer}, \& {Ehlert}}]{Sparre+20}
{Sparre}, M., {Pfrommer}, C., \& {Ehlert}, K. 2020, \mnras, 499, 4261,
  \dodoi{10.1093/mnras/staa3177}

\bibitem[{{Stone} {et~al.}(2020){Stone}, {Tomida}, {White}, \&
  {Felker}}]{Stone+20}
{Stone}, J.~M., {Tomida}, K., {White}, C.~J., \& {Felker}, K.~G. 2020, \apjs,
  249, 4, \dodoi{10.3847/1538-4365/ab929b}

\bibitem[{{Strickland} \& {Heckman}(2007)}]{Strickland&Heckman07}
{Strickland}, D.~K., \& {Heckman}, T.~M. 2007, \apj, 658, 258,
  \dodoi{10.1086/511174}

\bibitem[{{Sutherland} \& {Dopita}(1993)}]{Sutherland&Dopita93}
{Sutherland}, R.~S., \& {Dopita}, M.~A. 1993, \apjs, 88, 253,
  \dodoi{10.1086/191823}

\bibitem[{{Teng} {et~al.}(2013){Teng}, {Veilleux}, \& {Baker}}]{Teng+13}
{Teng}, S.~H., {Veilleux}, S., \& {Baker}, A.~J. 2013, \apj, 765, 95,
  \dodoi{10.1088/0004-637X/765/2/95}

\bibitem[{{Thomas} {et~al.}(2023){Thomas}, {Pfrommer}, \&
  {Pakmor}}]{Thosmas+23}
{Thomas}, T., {Pfrommer}, C., \& {Pakmor}, R. 2023, \mnras, 521, 3023,
  \dodoi{10.1093/mnras/stad472}

\bibitem[{{Uhlig} {et~al.}(2012){Uhlig}, {Pfrommer}, {Sharma}, {Nath},
  {En{\ss}lin}, \& {Springel}}]{Uhlig+12}
{Uhlig}, M., {Pfrommer}, C., {Sharma}, M., {et~al.} 2012, \mnras, 423, 2374,
  \dodoi{10.1111/j.1365-2966.2012.21045.x}

\bibitem[{{Veilleux} {et~al.}(2005){Veilleux}, {Cecil}, \&
  {Bland-Hawthorn}}]{Veilleux+05}
{Veilleux}, S., {Cecil}, G., \& {Bland-Hawthorn}, J. 2005, \araa, 43, 769,
  \dodoi{10.1146/annurev.astro.43.072103.150610}

\bibitem[{{Veilleux} {et~al.}(2020){Veilleux}, {Maiolino}, {Bolatto}, \&
  {Aalto}}]{Veilleux+20}
{Veilleux}, S., {Maiolino}, R., {Bolatto}, A.~D., \& {Aalto}, S. 2020, \aapr,
  28, 2, \dodoi{10.1007/s00159-019-0121-9}

\bibitem[{{Vijayan} {et~al.}(2020){Vijayan}, {Kim}, {Armillotta}, {Ostriker},
  \& {Li}}]{Vijayan+20}
{Vijayan}, A., {Kim}, C.-G., {Armillotta}, L., {Ostriker}, E.~C., \& {Li}, M.
  2020, \apj, 894, 12, \dodoi{10.3847/1538-4357/ab8474}

\bibitem[{{Wentzel}(1974)}]{Wentzel74}
{Wentzel}, D.~G. 1974, \araa, 12, 71,
  \dodoi{10.1146/annurev.aa.12.090174.000443}

\bibitem[{{Xu} {et~al.}(2022){Xu}, {Heckman}, {Henry}, {Berg}, {Chisholm},
  {James}, {Martin}, {Stark}, {Aloisi}, {Amor{\'\i}n}, {Arellano-C{\'o}rdova},
  {Bordoloi}, {Charlot}, {Chen}, {Hayes}, {Mingozzi}, {Sugahara}, {Kewley},
  {Ouchi}, {Scarlata}, \& {Steidel}}]{XuHeckman22}
{Xu}, X., {Heckman}, T., {Henry}, A., {et~al.} 2022, \apj, 933, 222,
  \dodoi{10.3847/1538-4357/ac6d56}

\bibitem[{{Yan} \& {Lazarian}(2002)}]{Yan&Lazarian02}
{Yan}, H., \& {Lazarian}, A. 2002, \prl, 89, 281102,
  \dodoi{10.1103/PhysRevLett.89.281102}

\bibitem[{{Zhang} {et~al.}(2017){Zhang}, {Thompson}, {Quataert}, \&
  {Murray}}]{Zhang+17}
{Zhang}, D., {Thompson}, T.~A., {Quataert}, E., \& {Murray}, N. 2017, \mnras,
  468, 4801, \dodoi{10.1093/mnras/stx822}

\bibitem[{{Zweibel}(2017)}]{Zweibel17}
{Zweibel}, E.~G. 2017, Physics of Plasmas, 24, 055402,
  \dodoi{10.1063/1.4984017}

\end{thebibliography}
\bibliographystyle{aasjournal}

\end{CJK*}
\end{document}